\documentclass[a4paper,11pt]{article}
\pdfoutput=1 

\usepackage{jheppub} 

\usepackage[T1]{fontenc} 

\usepackage{times}
\usepackage{graphicx}
\usepackage{amssymb}
\usepackage{xspace}
\usepackage{caption}
\usepackage{subcaption}
\usepackage{slashed}

\usepackage{soul}

\def\be{\begin{eqnarray}}
\def\ee{\end{eqnarray}}

\newcommand{\mt}[1]{\textrm{\tiny #1}}

\def\pt{{p_\mt{T}}}

\def\Raa{R_{\mt{AA}}}
\def\Rcp{R_{\mt{CP}}}

\def\aSC{{\kappa_{\rm sc}}}

\def\w{{\bf w}}
\def\v{{\bf v}}

\def\e{{\bf e}}

\newcommand{\pythia}{{\sc Pythia}\xspace}

\def\fig#1{{Fig.~\ref{#1}}}
\def\eq#1{{eq.~(\ref{#1})}}

\def\RAA{\Raa}

\def\RCP{\Rcp}

\def\Nob{\Psi^{\rm subleading}_\pt (r)}

\def\mpt{\slashed{p}^\parallel_\mt{T}}

\title{Angular Structure of Jet Quenching Within a Hybrid Strong/Weak Coupling Model}

\author[a,b]{Jorge Casalderrey-Solana,}
\author[c]{Doga Can Gulhan,}
\author[d,e,f]{Jos\'e Guilherme Milhano,}
\author[b]{Daniel Pablos,}
\author[g]{Krishna Rajagopal}

\affiliation[a]{
Rudolf Peierls Centre for Theoretical Physics, University of Oxford, 1 Keble Road, Oxford OX1 3NP, United Kingdom}

\affiliation[b]{
Departament de F\'\i sica Qu\`antica i Astrof\'\i sica \&  Institut de Ci\`encies del Cosmos (ICC), Universitat de Barcelona, Mart\'{\i}  i Franqu\`es 1, 08028 Barcelona, Spain}

\affiliation[c]{CERN, EP Department, CH-1211 Geneva 23, Switzerland}

\affiliation[d]{CENTRA, Instituto Superior T\'ecnico, Universidade de Lisboa, Av.~Rovisco Pais, P-1049-001 Lisboa, Portugal}
\affiliation[e]{Laborat\'orio de Instrumenta\c c\~ao e F\'isica Experimental de Part\'iculas (LIP), Av. Elias Garcia 14-1, P-1000-149 Lisboa, Portugal}
\affiliation[f]{Theoretical Physics Department, CERN, Geneva, Switzerland}

\affiliation[g]{Center for Theoretical Physics, Massachusetts Institute of Technology, Cambridge, MA 02139 USA}

\emailAdd{jorge.casalderreysolana@physics.ox.ac.uk}

\emailAdd{dgulhan@mit.edu}

\emailAdd{guilherme.milhano@tecnico.ulisboa.pt}

\emailAdd{dpablos@fqa.ub.edu}

\emailAdd{krishna@mit.edu}

\preprint{{\footnotesize ICCUB-16-033, CERN-TH-2016-201, MIT-CTP-4830}}

\abstract{
Within the context of a hybrid strong/weak coupling model of jet quenching, we study the 
modification of the angular distribution of the energy within jets in 
heavy ion collisions, as partons within jet showers lose energy
and get kicked as they traverse the strongly coupled plasma
produced in the collision.  
To describe 
the dynamics transverse to the jet axis,  we add 
the effects of transverse momentum broadening 
into our hybrid construction, introducing a parameter $K\equiv \hat q/T^3$
that governs its magnitude. We show that, because of the quenching
of the energy of partons within a jet, even when $K\neq 0$ the jets
that survive with some specified energy in the final state are narrower
than jets with that energy in proton-proton collisions. For this reason,
many standard observables are rather
insensitive to $K$.
We
propose a new differential jet shape ratio observable in which
the effects of transverse momentum broadening are apparent.
We also analyze the response of the medium to the passage
of the jet through it, noting that the momentum lost by the jet
appears as the momentum of a wake in the medium. After 
freezeout this wake becomes soft particles with a broad angular
distribution but with net momentum in the jet direction, meaning that
the wake contributes to what is reconstructed as a jet.  This effect must
therefore be included in any description of the angular structure
of the soft component of a jet.
We show that the particles coming from the response of the medium to the momentum
and energy deposited in it leads to a correlation between the momentum of soft particles
well separated from the jet in angle with the direction of the jet momentum, and find
qualitative but not quantitative agreement with experimental data on observables designed
to extract such a correlation.
More generally, by confronting the results that we obtain upon introducing transverse momentum
broadening and the response of the medium to the jet 
with available jet data, we highlight the importance of these processes for
understanding the internal, soft, angular structure of high energy jets.}

\begin{document} 
\maketitle
\flushbottom

\section{Introduction}
\label{sec:intro}

High energy heavy ion collisions provide a unique opportunity to explore the properties of 
hot, deconfined, strongly interacting matter, called quark-gluon plasma (QGP). The study of these collisions at the
Relativistic Heavy Ion Collider (RHIC) and at the Large Hadron Collider (LHC) 
has demonstrated that matter at temperatures 
above the crossover between hot hadronic matter
and hotter QGP 
exhibits
strong collective phenomena~\cite{Adcox:2004mh,Arsene:2004fa,Back:2004je,Adams:2005dq, Aamodt:2010pa,ATLAS:2011ah,Chatrchyan:2012ta}
which can be described successfully by  hydrodynamic simulations  of the rapid expansion and cooling of the initially lumpy droplets of matter
produced in the collisions~\cite{Huovinen:2001cy,Teaney:2001av,Hirano:2005xf,Romatschke:2007mq,Luzum:2008cw,Schenke:2010rr,hiranoLHC,Gale:2012rq,Shen:2014vra,
Shen:2014nfa,
Bernhard:2016tnd}.  
Such strong collectivity has also recently been observed in smaller colliding systems, including 
p-Pb, p-p or $^3{\rm He}$-Au~\cite{Khachatryan:2010gv,CMS:2012qk,Abelev:2012ola,Aad:2012gla,Adare:2013piz,Chatrchyan:2013nka,Khachatryan:2015waa,Adare:2015ctn,Aad:2015gqa,Khachatryan:2016ibd,Khachatryan:2016txc}, for which hydrodynamic simulations also seem to be successful~\cite{Bozek:2011if,Nagle:2013lja,Schenke:2014zha,Kozlov:2014fqa,Romatschke:2015gxa,Bozek:2015qpa,Habich:2015rtj,Koop:2015trj}.  The applicability of hydrodynamics from early times in the evolution and for small systems suggests that the matter formed in these 
ultrarelativistic collisions is a strongly coupled liquid. 
Support for this picture comes from analyses of collisions in strongly coupled gauge theories
with a dual holographic description which show that collisions of objects with transverse size $R$ produce
a droplet of strongly coupled liquid that can be described hydrodynamically as long as the
collisions are energetic enough that the temperature of the liquid at the time that it hydrodynamizes, $T_{\rm hydro}$,
satisfies $RT_{\rm hydro} \gtrsim 1$~\cite{Chesler:2015bba,Chesler:2016ceu} and 
that in these collisions and in the collisions of objects that are infinite in transverse extent~\cite{Chesler:2010bi,Casalderrey-Solana:2013aba,Chesler:2013lia} hydrodynamization occurs at a time
of order $1/T_{\rm hydro}$ after the collision.

The discovery that the QGP that filled the microseconds-old universe and that is recreated in collisions
at RHIC and the LHC is a strongly coupled liquid challenges us to understand how such a liquid
emerges from an asymptotically free gauge theory. When probed at very
short length scales, the strongly coupled QGP of QCD (unlike the strongly coupled
plasmas in theories with holographic descriptions) must consist of weakly coupled quarks and gluons.
This makes constraining the microscopic nature of QGP via studying
its interaction with energetic probes an important and interesting long-term goal. 
Some of the most salient examples of such probes are QCD jets. 
As a partonic jet shower propagates through the strongly coupled
plasma created in a heavy ion collision, it suffers a strong process of energy loss as a result of its interactions with the plasma.  The partons in the jet also pick up momentum transverse to their
direction of motion as they are jostled during their passage through the medium.
These interactions lead to a reduction in the  jet energy (or quenching) 
and to modifications  of the properties of jets produced in heavy ion
collisions relative to those of their counterparts produced in proton-proton collisions, that propagate in vacuum.
These interactions also result in the transfer of energy and momentum to the plasma: the jets
create a wake as they lose energy. 
By pursuing a large suite of jet measurements, the different LHC collaborations have observed 
strong modification of different jet observables  in heavy ion collisions~\cite{Aad:2010bu,Chatrchyan:2011sx,Chatrchyan:2012nia,Chatrchyan:2012gt,Chatrchyan:2012gw,Aad:2012vca,Raajet:HIN,Aad:2013sla,Chatrchyan:2013kwa,Abelev:2013kqa,Chatrchyan:2013exa,Chatrchyan:2014ava,Aad:2014wha,Aad:2014bxa,Adam:2015ewa,Adam:2015doa,Aad:2015bsa,Khachatryan:2015lha,Khachatryan:2016erx},  making jets promising probes for medium diagnostics. 
The first experimental constraints on jet quenching came from hadronic measurements at RHIC~\cite{Adcox:2001jp,Adler:2002xw,Adler:2002tq}. Analyses of
jets themselves and their modification 
are also being performed at RHIC~\cite{Ploskon:2009zd,Perepelitsa:2013faa,Adamczyk:2013jei,Jacobs:2015srw} and are one of the principal scientific goals of the
planned sPHENIX detector~\cite{Adare:2015kwa}.

To fully exploit their potential as tomographic probes, a detailed understanding of the interactions of jet showers 
with hot QCD matter is needed. 
A complete theoretical description of these processes is a challenging task, due to the multi-scale nature of jet probes. 
On the one hand, the production of jets and the processes via which an initial hard parton fragments into a shower
are governed by short distance physics that is weakly coupled.
On the other hand, the interaction of jets with the medium, as well as the dynamics of softer components
within jets, are sensitive to the strongly coupled dynamics of the plasma at scales of order its temperature.

 One class of approaches toward making progress on this difficult theoretical problem that has been pursued intensively
 starts from the premise that the dynamics of the plasma itself are weakly coupled, as if the temperature of the plasma were
asymptotically large, and from the premise that the interactions of energetic partons and jets  with the plasma are also governed
entirely by weakly coupled physics.  
(See Refs.~\cite{Jacobs:2004qv,CasalderreySolana:2007zz,Majumder:2010qh,Ghiglieri:2015zma,Blaizot:2015lma,Qin:2015srf} for reviews.)
Based on these approaches, 
Monte Carlo tools for analyzing jet observables are being developed~\cite{Zapp:2008af,Zapp:2008gi,Armesto:2009fj,Schenke:2009gb,Lokhtin:2011qq,Zapp:2012ak,Zapp:2013vla,Zapp:2013zya} 
and
many phenomenological studies of jets in medium have been confronted with LHC measurements
of a variety of jet observables~\cite{Vitev:2009rd,CasalderreySolana:2010eh,Qin:2010mn,Young:2011qx,He:2011pd,CasalderreySolana:2011rq,Renk:2012cx,Neufeld:2012df,Renk:2012cb,Dai:2012am,Apolinario:2012cg,Zapp:2012ak,Wang:2013cia,Ma:2013pha,Huang:2013vaa,Senzel:2013dta,Zapp:2013vla,Zapp:2013zya,Ramos:2014mba,Renk:2014lza,Perez-Ramos:2014mna,Chien:2015vja,Huang:2015mva,Chien:2015hda,Milhano:2015mng,Zhang:2015trf,Chang:2016gjp,Mueller:2016gko,Chen:2016vem}, including intrajet observables like those
that we shall focus on~\cite{CasalderreySolana:2010eh,CasalderreySolana:2011rq,Ramos:2014mba,Perez-Ramos:2014mna,Chien:2015hda,Zhang:2015trf,Chang:2016gjp,Mueller:2016gko,Chen:2016vem}.

However, the observation that QGP  is a strongly coupled liquid tells us that physics at scales of order its temperature
is governed by strong coupling dynamics.
This realization has lead to many fruitful connections between the physics of the QCD plasma 
and the gauge/gravity duality~\cite{Maldacena:1997re}. This technique allows us rigorous and quantitative 
access to nonperturbative, strongly coupled, physics --- including thermodynamics, transport coefficients, hydrodynamics,
thermalization, response to hard probes and other real time dynamics far from equilibrium  --- 
in a large family of non-abelian gauge theory plasmas
that have a dual holographic description in terms of a black hole spacetime in a gravitational theory with one higher dimension. 
Although the current formulation of the duality has not been shown to apply to QCD, the study of the plasmas in gauge theories that do have a holographic description has led to many insights into the dynamics of hot deconfined matter in QCD. (See 
Refs.~\cite{CasalderreySolana:2011us,DeWolfe:2013cua,Chesler:2015lsa} for reviews.) 
Within this context, there have been many interesting studies that address varied aspects of the interaction between high energy probes and strongly 
coupled plasma~\cite{Herzog:2006gh,Liu:2006ug,CasalderreySolana:2006rq,Gubser:2006bz,Liu:2006nn,Liu:2006he,Gubser:2006nz,Chernicoff:2006hi,CasalderreySolana:2007qw,Chesler:2007an,Gubser:2007ga,Chesler:2007sv,Hofman:2008ar,Gubser:2008as,Hatta:2008tx,Dominguez:2008vd,Chesler:2008wd,Chesler:2008uy,D'Eramo:2010ak,Arnold:2010ir,Arnold:2011qi,Arnold:2011vv,Chernicoff:2011xv,Chesler:2011nc,Arnold:2012uc,Arnold:2012qg,Chesler:2013urd,Ficnar:2013wba,Ficnar:2013qxa,Chesler:2014jva,Rougemont:2015wca,Chesler:2015nqz,Casalderrey-Solana:2015tas,Rajagopal:2016uip}.
However, in all the examples that are currently directly accessible via gauge/gravity duality, the gauge theory 
remains strongly coupled in the ultraviolet, which limits the direct applicability of these results for phenomenological applications.

To address the multifaceted dynamics of QCD jets in strongly coupled plasma more fully, in 
Refs.~\cite{Casalderrey-Solana:2014bpa,Casalderrey-Solana:2015vaa} we introduced a phenomenological hybrid 
strong/weak coupling approach to analyzing jet quenching. In this approach, we treat different physics processes of relevance for the interaction of developing jet showers with the quark gluon plasma differently. In our model, the production and evolution of the jet shower is treated perturbatively, 
because the physics governing these processes is expected to be weakly coupled.
And, we model
the interaction between each of the partons formed in the shower with the medium 
using the result for the rate of energy loss of an energetic quark in strongly coupled
plasma obtained via holographic calculations in Refs.~\cite{Chesler:2014jva,Chesler:2015nqz}. 
The Monte Carlo implementation of this hybrid model has been successful in describing 
LHC measurements of a large suite of single jet, dijet and photon+jet 
observables~\cite{Casalderrey-Solana:2014bpa,Casalderrey-Solana:2015vaa} and has
been used to make predictions for more such observables and for Z+jet observables~\cite{Casalderrey-Solana:2015vaa}.
To date, the model has contained only a single free parameter, with all its successes and predictions having been obtained
after fitting this parameter to a single measured data point.

In this paper, after reviewing the construction of our hybrid model briefly in Section~\ref{MC} we will extend 
the model in order to be able to use it to address the angular distribution of the energy within a jet and its modification via
its passage through the plasma, as well as the angular distribution of the energy lost by jets
during their passage through the plasma.
To do so, we will supplement our model with two important physics processes which 
were absent in our previous implementations. First, in Section~\ref{sec:Section2}  we add ``transverse momentum broadening'', namely 
the deflection to the direction of propagation of partons as a result of
the exchange of momentum transverse to their direction of motion between the parton and the medium.
We assume a Gaussian distribution of the magnitude of the momentum transfer and introduce
one further model parameter to specify the width of the Gaussian.
Second, in Section~\ref{sec:Section3}  we add a simplified model for the collective response of
the medium to the passage of the jet, namely the wake in the plasma
that carries the energy and momentum lost by the jet and deposited
in the plasma.  We shall frequently refer to this as the ``backreaction of the medium''.
We shall not actually focus on the form of the wake
itself, focusing instead on the soft hadrons in the final state that
result from the hadronization of the plasma including the wake and
that carry the net momentum and energy lost by the jet.
We make simplifying assumptions that allow us to add a representation of
the effects of the wake on the final state hadrons  that respects energy and 
momentum conservation, without introducing any new parameters
into our hybrid model.  Our principal assumption is that the energy
lost by the jet thermalizes quickly, in the sense that it quickly
becomes a hydrodynamic wake in the plasma that carries the ``lost''
energy and momentum, which in turn after freezeout becomes soft particles spread over
a calculable and wide range of angles relative to the jet.
This is an immediate and natural consequence of strong coupling dynamics~\cite{Chesler:2007an}.
Something similar can happen at weak coupling even though the energy is initially lost by
gluon radiation because these radiated gluons can experience a cascade of reinteractions that converts
the energy into soft particles at large angles~\cite{Blaizot:2013hx,Blaizot:2014ula,Kurkela:2014tla,Fister:2014zxa,Blaizot:2014rla,Blaizot:2015jea,Iancu:2015uja}.
The effects of both transverse momentum broadening~\cite{CasalderreySolana:2010eh,Qin:2010mn,Chang:2016gjp,Mueller:2016gko,Chen:2016vem} and the
backreaction of the medium~\cite{Neufeld:2011yh,Wang:2013cia,Tachibana:2014lja,Floerchinger:2014yqa,He:2015pra,Tachibana:2015qxa,Cao:2016gvr} on jet observables 
have also been
studied within the context of perturbative energy loss mechanisms.

After adding broadening and backreaction to our hybrid model, over the course of
Sections \ref{sec:Section2} and \ref{sec:Section3} we will assess their effects on measurements of
different jet and dijet observables that are sensitive to the angular structure of jets,
including the dependence of jet suppression on the reconstruction parameter $R$ which
controls the angular size of the jets that are found and reconstructed in a sample of events, 
several observables based upon the jet shape including a new
differential jet shape ratio that we introduce, the dijet acoplanarity, and
the balance of momenta along the dijet axis carried by particles
in the event with a given momentum.  We also return to the jet fragmentation
function, as we wish to see how it is modified by the soft particles coming
from the backreaction of the medium.

One of our significant findings is that jets with a given energy that were
produced in a heavy ion collision and so have had to traverse a droplet
of QGP have a narrower hard jet core than jets with the same energy
that were produced in vacuum, even if a substantial degree of transverse momentum
broadening is turned on. 
This is a consequence of the fact that wider jets 
 typically contain more, and less hard, partonic fragments and lose more energy than narrower jets with the same energy. 
 The same phenomenon has been observed in calculations of jet quenching that are entirely done at strong 
 coupling~\cite{Chesler:2015nqz,Rajagopal:2016uip} and 
 in Monte Carlo calculations of radiative energy loss that are entirely done at weak coupling~\cite{Milhano:2015mng}. 
 We find that this observation leads to two unexpected consequences. First, 
 jets with a larger reconstruction parameter $R$ 
 are more suppressed (the suppression factor $\RAA$ of jets is pushed downward)
 at least for moderate $R$.
 Second, the intrajet angular distribution of energy in
 the quenched jets is remarkably independent of the amount of transverse momentum broadening.
 Note that we are only able to see that energy loss causes jets with a given energy to be narrower in heavy ion collisions
 than in proton-proton collisions because we have incorporated event-by-event (maybe better phrased jet-by-jet) variations in the
 fragmentation pattern of jets; this physical effect, and its striking consequences, are absent in approaches based on analyzing how an average
 jet is modified by passage through the plasma, as for example in Refs.~\cite{Chang:2016gjp,Chen:2016vem}.
 
Although we find that both the standard jet shape and the dijet acoplanarity are remarkably independent of the amount of transverse
momentum broadening,
we are able to construct a new observable --- essentially a jet shape ratio that is partially differential in $\pt$  ---
 that  {\it is} sensitive to the amount of transverse momentum broadening that we introduce.  This points out a path toward
 detecting experimental evidence for the effects of transverse momentum broadening within a jet, which 
 would be a very important first step toward using jets to resolve the microscopic structure of
 quark-gluon plasma.

Our last significant finding is that when we implement the collective response of the medium to the passage of the jet, the
energy lost from the jet ends up in the form of soft particles separated from the jet axis by very large angles and is 
in qualitative agreement with experimental measurements of observables that are referred to as ``missing-$\pt$''
distributions which have recently been reported by CMS~\cite{Khachatryan:2015lha}.  
Careful comparison between our calculations of these and several other observables
and experimental data in Section~\ref{sec:Section3} indicates that our simplified treatment of the wake produces
slightly too many very soft particles ($\pt<2$~GeV)  at large angles and not enough particles 
with momenta in the $2-4$ GeV range. This is not entirely unexpected because
the approximations via which we treat the particles coming from the wake
are reliable only for particles with momenta that are not a lot larger
than the freezeout temperature, but it may also be an indication that the wake in the plasma does not actually thermalize as fully as we assume in our simplified treatment.

 



We close in Section~\ref{sec:discussion}  by discussing the various results of our analyses, identifying further improvements
of our implementation of the in-medium dynamics of jets in strongly coupled plasma for the future, and looking ahead at
the path toward using jets to resolve the microscopic properties of quark-gluon plasma.

\section{\label{MC} Brief Summary of the Hybrid Model}

In this Section, we provide a brief description of 
the hybrid model which we will employ to describe the modification of
jets produced in heavy ion collisions that propagate through a droplet of hot matter
relative to those produced in proton-proton collisions that propagate in vacuum.
A more detailed account of the model may be found in Refs.~\cite{Casalderrey-Solana:2014bpa,Casalderrey-Solana:2015vaa}.

The main motivation for introducing this model is to separate the strongly coupled dynamics of quark-gluon plasma itself
and of interactions between it and partons plowing through it
from the weakly coupled dynamics governing the production, 
showering and relaxation of virtuality of high energy QCD jets.
Since for any parton that showers and forms
a jet the initial virtuality of the parton is much larger than any
scale associated with the medium, of order its temperature $T$,
the first assumption of the model is that the evolution of the jet 
proceeds as in vacuum, with the branching of the parton shower 
unmodified by the presence of the strongly interacting plasma. 
Upon making this assumption, the modification of jet showers is 
only due to the interaction of each of the partons in the jet with
the strongly coupled medium.
After associating each parton in the jet 
with a life-time determined via a formation time argument~\cite{CasalderreySolana:2011gx,Casalderrey-Solana:2014bpa,Casalderrey-Solana:2015vaa}, 
we compute the energy lost by each parton as it propagates in the strongly coupled plasma. 
Since the interactions of each of these partons with the medium is sensitive to the medium scale, 
the rate of energy loss is controlled by strongly coupled dynamics.  The second assumption of the model is that the rate
at which a parton loses energy can be 
modeled by the rate of energy loss of light quark jets
in the strongly coupled plasma of  $\mathcal{N}=4$ supersymmetric Yang-Mills (SYM)
theory which has been computed via holography and is given by~\cite{Chesler:2014jva,Chesler:2015nqz}
\be
\label{CR_rate}
\left. \frac{d E}{dx}\right|_{\rm strongly~coupled}= - \frac{4}{\pi} E_{\rm in} \frac{x^2}{x_{\rm stop}^2} \frac{1}{\sqrt{x_{\rm stop}^2-x^2}}\,, 
\ee
with $x_{\rm stop}$ the distance over which the light quark jet would lose all of its energy if it 
propagated through plasma at a constant temperature $T$.  
In $\mathcal{N}=4$ SYM theory, jets with a given initial energy $E_{\rm in}$ can have
a wide range of initial opening angles, with the narrower jets having larger values of $x_{\rm stop}$~\cite{Chesler:2015nqz}.
There is a minimum possible initial opening angle, corresponding to the maximum possible $x_{\rm stop}$ 
for jets with a given initial energy~\cite{Chesler:2015nqz} that was computed holographically in
Refs.~\cite{Chesler:2008uy,Chesler:2011nc,Ficnar:2013wba}
and is given by 
\be
\label{CR_xstop}
\quad \quad x_{\rm stop}= \frac{1}{2\aSC}\frac{E_{\rm in}^{1/3}}{T^{4/3}}\ ,
\ee
with $\aSC=1.05 \lambda^{1/6}$.
In the hybrid model, we apply (\ref{CR_rate}) parton-by-parton to each parton in a QCD DGLAP shower as described by \pythia~\cite{Sjostrand:2007gs},
rather than attempting to use the ${\cal N}=4$~SYM jets that lose energy at the rate (\ref{CR_rate}) themselves as models
for QCD jets in plasma, as in Ref.~\cite{Rajagopal:2016uip}.  
Because we apply (\ref{CR_rate}) to individual partons, we use the form (\ref{CR_xstop}) for $x_{\rm stop}$ appropriate
for the skinniest possible ${\cal N}=4$ SYM jets.
We shall further assume that the most salient differences between the strongly coupled limit of ${\cal N}=4$ SYM theory and 
QCD can be incorporated via varying
the value of $\aSC$, which becomes the only fitting parameter 
of the hybrid model formulated in Refs.~\cite{Casalderrey-Solana:2014bpa,Casalderrey-Solana:2015vaa}.

We have implemented this hybrid 
model into a Monte Carlo simulation in which hard 
jets, showering as described by \pythia~\cite{Sjostrand:2007gs}, are embedded within a droplet of hot matter produced in 
a heavy ion collision, expanding and cooling as described by relativistic viscous hydrodynamics.
To generate the hard QCD jet shower, we employ \pythia 8.183~\cite{Sjostrand:2007gs}, allowing the 
DGLAP shower to evolve down to a minimum transverse momentum of $\pt^{\rm min}=1$~GeV. 
We distribute these hard events  in the transverse plane of the heavy ion collisions according the number of binary nucleon-nucleon collisions. 
The trajectory of each jet is tracked from the generation point as it propagates through the expanding cooling plasma 
until the jet reaches a region where the temperature has dropped below a  temperature 
$T_c$, below which we assume no further energy loss occurs. This $T_c$ is not sharply defined but
it should presumably be near the crossover between quark-gluon plasma and hadronic matter
and we therefore vary its value over the range $145< T_c< 170$~MeV.  Seeing how our results vary as we vary $T_c$ over this range  
serves as a gauge of some of the uncertainties in our model. 
The energy loss rate is computed via \eq{CR_rate}, with $x_{\rm stop}$ evaluated according to eq.~(\ref{CR_xstop})
at the local temperature 
as given by the hydrodynamic simulations of Refs.~\cite{Shen:2014vra,Shen:2014nfa}. 
Flow effects are taken into account by evaluating the rate of energy loss 
in the local fluid rest frame, as explained in Ref.~\cite{Casalderrey-Solana:2015vaa}. 
In Refs.~\cite{Casalderrey-Solana:2014bpa,Casalderrey-Solana:2015vaa}, we have fitted
the value of the parameter $\aSC$ to a single measurement, the suppression of the number of jets with
one transverse momentum in LHC heavy ion collisions with one centrality, and have then successfully
confronted this hybrid model with measurements of many single jet, dijet and $\gamma$-jet observables
as functions of jet transverse momentum and collision centrality and made predictions for many
further measurements of these types.


In the next Sections, we shall extend our implementation of the hybrid model  
to include two new physics processes, transverse momentum broadening in Section~\ref{sec:Section2} and the response of the medium to the jet in Section~\ref{sec:Section3}, 
and will evaluate their consequences for intra-jet observables, in particular those
related to the angular structure of jets.

\section{Transverse Kicks and Jet Broadening}
\label{sec:Section2}

Previously in Refs.~\cite{Casalderrey-Solana:2014bpa,Casalderrey-Solana:2015vaa}, all the effects of
the strongly coupled medium on the properties of the jets arise as a consequence of the energy lost
by the partons in the jet shower as they plow through the medium.  In this Section, we augment our hybrid model by adding a second
physical process, and hence a second free parameter, namely the kicks transverse to their direction of
motion that the partons in the jet receive as a consequence of plowing through the medium.  This process
has long been referred to as ``transverse momentum broadening'' based upon the expectation that the consequence
of the kicks in random transverse directions received by the many partons in a jet shower will be broadening of the jet.
In perturbative calculations, transverse momentum broadening arises from the
multiple soft exchanges of momentum that a parton suffers as it traverses a medium leading to a random change in its momentum 
and in particular providing the parton with some additional momentum perpendicular to its original
direction of propagation.

If 
the energetic parton suffers multiple soft exchanges as it traverses the medium, the distribution of the momentum transferred via this stochastic process
is well approximated as Gaussian.
As a consequence, the transverse momentum distribution of 
partons that have traversed a medium of length $L$ is approximately Gaussian 
with a width that scales with the medium length, $Q^2_\perp= \hat q L$. The quantity $\hat q$ that arises here is 
called the momentum broadening parameter; this property of the medium
codifies the typical squared momentum that the medium transfers to the probe per unit length.   It has dimension 3
and in a plasma in thermal equilibrium with temperature $T$ it is proportional to $T^3$, up to a possible logarithmic dependence
on the ratio of the parton energy to $T$.
In perturbative calculations of energy loss via gluon radiation, the medium parameter $\hat q$ also determines
the intensity of the gluon radiation induced by the medium, and hence is related to energy loss as well
as to momentum broadening.  In our model, we introduce $\hat q$ only as a way of parametrizing
momentum broadening.

In the strongly coupled plasma of ${\cal N}=4$ SYM theory, transverse momentum broadening has
been calculated holographically for both heavy quarks~\cite{CasalderreySolana:2006rq,Gubser:2006nz,CasalderreySolana:2007qw} 
and massless quarks~\cite{Liu:2006ug,D'Eramo:2010ak}.  Although there is no notion of scattering centers and no notion
of multiple discrete transfers of momentum in the strongly coupled limit, both heavy and massless quarks pick up
transverse momentum as they propagate through the hot strongly coupled liquid and the resulting transverse momentum
distribution is Gaussian with a width $Q^2_\perp= \hat q L$ with $\hat q \propto \sqrt{\lambda}T^3$, with $\lambda$ the 't~Hooft coupling.
However, unlike at weak coupling there is no strong correlation between the dynamics responsible for transverse momentum
broadening and that responsible for parton energy loss.

 
 We shall introduce broadening into our hybrid model by assuming a Gaussian distribution for the transverse momentum picked
up by each parton in the shower, with the momentum squared picked up per distance travelled given by
 \be
\label{eq:Kdef}
\hat q = K T^3 
\ee
with $T$ the local temperature of the medium at the location of a given
parton at a given time and with $K$ a theory-dependent constant that
we shall treat as a free parameter that should ultimately be determined
via fitting to data.
Since the medium is dynamical, with longitudinal and transverse expansion, $T$ and hence $\hat q$ varies
 with position and time. We shall obtain this dependence from the hydrodynamic simulation
 of the expanding cooling droplet of plasma in our hybrid model.

For massless or very energetic particles, in $\mathcal{N}=4$ SYM theory in the large number of colors ($N_c$) limit the value of $K$
has been calculated holographically~\cite{Liu:2006ug,D'Eramo:2010ak} and is given by
$K= K_{\mathcal {N}=4}\simeq 24$ for $\lambda=10$, a value of the 't Hooft coupling $\lambda=g^2 N_c$ that corresponds to $g^2/4\pi\simeq 0.27$ 
for $N_c=3$. 
$K_{\mathcal{N}=4}$ is proportional to $\sqrt{\lambda}$ as noted above.  In a large  class of conformal theories
$K_{\rm CFT}/K_{\mathcal{N}=4} = \sqrt{s_{\rm CFT}/s_{\mathcal{N}=4} }$~\cite{Liu:2006he}, $s$ being the entropy density. This suggests that
$K_{\rm QCD}$ is likely to be less than $K_{\mathcal{N}=4}$ since, at least at weak coupling, $s_{\rm QCD}/s_{\mathcal{N}=4}\simeq 0.40$.  
An alternative approach to gaining an expectation for the likely value of $K_{\rm QCD}$ is to start from
a perturbative analysis of parton energy loss, in which the value of $K$ controls energy loss via gluon
radiation and, via this relation, can be related to experimental observables that are sensitive to parton
energy loss like for example the suppression in the number of high-momentum hadrons in heavy
ion collisions as compared to proton-proton collisions.  The JET Collaboration has pursued this approach~\cite{Burke:2013yra};
the value of $\hat q$ that they have found corresponds to a value of $K$ given by $K_{\rm pert}\simeq 5$.
To date, nobody has extracted a value of $K$ via comparison to data on experimental observables
that are directly sensitive to transverse momentum kicks and jet broadening.  (A recent pioneering
attempt~\cite{Chen:2016vem} 
yields values ranging from 0 to several times larger than the value of $K_{\rm pert}$ obtained in Ref.~\cite{Burke:2013yra}.)
Our goal in this
Section is to introduce transverse momentum broadening into our hybrid model, treating $K$ as
a parameter that should ultimately be determined via comparison between calculations of the 
($K$-dependent) observable consequences of momentum broadening to experimental measurements
of observables that are directly sensitive to this physical process.

In this Section, we will use our hybrid model to analyze the consequences of transverse momentum broadening
for various different jet observables.  Based upon the discussion above, we expect that $K$ actually lies somewhere
around 5 to 20.  But, in order to better understand the consequences of broadening, we shall investigate the effects on 
observables of varying $K$ over the wide range $0\leq K\leq 100$.
Our principal conclusion, after investigating a suite of jet observables, will be that most observables, even those
tailored to measuring the angular structure of jets, are remarkably insensitive to broadening, showing little sensitivity
to $K$ over the full range that we explore.   
This conclusion becomes less surprising once we recall that even if every individual jet is broadened by its
passage through the medium, jets with a given energy can end up being narrower in heavy ion collisions
than in proton-proton collisions.  This happens if wider jets lose more energy than narrower jets (as is the case
in perturbative calculations in QCD~\cite{CasalderreySolana:2012ef} and in holographic calculations of jets in strongly coupled ${\cal N}=4$ SYM 
theory~\cite{Chesler:2015nqz}) and if the probability distribution
for the jet production absent any medium effects is a steeply falling function of jet energy, as is the case.
Recent weakly coupled Monte Carlo calculations of an ensemble of QCD jets in heavy ion collisions
as compared with those in proton-proton collisions~\cite{Milhano:2015mng} and recent holographic analyses of how
propagation through plasma modifies the energy and opening angle distribution of an ensemble
of ${\cal N}=4$ SYM jets with initial distributions as in perturbative QCD~\cite{Rajagopal:2016uip} both provide clear illustrations
of how jets with a given energy can end up narrower in heavy ion collisions even though every individual jet broadens as it propagates through
plasma.  We shall
find the same in our hybrid model.  The resulting insensitivity of jet observables to the value of  $K$ will make it
quite challenging to extract the value of this medium parameter from data.

In Section~\ref{sec:SensitiveObservable} we shall introduce a new observable
which {\it does} exhibit considerable sensitivity to in-medium transverse momentum broadening, 
proposing this observable as a possible route to using future experimental measurements to 
constrain the value of $K$.

\subsection{Introduction of Broadening into the Hybrid Model}
\label{sec:IntroBroadening}

We shall assume that each parton in a jet shower picks up some transverse momentum
as it propagates through the plasma for a time $dt$, with the transverse direction chosen
randomly and with the magnitude of the momentum chosen from a Gaussian distribution
with a width $\hat q \, dt$ with $\hat q$ specified in terms of the local
temperature of the plasma $T$ and the parameter $K$ according to (\ref{eq:Kdef}).
Many previous 
computations of in-medium energy loss and broadening of jets have been 
performed in the limit of a static fluid at rest. 
However, we shall study the interactions of jets with an expanding, cooling, droplet of plasma as described by hydrodynamics. 
As in our previous analysis of jet observables related to parton energy loss~\cite{Casalderrey-Solana:2015vaa}, 
we shall apply results appropriate to a static medium in the local fluid rest frame, meaning in the present case
that we are neglecting any effects of gradients in the fluid on transverse momentum broadening.
This prescription implies that all modifications to the momentum of the parton (loss of longitudinal momentum according to (\ref{CR_rate}) as previously and the
transverse momentum kick that we are introducing here)
are computed in the rest frame of the fluid at the location of the parton at a given time. 
Note that the transverse momentum kicks are transverse to the direction of motion of the parton in the local fluid rest frame, meaning that they
need not be transverse to its momentum in the collision center-of-mass frame.
We relegate the details of the transformation between the fluid frame and the collision frame to Appendix~\ref{tkk}.

Following our Monte Carlo approach, we perform a full simulation of an ensemble of jets. For the analysis performed in this Section, we studied 500,000 jet events,  generated and evolved by \pythia 8.183~\cite{Sjostrand:2007gs}. 
The point of origin in the transverse plane of the hard processes is distributed by sampling the binary collison probability density while initial transverse direction and rapidity are retained from \pythia.

We embed these jets in the hydrodynamic background of Refs.~\cite{Shen:2014vra,Shen:2014nfa}. 
We follow the trajectories of all the partons in the \pythia jet shower as they traverse the hot matter created in the collision, as described in the 
hydrodynamic simulation.   Between the time when each parton is created at a branching event and the
time when each parton itself branches,
we discretize its trajectory and at each point we add to its momentum a random transverse momentum 
chosen according to a Gaussian distribution of width $\hat q\, dt_F$ with $dt_F$ the length of the discretized time interval in the fluid frame. We have 
checked that our results do not depend on our choice of $dt_F$.

\begin{figure}
\centering 
\includegraphics[width=.5\textwidth]{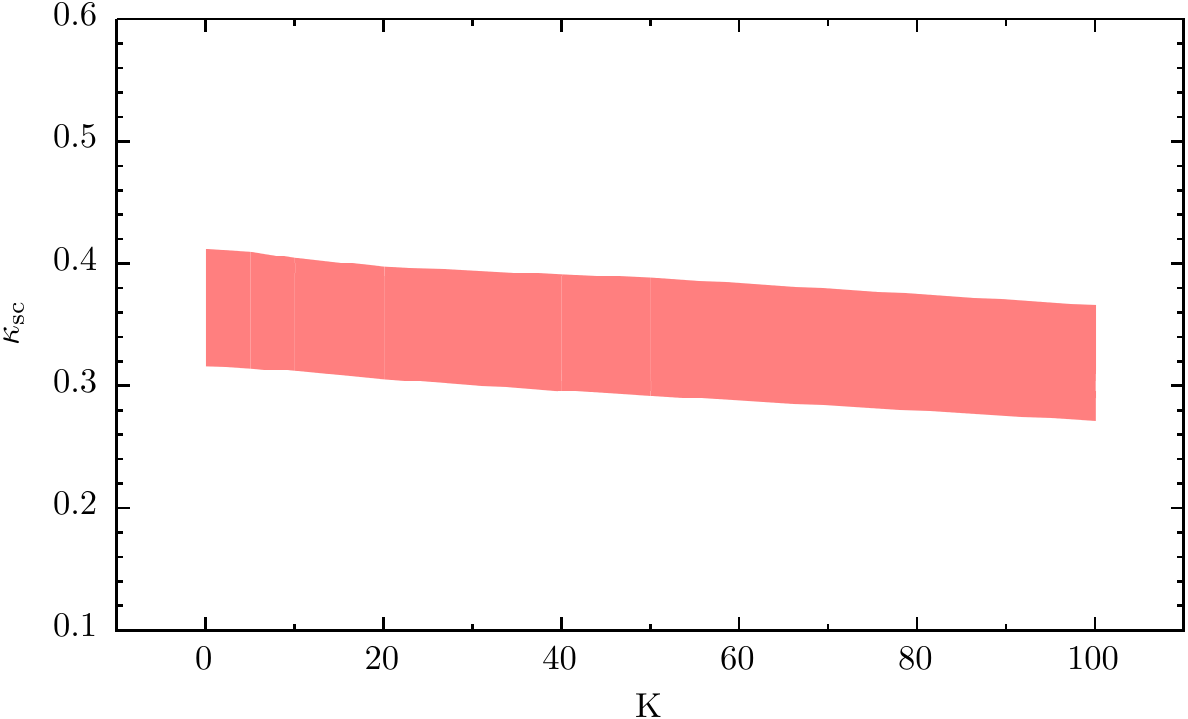}
\caption{\label{Fig:Kvskappa} When we introduce transverse momentum broadening via
a nonzero value of the broadening parameter $K$, this introduces a small increase in 
energy loss.  This means that for each nonzero value of $K$ we need to refit
the value of the 
energy loss parameter $\aSC$ 
to the measured value~\cite{Raajet:HIN} of $\RAA^{\rm jet}$ for
jets with 100~GeV$<p_{\rm T}< 110$~GeV and $-2<\eta<2$, as in Refs.~\cite{Casalderrey-Solana:2014bpa,Casalderrey-Solana:2015vaa}.
The resulting dependence of $\aSC$ on the broadening parameter $K$ is mild.  As in 
Refs.~\cite{Casalderrey-Solana:2014bpa,Casalderrey-Solana:2015vaa}, the width of the band of
values of $\aSC$ in this figure (and the consequent widths of the bands depicting our hybrid
model predictions in subsequent figures) comes both from the error bar on the experimentally measured data point used
to fix $\aSC$ and from
varying the crossover temperature $T_c$ as described in Section~\ref{MC} in order to get some sense of the
systematic uncertainties in the hybrid model. 
}
\end{figure}

As described in detail in Appendix~\ref{tkk}, we implement broadening 
in the local fluid rest frame by  assuming that in this frame the energy
and virtuality of a parton do not change when the parton is kicked; the
only thing that changes due to the tranverse
kicks it receives is the direction of the parton's momentum vector.
Again as described in detail in Appendix~\ref{tkk}, upon boosting
back from the local fluid rest frame to the collision center of
mass frame there will in general be a change to the collision frame energy
of the parton.  It turns out that more often than not in the collision frame
the partons lose a small amount of energy as a consequence of the
transverse kicks they receive. In addition, the transverse kicks they receive
may push some partons outside the jet. (Since the jet shape falls with distance from
the jet axis, although transverse kicks may also push partons into the jet this is less likely.)
These effects together mean that, summed over
the whole jet evolution, the transverse momentum kicks that we are adding 
will result in a slight increase to the overall jet energy loss at a fixed value of $\aSC$.
That is, the dynamics of broadening leads to a small increase in the quenching of jets. 
This means that for each nonzero value of $K$ we need to refit the value of the 
parameter $\aSC$ that, through (\ref{CR_rate}) and (\ref{CR_xstop}), determines the amount of energy loss in our 
hybrid model. We do so in \fig{Fig:Kvskappa}, finding that the effect is small.
For values of $K$ around 5 to 20, the effect of broadening implies a reduction of less than 5\% in the value of $\aSC$ relative to that
reported in our previous work. This effect is much smaller than the uncertainties represented by the width of the band in \fig{Fig:Kvskappa}. 
Even for the extreme value $K=100$, $\aSC$ is only reduced by about 10\%.  

\subsection{Insensitivity of Jet Observables to Broadening}

Having fixed the dependence of the quenching parameter $\aSC$ on the broadening parameter $K$, we can now begin our
exploration  of the effect of broadening on different observables. 
With an ensemble of events in hand, generated by \pythia and modified to include energy
loss as described in Section~\ref{MC} and broadening as described above,
the first step in the calculation of any jet observable is the finding and reconstruction
of jets in each of the events in the ensemble.  We do so using the anti-$k_t$ algorithm~\cite{Cacciari:2008gp} as implemented
in the FastJet package~\cite{Cacciari:2011ma}.  

The first observable that we consider is the suppression factor of jets (namely the ratio $\RAA^{\rm jet}$ of the number of jets with
specified kinematics in heavy ion collisions to the number of jets with the same kinematics  in proton-proton
collisions)
as a function of the jet reconstruction parameter 
$R$ that arises in the anti-$k_t$ algorithm~\cite{Cacciari:2008gp}.  
The anti-$k_t$ algorithm groups particles within $R$ of each other in the $(\eta,\phi)$ plane into what it defines as a single
jet whereas if $R$ were smaller it may reconstruct the same particles as several
smaller narrower jets.  Hence, choosing a larger $R$ translates into reconstructing an ensemble
of jets that tend to be wider in angle.
Because of this, the
dependence of the suppression factor $\RAA^{\rm jet}$ on $R$ is often considered a proxy
for modification of the angular structure of jets as a consequence of their interaction with the medium.
A naive expectation, then, would be that turning on transverse momentum broadening should make
jets broader and that this in turn should leave some imprint in the $R$-dependence of $\RAA^{\rm jet}$,
increasing the suppression for smaller $R$.  This is not at all what we observe.

\begin{figure}[t]
\centering 
\begin{tabular}{cc}
\includegraphics[width=.5\textwidth]{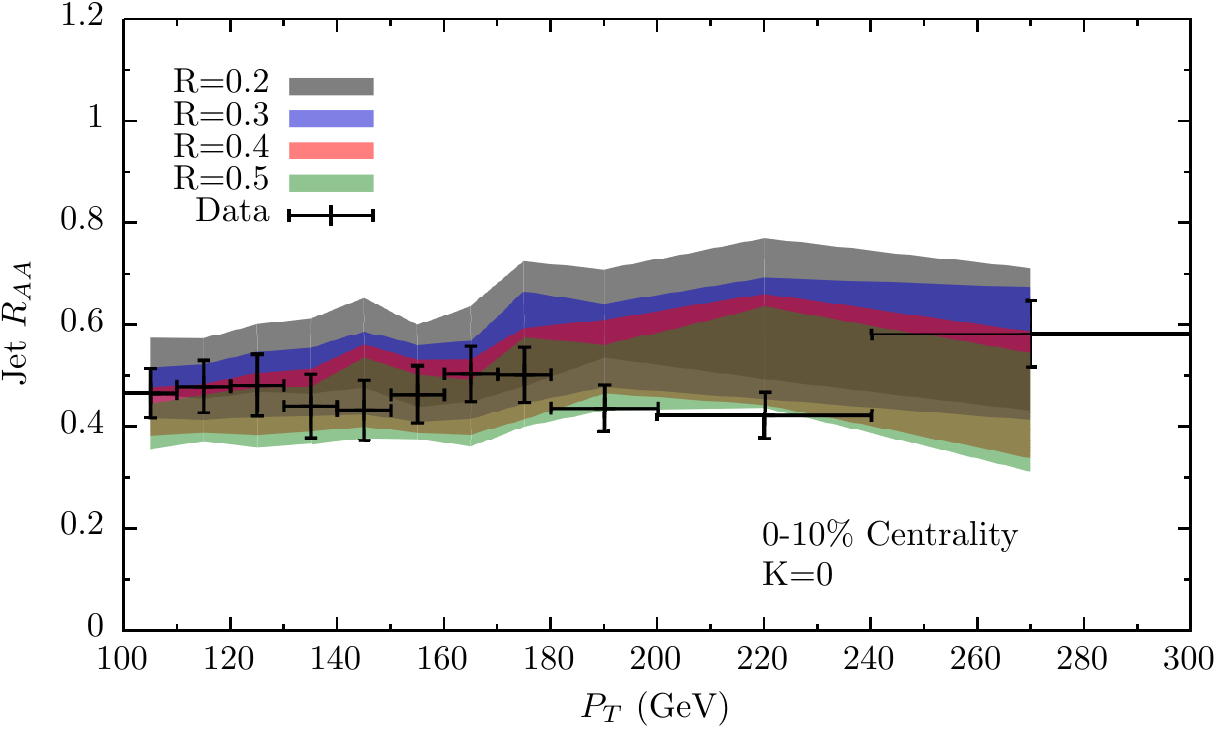}
&
\includegraphics[width=.5\textwidth]{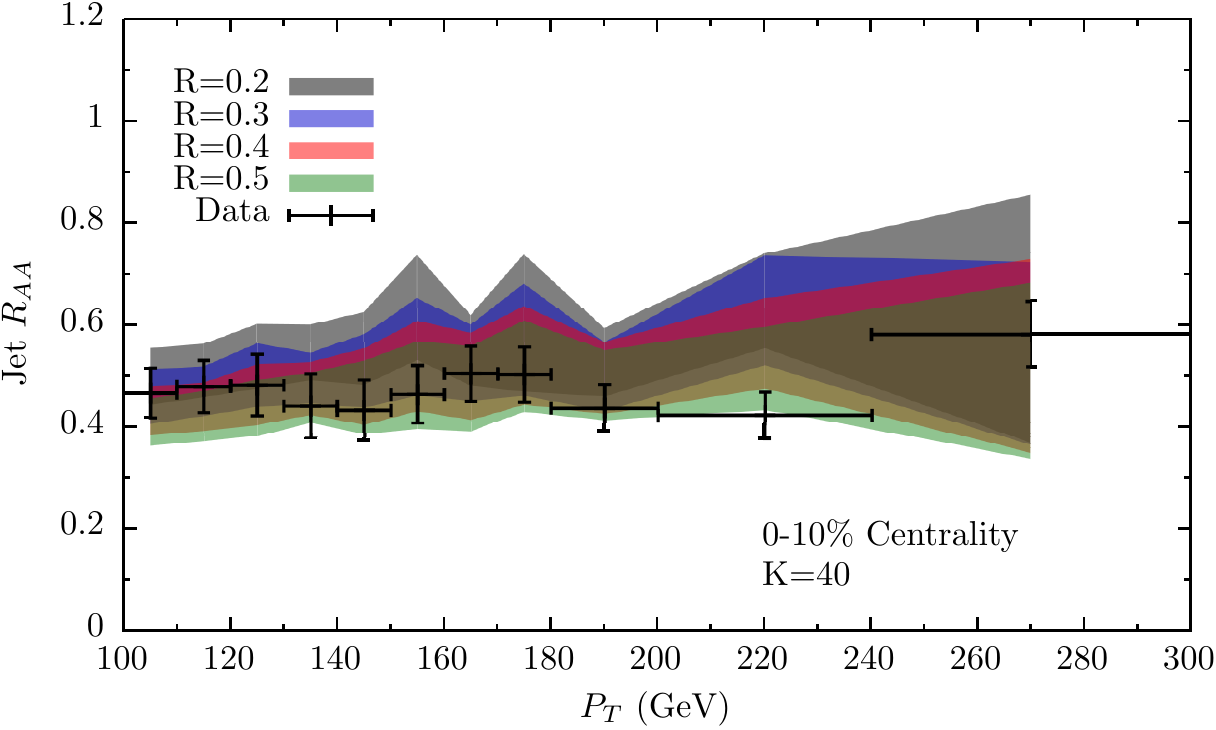}
\end{tabular}
\caption{\label{Fig:broRAAvsR}Dependence of $\RAA^{\rm jet}$ on the anti-$k_t$ jet reconstruction parameter $R$ for $K=0$ (no broadening, left panel) and $K=40$ (right panel). 
For comparison, we show the suppression of jets reconstructed using the anti-$k_t$
algorithm with $R=0.3$ as measured by CMS in the same interval 
of rapidity~\cite{Raajet:HIN}. 
}
\end{figure}

In \fig{Fig:broRAAvsR} we show the  $R$ dependence of the suppression factor $\RAA^{\rm jet}$ for $K=0$, {\it i.e.} no broadening, and for $K=40$
for jets  with $\left| \eta \right|<2$ as a function of $p_T$. The data points come from measurements of jets reconstructed with $R=0.3$; the colored bands are the results from our model for jets reconstructed with four different values of $R$.
The left-most data point in each panel 
is the one that we have used to constrain the value of $\aSC$, as we described in Section~\ref{sec:IntroBroadening}. 
The calculations shown in the two panels therefore have different values of $\aSC$, although the difference is small.  
%
In both panels, although the suppression factor shows only a very weak dependence on the reconstruction radius $R$, wide (larger $R$) jets tend to be somewhat more suppressed than narrow (smaller $R$) jets. 
This is a trend that we foreshadowed in the introduction to Section~\ref{sec:Section2}. When comparing jets at the same energy, 
wide jets contain more fragments (in our hybrid model, simply more partons) than narrow jets. 
Since the larger the number of partons traversing the medium the larger the lost energy, wide jets are naturally more suppressed. 
The same qualitative dependence of the suppression factor on the jet size has recently been observed in holographic 
computations~\cite{Chesler:2015nqz,Rajagopal:2016uip}. 
Measurements of the $R$-dependence of $\RAA$ for these high energy jets by the CMS collaboration show very little sensitivity to the value of 
$R$~\cite{Raajet:HIN},
as in our simulations, although the uncertainties in our calculations and in the measurements preclude a quantitative comparison at 
present.
We shall return to the $R$-dependence of $\RAA$ in Section~\ref{sec:Section3}.

In spite of the extreme transverse momentum broadening introduced by choosing $K=40$,  
the $R$-dependence of the observable plotted in \fig{Fig:broRAAvsR}
is almost identical in the two panels.
The origin of this lack of sensitivity to $K$ lies in the strong quenching of jets by the plasma, and
in particular in the fact that, as we have already noted, wider jets are more strongly quenched than narrower
jets. In particular, the softer partons within a wider jet that could serve to further broaden the
jet as they are kicked in transverse momentum instead lose almost all of their energy.
This means that
the jet sample that ends up dominating the inclusive jet spectrum ratio $\RAA^{\rm jet}$
is biased such that most jets in the sample contain only a 
few or even just one hard parton. For such jets, transverse momentum kicks, even with an
extreme value of $K$, serve only to change the direction of the jet axis, not to broaden the jet.



We turn now to the second of the three observables that we shall analyze in this Section,
one that we will use to look for exactly the change in the direction of the jet axis due to
transverse momentum broadening, namely due to the vector-summed effect of the transverse
momentum kicks felt by each of the partons in the jet.  If we consider a dijet pair, the change
in the direction of propagation of the two jets in the pair will in general differ, since broadening
is a stochastic process and also since the temperature as a function of time along the trajectory of each of the jets
will in general be different.
If all dijet pairs were produced back-to-back in azimuthal angle,
with $\Delta \phi=\pi$, deviations from $\Delta\phi=\pi$ due to different broadening-induced
kicks to the two jets in the pair could be used as a direct measurement of broadening.
Reality is not this simple.  The hard scattering processes that produce dijets often include radiative production of
more than two partons, and in many events where two jets are reconstructed there may in fact have
been a third or even fourth jet present also. This means that even in proton-proton collisions
there is a nontrivial distribution of $\Delta\phi$, centered around $\pi$ but with considerable width.
We shall start from this distribution, and then look at the effects on it due to propagation
through the plasma originating from energy loss and broadening.

\begin{figure}[t]
\centering 
\begin{tabular}{cc}
\includegraphics[width=.5\textwidth]{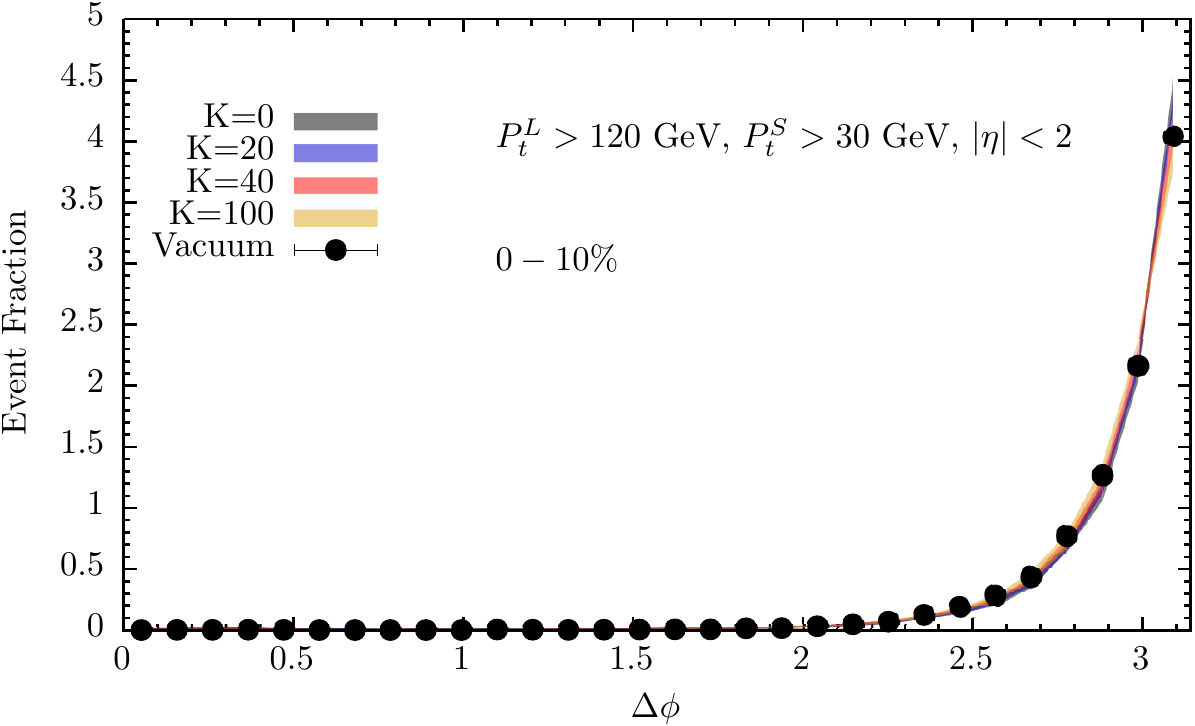}
&
\includegraphics[width=.5\textwidth]{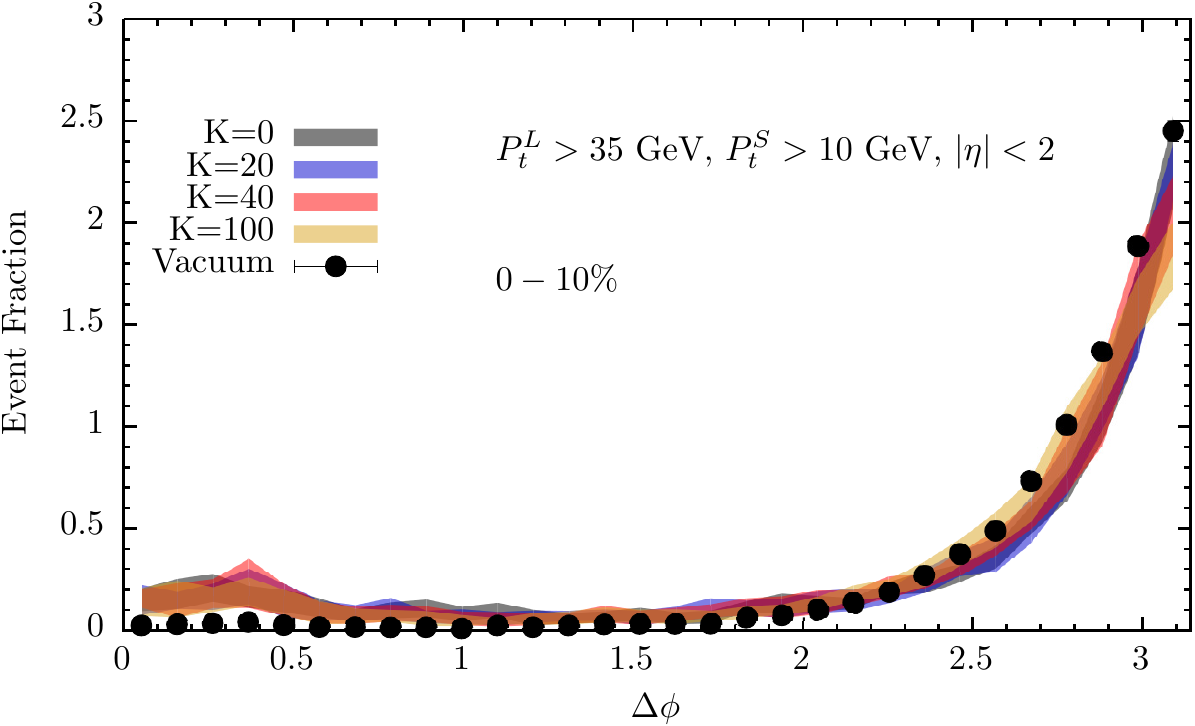}
\end{tabular}
\caption{\label{Fig:DeltaPhi} Dijet acoplanarity distribution  for high-energy (left) and low-energy (right) dijets in LHC heavy ion collisions with $\sqrt{s}=2.76$~ATeV for two different values of the 
broadening parameter $K$. For comparison, the black dots show the acoplanarity in proton-proton collisions as 
simulated by \pythia. 
}
\end{figure}

Deviations of $\Delta\phi$ away from $\pi$ are termed acoplanarity because the two jet axes and the
beam axis are not coplanar.
The black dots  in  \fig{Fig:DeltaPhi} shows the acoplanarity distribution for dijets in proton-proton collisions (propagating in vacuum)
for two different cuts on the transverse momentum of the dijet pair: $\pt^{\rm leading} > 120$ GeV and $\pt^{\rm subleading}>30$ GeV (left) and  $\pt^{\rm leading} > 35$ GeV and $\pt^{\rm subleading}>10$ GeV (right) as simulated by \pythia. 
The $\Delta \phi$-distribution is different in the two momentum regions displayed,  with the high-energy acoplanarity slightly narrower (closer to $\Delta \phi=\pi$) than the low energy one. 
This occurs because the fragmentation of higher energy jets leads to a narrower angular distribution of fragments than for
lower energy jets, and in a case where only two jets are produced the acoplanarity will be less if the jets are narrower.  

As already pointed out in Refs.~\cite{CasalderreySolana:2011rq,Mueller:2016gko,Chen:2016vem}, 
in the case of the high energy dijet pairs the effects of broadening on their acoplanarity
is much smaller than the width of the vacuum acoplanarity distribution.
Indeed, in the left panel of \fig{Fig:DeltaPhi} we see that 
our results with $K=0$ (no-broadening) and $K=100$ (extreme broadening) are both indistinguishable from the vacuum distribution. 

In the right panel of \fig{Fig:DeltaPhi}, we look at the acoplanarity distribution for
dijets with much lower energy, choosing dijets
with  $\pt^{\rm leading} > 35$ GeV and $\pt^{\rm subleading}>10$ GeV. 
It is challenging, perhaps prohibitively so, to measure jets with energies as low as this
in LHC heavy ion collisions, but even as an academic study the results are interesting.
First, our $K=0$ calculation apparently yields a narrower $\Delta \phi$ distribution than in vacuum. 
One contribution to the origin of this apparent effect lies in the increase in the number of dijet pairs
close to $\Delta\phi=0$; since what is plotted is a normalized probability distribution, this
tends to lower the curve near $\Delta\phi=\pi$.
The enhancement of almost collinear jet pairs is a consequence of energy loss and arises 
when a jet that had  been 
propagating somewhere near $\Delta\phi=\pi$ loses so much energy that its energy drops below that
of a third jet produced close to $\Delta\phi=0$, meaning that what is reconstructed is a dijet with
$\Delta\phi\sim 0$.
We have checked that if we restrict the dijet distribution to pairs of jets moving in opposite hemispheres, 
the $K=0$ and vacuum acoplanarity distributions are much more similar.
The second reason why the distribution of $\Delta\phi$ around $\pi$ is slightly narrower in heavy
ion collisions than in vacuum is that a quenched jet seen in a heavy ion collision with a given energy
began with a larger energy, and the acoplanarity distribution for higher energy dijets as produced in vacuum 
is narrower.

We can now look at the acoplanarity distribution with $K=100$ for the
low energy dijets in the right panel of \fig{Fig:DeltaPhi}.  We see that introducing this 
extreme degree of broadening does make the acoplanarity distribution very slightly wider than
the $K=0$ distribution, perhaps by coincidence bringing it back into agreement with the
vacuum acoplanarity distribution.
The principal conclusion, though, is that in LHC heavy ion collisions, even for dijets
with very low energies and even with substantially more broadening than is expected,
the effects of broadening on the acoplanarity distribution are very small.

 \begin{figure}[t]
\centering 
\includegraphics[width=.5\textwidth]{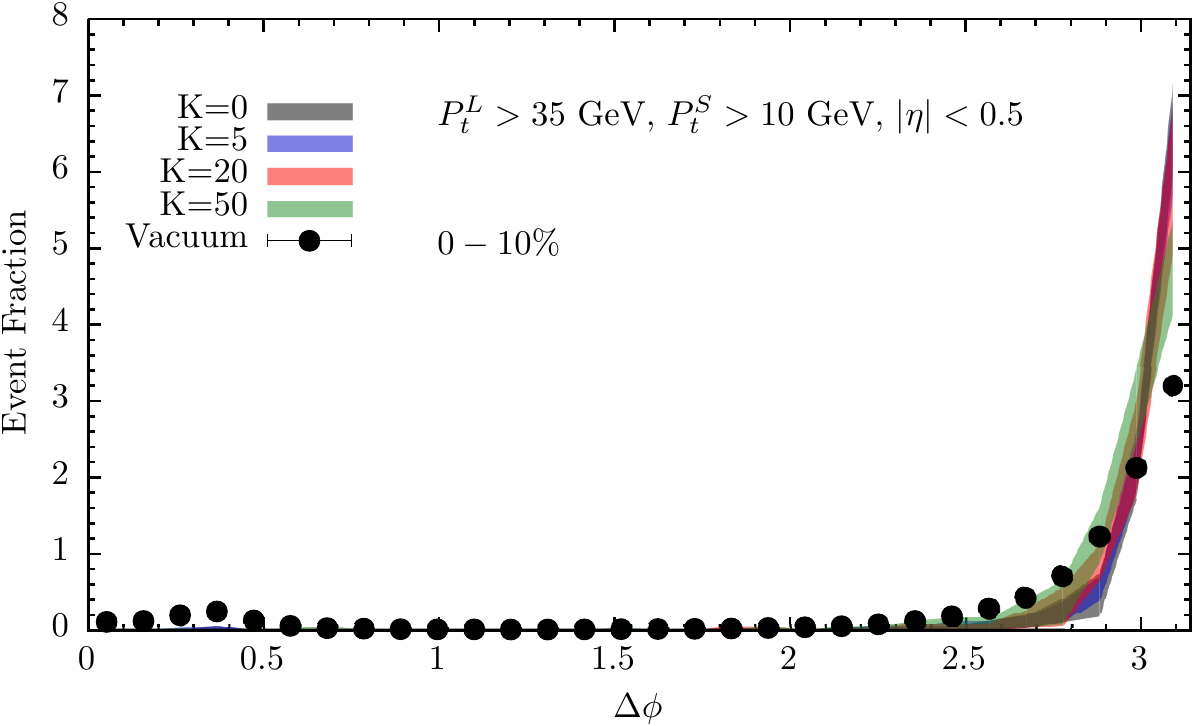}
\caption{\label{Fig:AcoRHIC} Dijet acoplanarity  distribution for dijets in heavy ion collisions at RHIC for different values of the broadening parameter $K$. 
For comparison, the black dots show the acoplanarity distribution in proton-proton collisions as 
simulated by \pythia. 
}
\end{figure}

As noted in Refs.~\cite{Mueller:2016gko,Chen:2016vem}, the effects of broadening on acoplanarity distributions are larger for low energy jets at RHIC, where the smaller soft background makes these measurements more feasible.   
Furthermore, the vacuum acoplanarity distribution is much narrower in RHIC heavy ion collisions than at the LHC 
for dijets with a comparable range in transverse momentum.  As we now explain, this can be attributed
to the fact that the jet spectrum at a given jet energy
is more steeply falling in lower energy RHIC collisions than it is at the LHC. 
One way in which dijets with a given energy pick up acoplanarity is if the subleading jet
in the pair started out at a higher energy and radiated a relatively hard gluon, which became
a third jet that balances the acoplanarity in the momenta of the reconstructed dijet pair.  
If at a given energy the spectrum is a more steeply falling function of energy this means that there are
fewer initially higher energy jets that could radiate and end up at the given energy.  Thus, a steeper spectrum 
as in RHIC collisions yields
a narrower acoplanarity distribution.
This is illustrated by the black dots in \fig{Fig:AcoRHIC} that show the acoplanarity distribution in RHIC heavy ion collisions
for dijet pairs with  $\pt^{\rm leading} > 35$ GeV and $\pt^{\rm subleading}>10$ GeV, the same dijet energies as in the right
panel of \fig{Fig:DeltaPhi} although here we have chosen dijets with $|\eta|<0.5$. 
Note that here the vacuum distribution 
shows a small accumulation of events at $\Delta \phi \lesssim 0.3$. This is a result of  
 the narrow rapidity coverage $|\eta|< 0.5$: in a small fraction of events, the jet that balances most of the transverse 
 momentum of the leading jet falls outside the accepted $\eta$ range and what is reconstructed as a dijet is the
 leading jet and a subleading jet pointing in a similar direction.
 
We see in \fig{Fig:AcoRHIC} that the RHIC acoplanarity distribution becomes visibly narrower in heavy
ion collisions if we neglect broadening, setting $K=0$.  This effect is perhaps hinted at in
the right panel of \fig{Fig:DeltaPhi} but becomes clearly visible at RHIC (energy loss, and therefore this effect, are not included
in the calculations of Refs.~\cite{Mueller:2016gko,Chen:2016vem}). It is a consequence
of strong energy loss and a steeply falling jet spectrum in sum.  Wider jets with a given energy loses more energy than narrower jets,
meaning that the jets that start out at in a given energy bin and stay in that bin are the narrow jets.  Because of the  
steeply falling spectrum, there are not many jets that originate with higher
energies, lose energy, and end up in the given energy bin.
The result is a narrowing of
jets that remain in a given energy bin, something that has been seen previously in
both perturbative~\cite{Milhano:2015mng} and holographic~\cite{Chesler:2015nqz,Rajagopal:2016uip} analyses.
And, we see from the $K=0$ results in \fig{Fig:AcoRHIC}  that narrower dijets are less acoplanar.
When we now turn on broadening, in addition to energy loss, the jets get broader and the acoplanarity
increases.  The effect of broadening is small for realistic values of $K$ in the 5-20 range; it takes unrealistically
large values of $K$ to broaden the jets sufficiently that the acoplanarity distribution becomes as wide as in vacuum.

The lesson here is that acoplanarity is to some degree 
sensitive to both energy loss and broadening, more so for
lower energy jets in lower energy collisions where the spectrum is more steeply falling.  But, even in
the best case, this observable exhibits little sensitivity to broadening, with the narrowing of the acoplanarity
distribution due to energy loss
being greater than the broadening of the acoplanarity distribution due to momentum 
broadening with realistic values of $K$.



\begin{figure}[t]
\centering 
\begin{tabular}{cc}
\includegraphics[width=.5\textwidth]{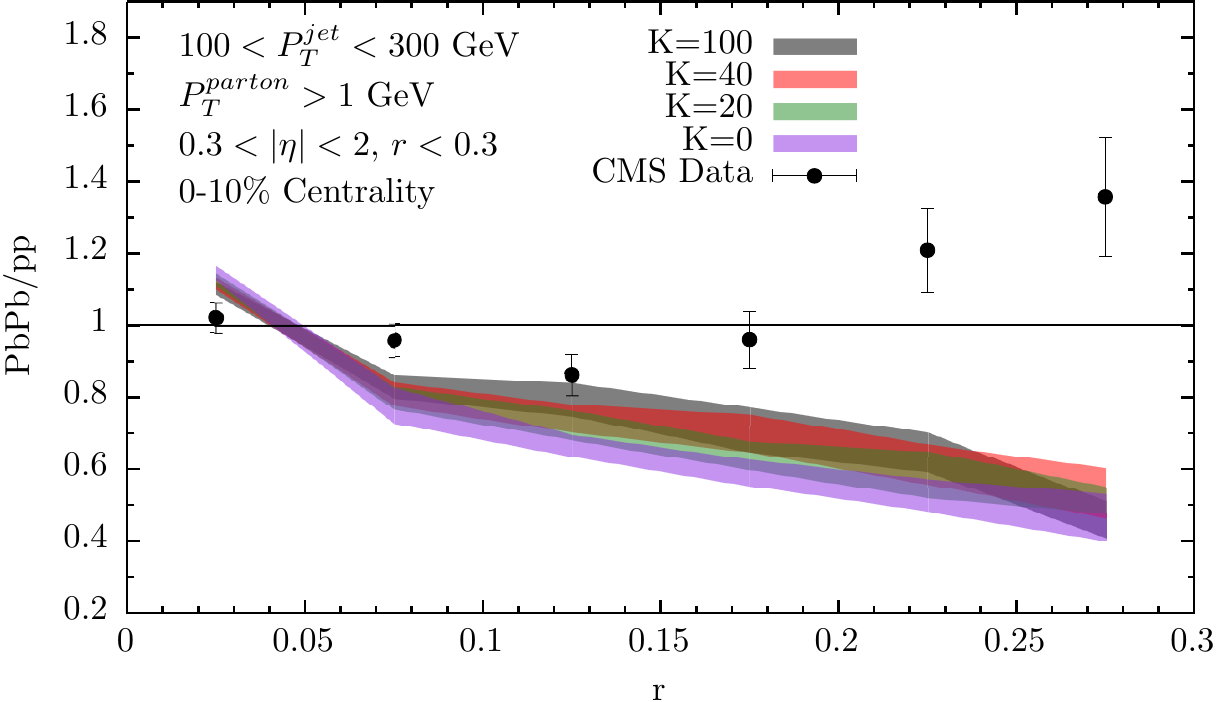}
&
\includegraphics[width=.5\textwidth]{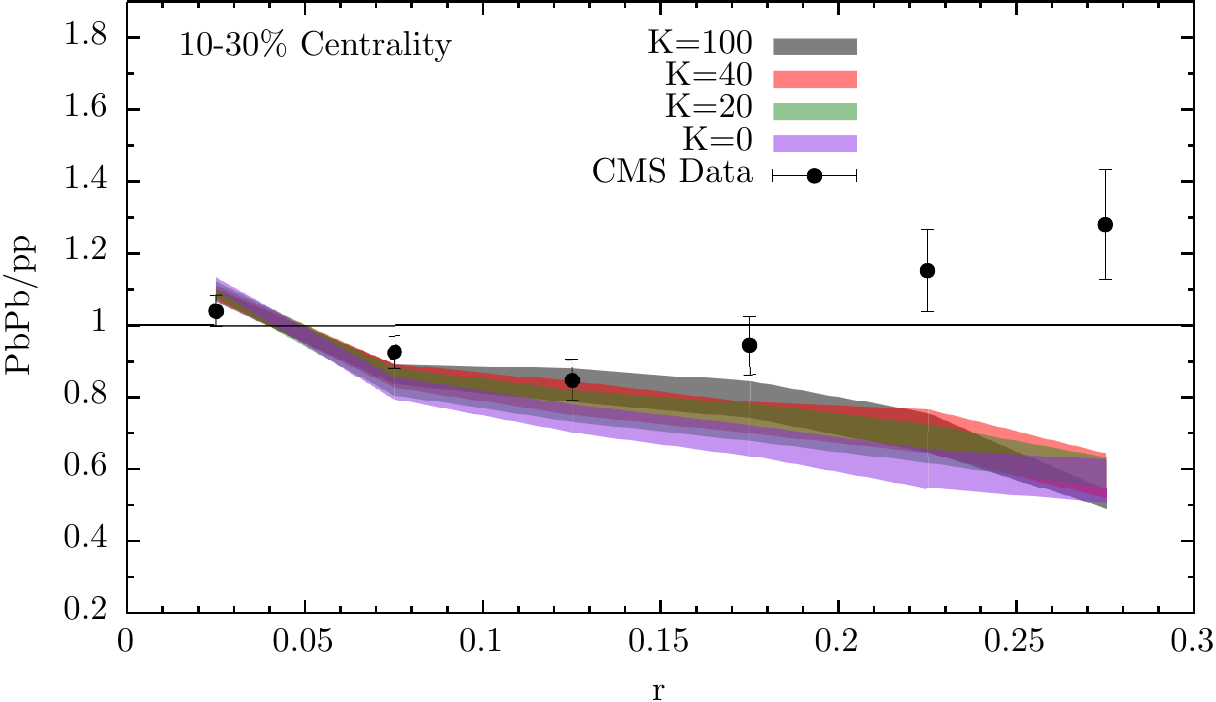}
\end{tabular}
\caption{\label{Fig:broShapes}Left: Ratio of the jet shape in PbPb collisions to that in proton-proton collisions for different values of the 
broadening parameter $K$ as compared to CMS data from heavy ion collisions with $\sqrt{s}=2.76$~ATeV~\cite{Chatrchyan:2013kwa} 
and 0-10\% centrality (left) or 10-30\% centrality (right).
}
\end{figure}

The third observable that we shall analyze is called the jet shape and is an intrajet observable that is a measure of the angular
distribution of the energy within a jet.
The jet shape is defined as the fraction of the jet energy in jets reconstructed with 
a given anti-$k_t$ parameter $R$ that is contained within an annulus of radius $r$ and width $\delta r$ (in $\eta-\phi$ space) 
centered on the jet axis.  
Following the analysis in Ref.~\cite{Chatrchyan:2013kwa}, we define the differential jet shape as
\begin{equation}
\rho(r)\equiv \frac{1}{N_{\textrm{jets}}}\frac{1}{\delta r}\sum\limits_{\textrm{jets}} \frac{\sum\limits_{i \, \in \,  r\pm\delta r/2}
p_t^{i,\textrm{track}}}{p_t^{\textrm{jet}}} 
\label{eq:JetShapeDefn}
\end{equation} 
for $r<R$,
where the tracks in the sum don't necessarily have to belong to the jet constituents defined through the anti-$k_t$ clustering. For this reason the final jet shape distribution is multiplied by the event averaged factor $\left\langle p_t^{\textrm{jet}}/\sum_{i=0}^{N_{\textrm{bins}}} p_t^{\textrm{track}}(r_i) \right\rangle$ so that it is normalized to one. 
We show the result of this analysis in   \fig{Fig:broShapes}, where we compare the ratio of 
the jet shape for the quenched jets in PbPb collisions to that for the unquenched jets in proton-proton collisions. 
In this analysis, the sum over jets in (\ref{eq:JetShapeDefn}) includes all jets with $p_t^{\textrm{jet}}>100$ GeV and $0.3<|\eta|<2$.
For reference, we also show the experimental results for this ratio, as measured by the CMS collaboration~\cite{Chatrchyan:2013kwa}.

The hybrid model calculations with $K=0$ shown in  \fig{Fig:broShapes} provide clear confirmation that, as we have
discussed, energy loss serves to narrow the angular size of jets with a given energy in heavy ion collisions
relative to that of jets with the same energy in proton-proton collisions.  Again, this arises because wider
jets lose more energy than narrower jets, leaving behind a sample that is dominated by narrower jets.  Adding transverse momentum broadening by turning on 
a nonzero value of $K$ serves to broaden the jets in the sample, slightly.  The effect of broadening is very small
even for the unrealistically large choice $K=100$.\footnote{In the results of Ref.~\cite{Chang:2016gjp}, introducing broadening results in larger effects on the jet shapes.
There are at least two reasons for this.  First, the analysis of Ref.~\cite{Chang:2016gjp} focuses on the modification of average jets. Leaving out jet-by-jet fluctuations means
that this analysis cannot see that wider jets lose more energy than narrower ones, and so does not include the consequent narrowing of jets with a given energy in PbPb collisions relative to jets with the same energy in proton-proton collisions. And, second, in Ref.~\cite{Chang:2016gjp} all the partons present in a \pythia shower just before hadronization are assumed to have passed through the medium, and felt the effects of broadening, whereas in our calculation of the development of the  jet shower in spacetime we see that many partons are produced at splittings that occur after the jet has already departed from the medium, and therefore do not feel any transverse momentum kicks.} (Note that transverse momentum picked up by
the one or few hardest partons in the jet serves to deflect the angle of the reconstructed jet,
affecting the acoplanarity.  This has little effect on the jet shape since $r$ is measured relative
to the center of the reconstructed jet, not relative to whatever the original direction of its parent parton was.)
%
%
It is apparent in \fig{Fig:broShapes} that our analysis in this Section does not do a good job of describing
the jet shape ratio measured in experimental data~\cite{Chatrchyan:2013kwa}, in particular
at larger values of $r$. As we shall discuss in Section~\ref{sec:Section3}, at larger and larger $r$  the
partons in the reconstructed jet are softer and softer.  Our hybrid model fails to describe
a medium-induced enhancement in the production of soft particles at large angles relative to the jet
direction seen in heavy ion collisions relative to proton-proton collisions.
This enhancement, which does not contribute much to the overall jet energy,
points to the existence of soft modes moving in the same direction
as the jet, which is to say a moving wake in the plasma that the jet itself
excites as it loses energy and momentum to the plasma.  In our implementation
of the hybrid model to this point, no effects of such a wake are included. We are in effect 
making the assumption that all the energy lost by the jet is so fully thermalized that it
ends up as a little bit more plasma or a little bit hotter plasma, becoming a
distribution of soft thermal particles in the final state that is isotropic, uncorrelated
with the jet direction.
This cannot be the whole story since, after all, momentum
is conserved and the jet loses momentum to the plasma as well as energy.  We shall return to this 
in Section~\ref{sec:Section3}.


\subsection{An Observable that {\it is} Sensitive to Broadening}
\label{sec:SensitiveObservable}

Before returning to the question of where the momentum and energy lost by the jet ends up and how this
affects angular observables like the jet shape, we close this Section by identifying a more differential
observable that {\it is} directly sensitive to transverse momentum broadening, meaning to the 
value of the broadening parameter $K$.

Let us recapitulate why the observables that we looked at above are not sensitive to broadening.
Kicks received by the few highest momentum
partons in a jet contribute to the acoplanarity, as we have discussed, but because these partons
have such a high momentum the effects on the acoplanarity are quantitatively very small.
The conventional jet shape observable at larger values of $r$ is dominated by very soft
particles.  In our hybrid model as formulated in this section, these softest particles tend to originate
from the last fragmentation events in the shower, as harder partons fragment after they have
already left the medium.  Soft partons that are produced earlier, in the medium, rapidly lose
all their energy.  Soft partons at large $r$ that are produced outside the medium cannot
be affected by broadening.  For different reasons, therefore, both the acoplanarity (dominated
by the few hardest partons) and the jet shape at large $r$ (dominated by soft partons which survive because they were produced
late, after the shower exits the plasma) are insensitive to transverse momentum broadening.

The key is to focus on the angular distribution of the energy in the jet that is carried
by particles in an intermediate interval of transverse momentum.  We must focus
on semi-hard partons that are sufficiently soft that they can be deflected significantly
by the transverse momentum kicks that they receive from the medium but that are
sufficiently hard that they survive propagation through the plasma and emerge from it.
%
%
The new observable that we define is, in essence, a more differential version of the jet shape. Instead of determining the $r$ distribution of all the jet energy, we focus on the distribution of jet energy carried by particles in a certain interval of transverse momentum. And, we construct this observable using only the subleading jets in dijet pairs,
specifically subleading jets with $\pt^{\rm subleading}>30$~GeV in a dijet whose leading jet has $\pt^{\rm leading}>120$~GeV and where the two jets are separated by $\Delta\phi>5\pi/6$.  We make this choice because jets in a sample of subleading jets have on average lost more energy than in a sample of leading or inclusive jets.
(We have checked that if we use a sample of leading or inclusive jets, we do see the effects that we shall describe below but they are smaller in magnitude than in the sample of subleading jets.)
We denote our new observable by $\Nob$, with
\be
\Nob\equiv \frac{1}{N_{\textrm{subleading jets}}}\frac{1}{\delta r}\sum\limits_{\textrm{subleading jets}} \frac{\sum\limits_{i \, \in \,  r\pm\Delta r/2 ; \, \pt^{i,\textrm{track}} \,\in\, \textrm{range}}
\pt^{i,\textrm{track}}}{\pt^{\textrm{jet}}} \, ,	
\ee
where we take the particles that we include to be only those that have a transverse momentum in a specified range. We take this range to be 10~GeV~$<\pt< 20$ GeV, a choice whose motivation we describe below.
(Note that we first find and reconstruct jets using all particles, in the standard fashion, and it is from the fully reconstructed jets we obtain $\pt^{\textrm{jet}}$ for the leading and subleading jets.  After we have a sample of reconstructed subleading jets, we evaluate $\Nob$ using only the particles within the specified range 10~GeV~$<\pt< 20$ GeV.)
We shall look at the observable $\Nob$ for angles up to $r<1$.
Recent preliminary results regarding somewhat similar jet shape observables, shown in Fig. 12 of Ref.~\cite{CMS:2015kca},
indicate that the measurement of $\Nob$ will require background subtraction. We have chosen to follow the procedure used
in Ref.~\cite{CMS:2015kca}.
We discretize the event in $(\Delta \eta,\Delta \phi)$ space, where $\Delta \eta$ and $\Delta\phi$ are distances from the jet axis in rapidity and azimuthal angle, with $|\Delta\eta|<2.5$, in bins of 
width $0.025 \times 0.025$, building a two-dimensional energy density distribution using only tracks with $\pt$ in the desired range.
We take reconstruction and background subtraction effects into account by smearing the jet energy, computing the $\phi$-dependent average energy density 
far away from the jet axis in rapidity by summing over all bins with  $1.5<|\Delta\eta|<2.5$, for each value of $\phi$, in each event, and then subtracting this average long-range energy density from the energy density in each of the bins with $|\Delta\eta|<1.5$, for each value of $\phi$, in each event.  After this background subtraction, we then construct the observable $\Nob$
by summing the energy lying in the annulus at a distance $\Delta r=\sqrt{\Delta\eta^2+\Delta\phi^2}$ from the jet.
Finally, to maintain consistency with the definition of the standard jet shape
we multiply $\Nob$ by the factor $\left\langle \pt^{\textrm{jet}}/\sum_{i=0}^{N_{\textrm{bins}}} \pt^{\textrm{track}}(r_i) \right\rangle$ in order to ensure that $\Nob$ is normalized to one.

\begin{figure}[t]
\centering 
\begin{tabular}{cc}
\includegraphics[width=.9\textwidth]{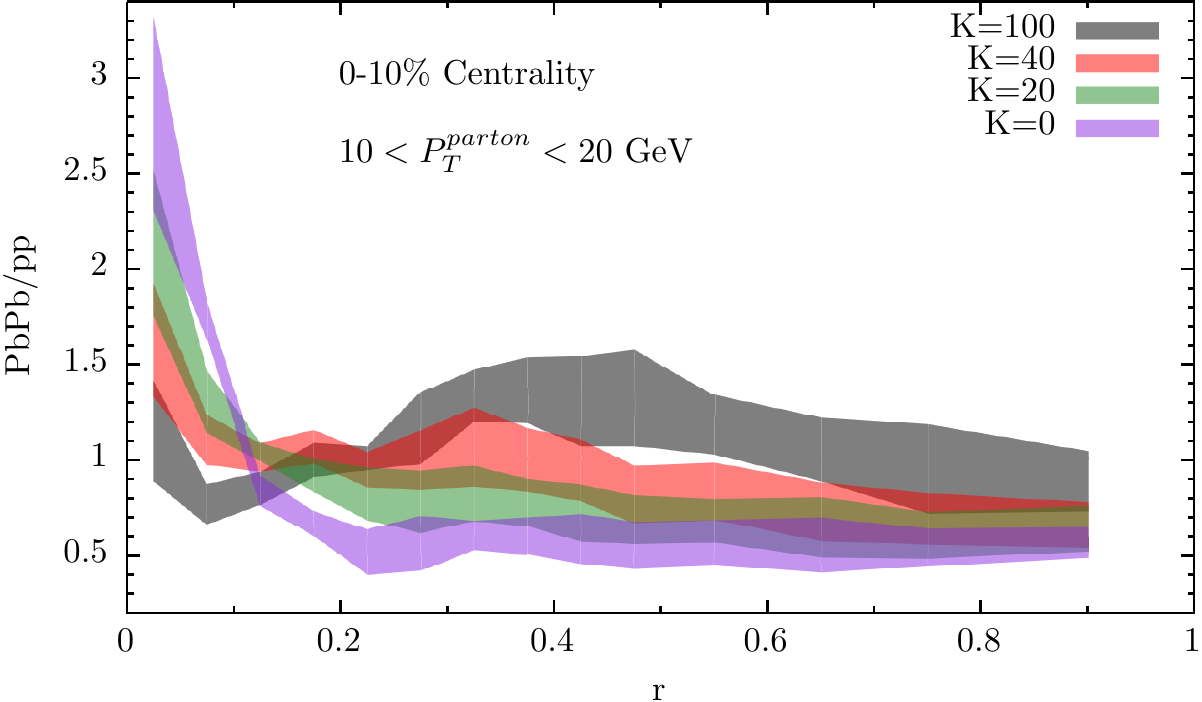}
\end{tabular}
\caption{\label{Fig:NewObs} Differential jet shape ratio constructed from the subleading jets in dijet pairs satisfying $\pt^{\rm leading}>120$ GeV, $\pt^{\rm subleading}>30$ GeV, and $\Delta\phi>5 \pi/6$. The analysis only includes partons whose $\pt$ lies within the intermediate range $10<\pt<20$ GeV.  The effect of
broadening on this observable is apparent.
}
\end{figure}

In \fig{Fig:NewObs} we show the prediction of our hybrid 
model for the ratio of the new observable $\Nob$ in PbPb collisions to
that in proton-proton collisions for several values of the broadening parameter $K$.
We show results for $\Nob$ with $\pt$ in the interval 10~GeV~$ <\pt< 20$~GeV.
The selection of these cuts follows from several requirements. We want the partons to be hard with respect to the medium temperature to 
ensure that no possible collective effects (including those we will focus on
in Section~\ref{sec:Section3}) can affect the measurement; this is safely achieved by the lower $\pt$ cut of 10 GeV. 
We also want the partons to be energetic enough to survive propagation through enough plasma that they pick up
some transverse  momentum kicks, which is to say so that they can exhibit sensitivity to the
broadening parameter $K$.  This motivates trying even somewhat larger values of the lower $\pt$ cut.
From the other side, we want to choose an upper $\pt$ cut so that the partons used in the
definition of $\Nob$ do have their direction of propagation significantly altered by the transverse
momentum that they pick up from the plasma.  We have found that the range 
10~GeV~$ <\pt< 20$~GeV serves our purposes well.  However, we have also investigated
$\Nob$ with 20~GeV~$ <\pt< 40$~GeV and this works almost as well.  
%
%
 Unlike for the less differential observables described previously, we see from \fig{Fig:NewObs} that $\Nob$ 
 shows significant sensitivity to the value of $K$. Indeed, as $K$ grows, the partons in this momentum range are more and more kicked 
 out to angles further away from the center of the jet, populating the large $r$ region  and depleting the region nearest to the jet axis ($r=0$).

The main features of the results plotted in  \fig{Fig:NewObs} may be understood as coming from the dynamics of broadening and parton energy loss.
Let us begin by looking at the $K=0$ curve.
In the absence of any transverse momentum kicks, the dynamics of energy loss in the hybrid model depletes the number of partons at large angles $r$.  This happens because these partons are produced early and are on average softer, and for both reasons they lose more energy in the plasma. This pushes the PbPb jet shape down at large $r$. Because the jet shape is normalized, it is pushed up at small $r$.   Since there is no parton energy loss for the jets in pp collisions, the modification of the numerator in the jet shape ratio shows up directly in the $K=0$ jet shape ratio itself, plotted  in \fig{Fig:NewObs}.  Now let us ask what happens when we turn on $K\neq 0$, adding transverse kicks felt by all the partons in the jet as they propagate through the medium, namely broadening.
As mentioned above, the depletion of partons at small
angles $r$ from the jet axis due to broadening
results in the reduction of the jet shape at small $r$ in PbPb collisions relative to that in pp collisions with increasing $K$.
Since the shape of jets in vacuum is a rapidly falling function of $r$, kicking some partons from smaller $r$ to larger $r$ serves to enhance the jet shape in PbPb collisions, and hence pushes the jet shape ratio plotted in \fig{Fig:NewObs} upwards
at larger $r$, again as a function of increasing $K$.   Furthermore, as $K$ increases partons that are kicked from smaller $r$ can end up at larger and larger values of $r$, meaning that at larger and larger values of $K$ the increase in the jet shape ratio seen in \fig{Fig:NewObs} extends further and further to the right.
These $K$-dependent effects are apparent in 
\fig{Fig:NewObs}, and they are of course the reason why we have selected and are highlighting this observable.

\begin{figure}[t]
\centering 
\begin{tabular}{cc}
\includegraphics[width=.9\textwidth]{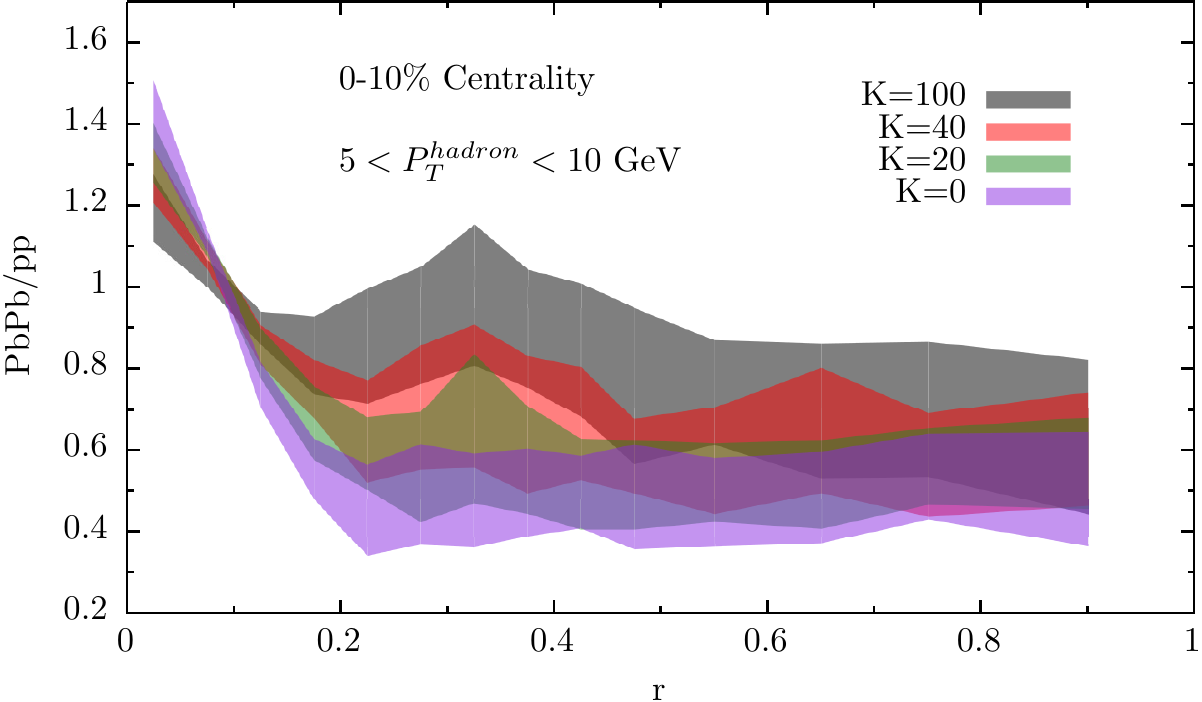}
\end{tabular}
\caption{\label{Fig:NewObsHad} Differential jet shape ratio $\Nob$ constructed from charged {\it hadrons} with 5~GeV$<\pt<$10~GeV, for
subleading jets that satisfy the same cuts as in \fig{Fig:NewObs}.
}
\end{figure}

In experiment, of course, the
range in $\pt$ employed in the definition of $\Nob$ will need to be a range of  momenta of final state hadrons, not a range of momenta
of partons.  Further investigation of the effects of hadronization, which are not under good theoretical control, is clearly important.
Nevertheless, the observed sensitivity to transverse momentum broadening that we see in Fig.~\ref{Fig:NewObs}, and that we have not seen 
in any of the other observables we have studied, encourages us to explore hadronic versions of 
this observable that are sensitive to the angular distribution of the partons in jets with $10<\pt<20$ GeV that contribute in \fig{Fig:NewObs}, with the goal of a direct extraction of the broadening parameter $K$, a key characteristic of the medium. In \fig{Fig:NewObsHad} we present the prediction from our hybrid model for a hadronized 
version of \fig{Fig:NewObs}, namely the observable $\Nob$ computed for hadrons rather than partons, with the specified range for the $\pt$ of the hadrons
in the analysis taken to be $5<\pt<10$ GeV.  (We use the hadronization prescription described at the end of
Section \ref{sec:bkrImpl}.)  We have selected a lower momentum range for the hadrons entering the analysis in \fig{Fig:NewObsHad} than for the partons 
in the analysis in \fig{Fig:NewObs} simply because hadronization turns partons into softer hadrons.  
Other choices of momentum range can be investigated.

The main features observed in \fig{Fig:NewObs} are also observed in  \fig{Fig:NewObsHad}.  As in the partonic case, 
broadening 
means that as $K$ increases we see a decreasing jet shape ratio at small $r$ and an enhancement in the intermediate $r$ region, with the region where enhancement is seen extending to larger $r$ as $K$ increases further.
However, in the hadronic case, the separation between the curves with different values of  $K$ values is less pronounced than in the partonic case. This is due, in part, to the fact that hadrons within any given range in momentum originate from partons with a wider range of momenta. The behavior of the  $\pt$-dependent partonic jet shape $\Nob$ is different in different $\pt$ regions, and in particular it becomes less sensitive to $K$ for partons with very large momenta. Since some of the hadrons with momenta in the 5~GeV$<\pt<$10~GeV range that we have used to construct the hadronic differential jet shape ratio in  \fig{Fig:NewObsHad} come from the hadronization of partons with much higher momentum, this hadronic observable plotted in  \fig{Fig:NewObsHad} shows less sensitivity to broadening than that seen in 
 \fig{Fig:NewObs}.
Because hadronization can turn a single hard parton at very small $r$ into several softer partons that are more spread out in $r$, it tends to spread the largest changes seen at very small $r$ in \fig{Fig:NewObs} over a wider range of $r$ in \fig{Fig:NewObsHad}, reducing their amplitude in the plot.
Despite these effects which serve to dilute the $K$-dependence seen in the partonic observable, the hadronic observable in  \fig{Fig:NewObsHad} displays sensitivity to transverse momentum broadening dynamics.

Although it is of course necessary to do further investigations of hadronization, it seems promising that by selecting hadrons from subleading jets in an intermediate $\pt$-range,  5~GeV$<\pt<$10~GeV in \fig{Fig:NewObsHad}, the effects of the transverse momentum kicks on the partons in the jet on the observable are enhanced, and are considerable, relative to what we have seen for 
less differential observables such as the usual jet shapes in \fig{Fig:broShapes}.  The further investigation of observables that, like the $\pt$-differential jet shape ratio we have constructed from subleading jets, focus on the transverse dynamics of partons in the jet that are neither very soft (as in that regime other effects that we investigate in the next Section dominate) nor very hard (as in that regime the effects of transverse kicks received from the medium are hard to see) holds considerable promise.  Such observables  represent the only path that we can see at present toward an experimental determination of the value of $K$ directly from its definition in terms of momentum broadening rather than indirectly via inference from measurements of energy loss. And, $K=\hat q/T^3$ is a key parameter in our model or in any model as it quantifies a central property of the medium that is related to how strongly coupled the fluid is.  We may not yet have hit upon the precise definition of the optimal observable, as doing so requires balancing choices of jet selection, $\pt$ cuts, jet shape measure, and background subtraction in the face of potentially competing goals: maximizing sensitivity to $K$ while at the same time optimizing the statistics, signal-to-noise, and utility of the analysis of any specific data set.  We look forward to seeing this done in consultation between experimentalists and theorists. Our results in \fig{Fig:NewObsHad} provide strong motivation for doing so.

\section{Medium Backreaction: Observable Consequences of a Wake in the Plasma}
\label{sec:Section3}

In this Section we will take the first steps toward incorporating another important
feature of jet quenching dynamics in plasma into our hybrid model: the conservation
of the momentum and energy lost by the jet.
In our implementation of the model up to now, we have assumed that the energy lost by the energetic
partons in the jet thermalizes to such a complete extent that, after hadronization, it becomes particles
moving in random directions that are completely uncorrelated with the jet direction.
The assumption of rapid thermalization is well motivated by the many lines of evidence
indicating that quark-gluon plasma is a strongly coupled liquid.  It is also motivated
by the holographic analysis of the stress tensor that describes the response of the
${\cal N}=4$ SYM fluid to the passage of an energetic parton through it~\cite{Chesler:2007an,Chesler:2007sv} 
which showed that after a short time of order $1/T$ all the energy dumped into the medium was rapidly converted 
into a hydrodynamic excitations of the system. 
The possible hydrodynamic collective response of the plasma to the passage of an energetic parton through it was 
characterized earlier in Refs.~\cite{Stoecker:2004qu,CasalderreySolana:2004qm,Satarov:2005mv,CasalderreySolana:2006sq};
the holographic computation in Refs.~\cite{Chesler:2007an,Chesler:2007sv} provided a concrete realization in a full quantum
field theoretical calculation, confirming that the anticipated hydrodynamic response could indeed be excited via the passage
through the fluid of an energetic probe whose size is much smaller than $1/T$.
This notwithstanding, it cannot be correct to assume that the energy lost
from the jet, after thermalization, has no memory of the direction of motion of the jet.
After all, momentum is conserved and the jet loses momentum to the plasma, as well as energy.
And indeed, the hydrodynamic response of the plasma includes both a wake and sound waves (a Mach cone)
that carry momentum in the direction of the jet, as well as energy.  
In this Section, we will provide an initial account of this collective response 
by providing a simplified description of the backreaction of the medium to the passage
of the jet that respects energy and momentum conservation without introducing any additional parameters into our hybrid model. 
We leave a full treatment (which would involve sourcing, propagating, and hadronizing a hydrodynamic wake in
the expanding cooling hydrodynamic fluid) to the future.

The mechanism we will describe is 
not exclusive to strong coupling, although this scenario provides a completely 
natural realization for this collective response. 
Even in perturbative QCD analyses in which a hard parton loses energy by radiating gluons
the radiated gluons themselves interact with the medium and radiate further softer gluons, with
the result being 
a rapid degradation of the emitted gluon momenta. This degradation may be viewed as a rapid transfer of energy from hard (jet) modes to soft (medium) modes and it has been described recently in detail in Refs.~\cite{Blaizot:2013hx,Kurkela:2014tla,Blaizot:2014ula,Fister:2014zxa,Blaizot:2014rla,Blaizot:2015jea,Iancu:2015uja}. 
From this perspective, it is possible that the medium modes excited by the jet may remember more  than
just the energy and momentum that the jet lost and they gained, but keeping track of their momentum and energy
as we shall do is certainly a good first step.  One circumstance in which our analysis in this Section
would not be relevant is for a medium that is sufficiently thin and sufficiently weakly coupled that gluons
radiated by the hard parton do not interact after they are radiated, but this is an extreme scenario.


Following this discussion, in Section~\ref{sec:PerturbedSpectrum} we will describe a simple implementation of the 
backreaction of the plasma to the passage of a jet of energetic partons through it. 
We will refer to the backreaction of the plasma in generic terms as the wake of the jet, as
we shall not need to focus on the distinction between a diffusive wake (moving and
perhaps heated fluid) and propagating sound
waves since upon making the approximations that we describe below the perturbations to the spectrum
of hadrons in the final state that we compute cannot distinguish between diffusive and sound modes.
We will characterize the medium response to the transfer of momentum to it from the jet, that is the wake, by analyzing the induced velocity and temperature  
variations of the hydrodynamic behavior of the quark-gluon plasma fluid, upon making several simplifying assumptions. 
Since the total energy deposited in the medium by a typical jet in heavy ion collisions ($\mathcal{O}\sim 20$~GeV)
is small when compared to the total energy per unit rapidity in the event $(\mathcal{O}\sim 1$~TeV), as in Refs.~\cite{Stoecker:2004qu,CasalderreySolana:2004qm,Satarov:2005mv,CasalderreySolana:2006sq} we 
shall treat the additional momentum acquired by the medium as a consequence of the passage
of the jet
as a small perturbation.
(For studies of the non-linear response of the plasma, see Refs.~\cite{Betz:2008js,Betz:2008ka,Tachibana:2014lja,Tachibana:2015qxa}). 
We shall also assume that the small perturbation of the velocity and temperature of the medium translates into small perturbations to the resulting distribution of particles at all momenta.  These physical assumptions 
can be described somewhat  loosely by saying that we are assuming that the energy deposited into the medium as a wake thermalizes to the maximum degree allowed by conservation of momentum and energy, turning into a perturbation to the spectra of the hadrons in the final state that remembers the energy and momentum deposited by the jet into the medium but nothing else about its origin. As we have discussed, this is natural at strong coupling, or at weak coupling if the gluons radiated by the hard partons in the jet themselves radiate many times further, but would not be valid if the energy lost by the jet is carried by only a few particles.

For simplicity, we will also assume that the unperturbed fluid is well described by a boost invariant flow.
And, although in our simulation of the amount of energy that the jet loses we do take the transverse
flow of the medium into account and we do use a background obtained via solving viscous hydrodynamics, in this initial study of the wake we will 
for simplicity neglect the transverse flow of the unperturbed fluid and employ ideal hydrodynamics.
All that said, our aim will actually not be a description of the perturbations of the hydrodynamics {\it per se}: we wish to focus
instead directly on the modification to the spectrum of hadrons produced after the perturbed hydrodynamic fluid freezes out.
The last simplifying assumption that we make is that the plasma freezes out along a constant proper time hypersurface.

After making these various approximations, in Section ~\ref{sec:PerturbedSpectrum} we will derive 
a simple expression for the modification to the spectrum of hadrons formed
as the hydrodynamic fluid perturbed by the passage of a jet through it freezes out, 
an expression that is determined
solely by the amount of energy and momentum lost by the jet. 
In subsequent subsections, we will describe how we implement this expression in our hybrid
model, and look at its consequences for a number of jet observables.

It is inevitable that when a jet reconstruction algorithm is used to find and reconstruct jets
in  heavy ion collisions, some of the energy and momentum that is counted as part of
a jet in fact comes from hadrons formed from the plasma as it freezes out, given that
the plasma includes a
moving heated wake that, by momentum conservation, is flowing in the same direction as 
the jet~\cite{Wang:2013cia,Tachibana:2014lja,He:2015pra,Tachibana:2015qxa,Cao:2016gvr}. 
Furthermore, since
any background subtraction procedure involves comparing events with a jet or jets to
events that do not contain jets, and since events that do not contain jets also
do not contain wakes, the particles from the hadronization of the wake that end up
reconstructed as part of a jet will not be removed by background subtraction.  Since there
is no way for experimentalists to remove them from the jets they reconstruct, theorists
must add them to the jets in their calculations.  This is our goal in this Section.

\subsection{The Spectrum of Hadrons from a Medium Perturbed by the Passage of a Jet}
\label{sec:PerturbedSpectrum}

In a boost invariant fluid with no transverse velocity,  the wake associated with the passage of a 
jet 
may be characterized by a perturbation to the velocity field with the form
\be
\delta u^\mu= \left(0,\delta u^i, \delta u^\eta \right), 
\label{deltauDefn}
\ee
where $\delta u^i$ with $i=1,2$ 
and $\delta u^\eta$ are the variations of the velocity field in the transverse plane and in the space-time rapidity direction. 
The disturbance will also be associated with  a change in the temperature of the plasma $\delta T$. 
These perturbations are functions of time and of all the space coordinates, and need not be boost invariant.
(Note that because $u^\mu+\delta u^\mu$ is normalized, the $\delta u^\tau$ component in (\ref{deltauDefn})  is fixed
by, and quadratic in, $\delta u^i$ and $\delta u^\eta$ and we have therefore neglected it.)
At a fixed proper time $\tau$,  the total momentum stored in the perturbation of the plasma  is 
\be
\label{eq:momentumL}
\Delta P_{\perp}^i=w \, \tau \int d^2x_{\perp} \, d\eta \, \delta u^i_{\perp}\,, \qquad \Delta P^{\eta}= w \,\tau  \int d^2x_{\perp} \, d\eta \, \delta u^{\eta}\,,
\ee
where we have used  the fact that, to leading order in the perturbation, 
the variation of the stress tensor of ideal hydrodynamics takes the form 
$\delta T^\tau_a = w \delta u_a$ for $a=1,2,\eta$
with $w=\varepsilon+P$ is the enthalpy of the unperturbed fluid,  
which is related to the entropy density of the unperturbed fluid through $w=Ts$ and which is constant on a fixed-$\tau$ surface in a fluid which is boost invariant and has no transverse expansion.
Because we are simplifying this analysis by assuming ideal hydrodynamics,
we are neglecting any production of entropy during the hydrodynamic
evolution, for example as the sound waves excited in the
plasma are damped by viscosity.  The only entropy dumped into the plasma
is that dumped into the wake initially by the jet itself, and this entropy is then
equal to the entropy associated with the perturbation computed at late time
by computing the flux of the entropy current across a fixed $\tau$ hyper-surface 
\be
\label{eq:entropyL}
\Delta S= \frac{s\,\tau  }{c_s^2} \int d\eta \, d^2x_{\perp} \, \frac{\delta T}{T} \, ,
\ee
where $c_s$ is the speed of sound of the unperturbed fluid which, within the boost invariant assumption, only depends on the proper time $\tau$.

Both eqs. (\ref{eq:momentumL}) and (\ref{eq:entropyL}) 
are valid on any fixed proper time hypersurface including, in particular, on the freeze-out hypersurface. 
These expressions
describe all the momentum and entropy that was dumped into the plasma by 
the disturbance, as they are at the freeze-out time.
Following our assumption that all the momentum lost by the jet 
is incorporated into the plasma as a wake,
$\Delta P_{\perp}^i$ and $\Delta P^\eta$ are given by the transverse momentum  and momentum rapidity 
lost by the jet as it traverses the plasma, respectively. 
Since in our 
implementation of energy loss we have assumed that the rapidity of the jet remains constant 
(or almost constant if we turn on transverse momentum broadening) we set $\Delta P^\eta=0$ and use 
our hybrid model calculation to give us the $\Delta P_{\perp}^i$ injected into
the hydrodynamic fluid by each of the jets we analyze.
We shall see below that the entropy production $\Delta S$ associated with the jet passage through the medium 
may also be constrained by the total energy carried by the particles produced by the perturbation. 

Our goal is to determine the effect of the perturbations $\delta u^a$ and $\delta T$ that
describe the wake in the fluid that the passage of the jet creates on the spectrum of particles
produced when the fluid freezes out. To do so, 
we will employ the standard Cooper-Frye prescription~\cite{Cooper:1974mv} 
\be
\label{eq:CF}
E\frac{dN}{d^3p} = \frac{1}{(2 \pi)^3} \int d\sigma^{\mu} \, p_{\mu} \, f(u^{\mu}p_{\mu})  \, ,
\ee
where the $\sigma_\mu$ integral is over the freeze-out hypersurface 
and where for simplicity we shall assume a 
Boltzmann distribution $f(E)=\exp\left(-E/T\right)$. 
Expanding to leading order in the perturbation, we obtain the expression
\begin{equation}
\begin{split}
\label{eq:deltaNG}
E\frac{d\,\Delta N}{d^3p}=&\frac{\tau}{(2 \pi)^3} \int d^2x_{\perp} \, d\eta \, m_T \, \rm{cosh}(y-\eta) \, \exp\left[ -\frac{m_T}{T}\cosh (y-\eta) \right] \\
 &\times \Bigg\{ p^i_{\perp}\frac{\delta u^i_{\perp}}{T}+\tau^2 p^{\eta}\, \frac{\delta u^{\eta}}{T}+\frac{m_T}{T}\frac{\delta T}{T}\rm{cosh}(y-\eta) \Bigg\},
\end{split}
\end{equation}
where $m_T\equiv \sqrt{m^2+ p_T^2}$ is the transverse mass of the emitted thermal particle.
Note that this perturbative 
expression is only valid for particles whose momenta $p$ are comparable to the temperature $T$ on the freeze-out surface,
where the small perturbations in the hydrodynamic quantities results in a small perturbation on the resulting 
distribution of particles.  For momenta far above $T$, where the thermal distributions are exponentially small,
the perturbations in the hydrodynamic quantities can have large relative effects; in this regime, the
expression (\ref{eq:deltaNG}) is not valid and, because it is based on a linear approximation
to an exponential, it in fact underestimates
the particle production from the wake. However, 
this regime of the spectrum contributes little to overall yields.


The expression in \eq{eq:deltaNG} is general and independent of the space-time dependence of the perturbed hydrodynamic fields. 
To proceed further we will assume that during the space-time
evolution of the perturbation over the boost invariant background, 
the space-time rapidity of the disturbance remains approximately constant. Since high-energy jets propagate at a fixed space-time rapidity $\eta_j$ equal 
to their momentum rapidity $y_j$, $\eta_j= y_j$, this assumption implies that the perturbation is narrow 
around the momentum rapidity of the jet, which allows us to perform the $\eta$ integration in \eq{eq:deltaNG} by replacing 
$\eta\rightarrow y_j$. We perform this integration, 
use eqs. (\ref{eq:momentumL}) and (\ref{eq:entropyL}) to relate the three terms in \eq{eq:deltaNG}
to $\Delta P_\perp^i$, $\Delta P^\eta$ (which vanishes) and $\Delta S$,
and then impose that the energy and 
momentum of the emitted particles equals the energy and momentum lost by the jet
\be
\Delta E = \int d^3p \, \frac{d\Delta N}{d^3p} \, E \,, \qquad \Delta P_{\perp,i} = \int d^3p \, \frac{d\Delta N}{d^3p} \, p_{\perp,i} \,,
\ee
where the integration of the first of these equations leads to the relation $\Delta S=\Delta E/(T \cosh y_j)$.
After these
manipulations, we are able
to express the spectrum of particles \eq{eq:deltaNG} emitted from the boosted, heated up, wake in the  fluid  as
\begin{equation}
\label{onebody}
\begin{split}
E\frac{d\Delta N}{d^3p}=&\frac{1}{32 \pi} \, \frac{m_T}{T^5} \, \rm{cosh}(y-y_j)  \exp\left[-\frac{m_T}{T}\rm{cosh}(y-y_j)\right] \\
 &\times \Bigg\{ p_{\perp} \Delta P_{\perp} \cos (\phi-\phi_j) +\frac{1}{3}m_T \, \Delta M_T \, \rm{cosh}(y-y_j) \Bigg\}.
\end{split}
\end{equation}
where $p_T$, $m_T$, $\phi$ and $y$ are the transverse momentum, transverse mass, azimuthal angle and rapidity of the emitted thermal particles whose distribution we have obtained,
and where $\Delta P_T$ and $\Delta M_T=\Delta E/\cosh y_j$ are 
the transverse momentum and transverse mass transferred from the jet (whose azimuthal angle and rapidity
are  $\phi_j$ and $y_j$) to the wake in the fluid.   
Note that the distribution (\ref{onebody})  is a small correction that must be added to the one-body distribution of particles that the unperturbed hot plasma
would have emitted. 
In particular, the distribution (\ref{onebody}) 
may be negative.  For example, this occurs for particles emitted in the direction opposite
to the direction of the jet.  Negative values simply mean that
the perturbed thermal fluid emits less particles 
in the direction opposite to the direction in which the jet was propagating 
than the unperturbed fluid would have.   This is a direct consequence of the
fact that the jet loses momentum to the fluid,
exciting a wake of fluid moving with net momentum along the direction of the jet.


The closed form expression (\ref{onebody}), 
which only depends on the momentum lost by the jets in the plasma and on the kinematics of the jet, will be the basis for our analysis of 
the observable consequences of the wake in the plasma, which is to say of the
backreaction of the medium. 
Subject to the assumptions employed in its derivation, the spectrum (\ref{onebody}) 
will allow us to provide estimates of the observable effects of the 
collective response of the plasma to the passage of the jet through it
without having to model the complicated processes (pre-hydrodynamic
and hydrodynamic) via which the energy
lost by the jet relaxes.
As stated above, even within the assumptions employed in
its derivation the expression (\ref{onebody}) 
is only valid for particles emitted with a momentum comparable to the mean thermal
momentum in a fluid cell at freeze-out; it need not be valid for semi-hard particles produced in the plasma. 
Perhaps the most important assumption  in its derivation is
the assumption that the wake in the plasma is fully thermalized, with the only aspects of
its origins that it remembers being
its total energy and momentum.   This too need not be valid for semi-hard particles, some
of which will originate from the energy lost by the jet either near the edge of the plasma
or shortly before freezeout that do not thermalize.
Nevertheless, we will use the spectrum (\ref{onebody}) at all momenta to obtain first estimates
of the observable consequences of the presence of a wake in the plasma in events with jets. 
In the next subsection, we shall explain how we have implemented this spectrum 
in our hybrid model analysis of jets propagating within hot QCD plasma.

\subsection{\label{sec:bkrImpl}Implementation of Backreaction, Background Model, and Jet Hadronization}

The implementation of the simplified expression (\ref{onebody}) for the spectrum of particles resulting
from the wake that is the backreaction of the medium to the presence of the jet
demands further modelling for a proper description in heavy ion collisions. Three aspects that we will need to 
incorporate into our description are: (i) the effect of the radial flow and chemical composition of the unperturbed
fireball on the particles resulting from the backreaction perturbation;
(ii) adding a background of particles coming from the freeze-out of the unperturbed fireball to
our hybrid model in order to be able 
to properly account for the negative contribution from the perturbation (\ref{onebody});
and (iii) the generation of particles consistent with the one-body distribution \eq{onebody}. 
Furthermore, since the particles produced after decoupling are hadrons, we will also need to consider the 
hadronization of our quenched jets, which we have not needed to include 
in our previous implementation 
of the hybrid model. We will discuss these four aspects sequentially in this subsection.

As we have stated, our derivation of \eq{onebody} neglects the effect of transverse flow in the
unperturbed fluid, meaning that we have neglected radial flow, as well as elliptic flow and higher
azimuthal harmonics of the transverse flow.  The effects of elliptic flow and higher harmonics 
are small in the most central collisions, but radial flow cannot be neglected
as it has important consequences for the spectrum of particles produced by the fireball. 
The radial boost in the spectrum of particles due to the radial flow in the fluid from which
they are formed yields a blue-shifted spectrum which is harder than in the absence of the boost.
Since heavier particles pick up more momentum than lighter particles when all are boosted by
the same velocity, 
another consequence of radial flow is differing spectra for particles with different masses, with
heavier hadrons getting harder spectra than lighter hadrons.
The result (\ref{onebody}) was derived as a perturbation on a background that
does not include either effect.  
As a crude step towards including both effects, we will employ
the spectrum \eq{onebody} but instead of using the temperature $T$ at the time
of freezeout we will use species-dependent, momentum-dependent, empirical expressions for $T$ that
provide a good description of the measured particle spectra in Ref.~\cite{Abelev:2013vea}
upon fitting the
measured spectra to ``thermal'' spectra without radial flow.
%
Specifically, we assume a proton to pion ratio of $0.05$, neglect hadrons other than protons and pions, and use the following
momentum-dependent ``temperatures'' in \eq{onebody}: 
\begin{eqnarray}
T_{\pi}(\pt)&=&\begin{cases} 0.19~{\rm GeV} &\mbox{if }\  \pt<0.7~{\rm GeV}\\
0.21 \, \left(\frac{\pt}{\rm{GeV}}\right)^{0.28} \, \rm{GeV} &\mbox{if }\  \pt>0.7~{\rm GeV} \end{cases}\\
T_p(\pt)&=&\begin{cases} 
0.15~{\rm GeV} &\mbox{if }\  \pt<0.07~{\rm GeV}\\
0.33 \, \left(\frac{\pt}{\rm{GeV}}\right)^{0.3} \, \rm{GeV} &\mbox{if }\  0.07~{\rm GeV} < \pt<1.9~{\rm GeV}\\
0.4~{\rm GeV} &\mbox{if }\  \pt>1.9~{\rm GeV} \end{cases}
\ee  
These empirical expressions
provide a good description of both the proton and pion spectra in central heavy ion collisions with $\sqrt{s}=2.76$~TeV at the LHC for $\pt<3$~GeV.
(The largest deviations, around 10\%, occur for $\pt \lesssim 0.5$~GeV, where the measured pion yields contain large contributions from resonance decays.)  We shall use them in \eq{onebody} for $\pt<5$~GeV.
This approach is admittedly crude, but it is of value in this exploratory investigation as it
allows us to use the closed form expression (\ref{onebody}) rather than attempting a full hydrodynamic
calculation of the wake induced by each of the jets in our ensemble of events.

As we have already discussed, a characteristic  feature of \eq{onebody} is that 
the spectrum of particles coming from the perturbation, from the wake in the fluid,
can become negative at large azimuthal angles with respect to the jet direction,
particulary for particles with small $m_T$ or
for jets with small $\Delta  M_{T}$.
This reflects the fact that the wake is made up of fluid moving with a net momentum
in the jet direction, with
the negative contribution to the perturbation representing a depletion of the momentum
in directions opposite to that of the jet.
This means that in order to implement \eq{onebody} in our hybrid model, we need
to embed our jet sample in a background whose purpose is to provide sufficient
thermal particles such that where the perturbation (\ref{onebody}) is negative
we have some thermal particles that we can remove.
In previous implementations of our hybrid model in which we neglected the back
reaction of the medium in response to the presence of a jet we ignored the background on
the basis that in experimental analyses of jet data it would be subtracted.
Contributions from the wake cannot be subtracted, which is why we are now
adding them.  Since these perturbations can be negative, we now need a baseline background too.
The background that we use is oversimplified. It is constructed by generating an ensemble of 
pions and protons which is flat in $\phi$ and $\eta$ that reproduces the measured  
particle yields and spectra~\cite{Abelev:2013vea}.
The addition of this contribution will force us to introduce a background subtraction 
procedure in our analysis of in-medium jets. We will describe this procedure in detail in Subsection~\ref{sec:obs} and
Appendix~\ref{app:bgsubs}

For each event in our sample, we determine the momentum lost by each of the partons in the jet shower as
well as in initial state radiation following our hybrid strong/weak coupling model as described in Section~\ref{MC}. 
(For the backreaction analysis, we include all partons in the jet shower and in initial state radiation, 
whether or not they end up being reconstructed in a jet.)
Since each of the propagating partons  loses energy into the medium, each of them generates its own wake induced by its lost momentum. 
In the linearized approximation that we have employed, the multiple wakes do not alter each other, and the final spectrum is the 
superposition of the spectra generated by the wake of each propagating parton. 
At hadronization, each of the induced wakes generates an ensemble of particles with the one body distribution (\ref{onebody})
computed from the kinematics  of each parton and its lost momentum. 

In order to incorporate the effect of fluctuations in the reconstructed jets, we generate
the ensemble of particles coming from the wake in the medium
via a simple Metropolis algorithm designed to satisfy the conservation of the lost jet energy while drawing
particles from the distribution (\ref{onebody}).
First we generate an independent list of particles, including protons with 5\%  probability and pions with 95\% probability, from the one body distribution (\ref{onebody}),  
until the sum of their energies reaches the lost energy. 
This ensemble generally contains particles in the region of azimuthal angle in which  \eq{onebody} becomes negative, which we will call negative particles. 
Whenever a negative particle is produced, its contribution to the net energy and momentum of the ensemble is negative.  
We later neutralize these negative particles by removing a particle from the background which is sufficiently close in $(\eta,\phi)$ space and in 
transverse energy.\footnote{Negative particle neutralization proceeds as follows.  We identify the background particles within an angle of $\Delta r=0.3$
in $(\eta,\phi)$ space of the negative particle we wish to neutralize.
Among the candidate background particles, we choose the best candidate in terms of $E_T$ and angular position in $\eta-\phi$ plane, which means we 
minimize $\Delta E_T$ and $\Delta r$, the difference in transverse energy and angular position with respect to that of the negative particle, respectively. 
We do the minimization by starting with any one of the candidate background particles and then checking each one of the other candidates
to see whether choosing it instead reduces $\Delta E_T$ while not increasing $\Delta r$ by more than 0.05
or reduces $\Delta r$ while not increasing $\Delta E_T$ by more than 0.1~GeV. 
We have checked that after using this procedure we end up with 90\% of the negative particles neutralized
to better than 0.1~GeV in transverse energy via subtraction of a background particle that was within
$\Delta r=0.13$ of it.}
From this initial ensemble, whose four-momentum in general will not coincide with the momentum  lost by the jet, 
we randomly select a particle which we replace by a new particle drawn from the distribution (\ref{onebody}). 
If the change improves four-momentum conservation, it is accepted. Otherwise, the change may be accepted with a probability distribution
 \be
 W(p^\mu_{\rm new \,  ensemble}) = \frac{e^{-\left(p^{\mu} _{\rm new \, ensemble}-\Delta P^\mu\right)^2}}{e^{-\left(p^\mu _{\rm ensemble}-\Delta P^\mu\right)^2} }\, , 
\ee
where $p^\mu_{\rm new \, ensemble}$ is the four momentum of the candidate ensemble including
the newly drawn particle, $p^\mu_{\rm ensemble}$ the four-momentum of the previous ensemble, and  $\Delta P^\mu$ is the 
four-momentum lost by the jet.
We repeat the procedure until each of the four components of $p_{\rm ensemble}$ is within $0.4$~GeV of 
the momentum lost by the jet. (We have explicitly checked that changing this threshold does not significantly change our results.) 
The ensemble generated after this procedure conserves energy and momentum (within the tolerance above) and possesses a one-body 
distribution identical to \eq{onebody}, something that we have checked explicitly.

Since the medium response manifests itself in the form of modified distributions for hadrons, namely pions and protons in our approach, we are forced to consider the hadronization of jet showers in order to properly incorporate the particles from
the jet itself and those from its wake in each event before then reconstructing jets and calculating
observables.
Hadronization leads, generically, to a softening of the typical jet fragments.
Because of its nonperturbative nature, 
hadronization even in vacuum remains a fundamental  problem and 
presents serious challenges to phenomenological modelling. 
Furthermore,
it is not presently understood how
the phenomenological hadronization models that have been applied successfully
to QCD processes in vacuum should be modified 
due to  the presence of a heavy ion environment. As an example, changes in how color flows in the jet shower 
resulting from
soft exchanges between partons in the shower and the medium 
lead to significant modifications to subsequent 
hadronization in certain regions of phase space~\cite{Beraudo:2011bh,Beraudo:2012bq}, but the overall 
description of these effects 
remains to be determined. 
Because of all these uncertainties, in this work we will adopt a simplified model for the in-medium hadronization, as in much of
the literature. We simply assume that hadronization of high energy jets occurs in heavy ion collisions
as in vacuum, in particular keeping the same color correlations between partons in the shower
even though we know that in reality these must change as the partons interact with the medium.
Although several different prescriptions for hadronization in vacuum 
exist, in this work we will employ the Lund string model as implemented in \pythia,
feeding the showers to this hadronization model after they have been modified according to our hybrid
approach to energy loss and broadening. 
We defer comparisons between different models for hadronization to future work.

\subsection{\label{sec:obs}The Effect of Backreaction on Jet Observables}

After the implementation of the energy loss suffered by the partons in jet showers produced in hard processes
as they propagate through the hydrodynamic medium
via the hybrid model described in Section~\ref{MC}, the incorporation of the effects of transverse momentum
broadening described in Section~\ref{sec:Section2} if we choose $K\neq 0$, the incorporation of a thermal-like 
background of particles and the perturbation to that background corresponding to the effects of the wake
in the hydrodynamic medium as it responds to the passage of the jet
%
as described in subsection~\ref{sec:bkrImpl}, we now have 
a full event simulation from which to extract calculations of medium-modified jet observables.  This will be our goal in this
and the next subsections.
In contrast to our previous publications, the inclusion of a background forces us to implement a background subtraction procedure, making our 
analyses of quenched events more similar to the actual analysis of jet data at the LHC. 
The fact that the medium responds to the passage of the jet, and in particular the fact that the medium picks up momentum in the jet direction,
makes this complicated procedure absolutely necessary. 

For hard jets 
produced together with a soft background that is completely uncorrelated with the jet direction, 
there are a number of established techniques that allow for systematic removal of the effects of background
particles from jet observables.  (See, for example, Refs.~\cite{Cacciari:2007fd,Cacciari:2008gn,Cacciari:2010te}.) 
These procedures, generically referred to as background subtraction, are routinely applied to jet measurements at the LHC and, at 
least in proton-proton collisions, they efficiently remove  the effects of soft (non-perturbative) 
backgrounds that may be large but that are uncorrelated with the jet, allowing the measurement of theoretically controlled hard processes. 
However, in heavy ion collisions the fact that the medium includes a wake that carries momentum in the jet direction
means, in effect, that a component of the background is correlated with the jet direction.
This makes it impossible for a background subtraction procedure to separate the jet (which has been
modified, via energy loss and broadening) 
from the medium (which has been modified, via the wake).  In order to compare to experimental measurements, therefore,
we have added a background and a wake and must now perform a background subtraction as if the background
were uncorrelated with the jet direction, followed by
jet reconstruction, just as in an experimental analysis. 
This procedure is not necessary for jet observables that are dominated by the harder components of a jet.  This procedure
is important for the softer components, since the softer components of what is reconstructed as a jet will include
contributions from the jet itself and from the wake in the medium.  In particular, this procedure is critical to gauging
the effects of the wake on observables.
%
%
We have implemented a full background subtraction procedure to analyze the events produced within our framework. 
In particular, we have implemented a version of the so called noise/pedestal background subtraction procedure~\cite{Kodolova:2007hd,Aad:2012vca} and then done a jet energy scale correction; 
the details of our implementation can be found in Appendix~\ref{app:bgsubs}.

\begin{figure}[t]
\centering 
\begin{tabular}{cc}
\includegraphics[width=.5\textwidth]{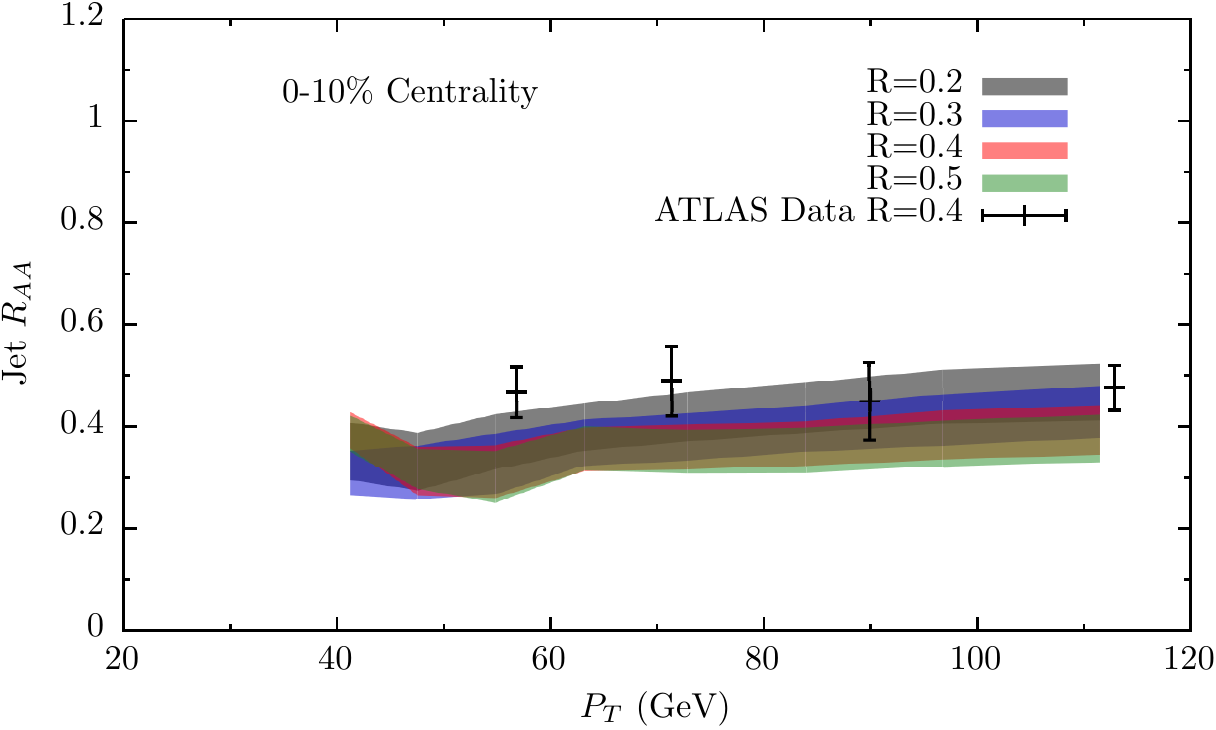}
&
\includegraphics[width=.5\textwidth]{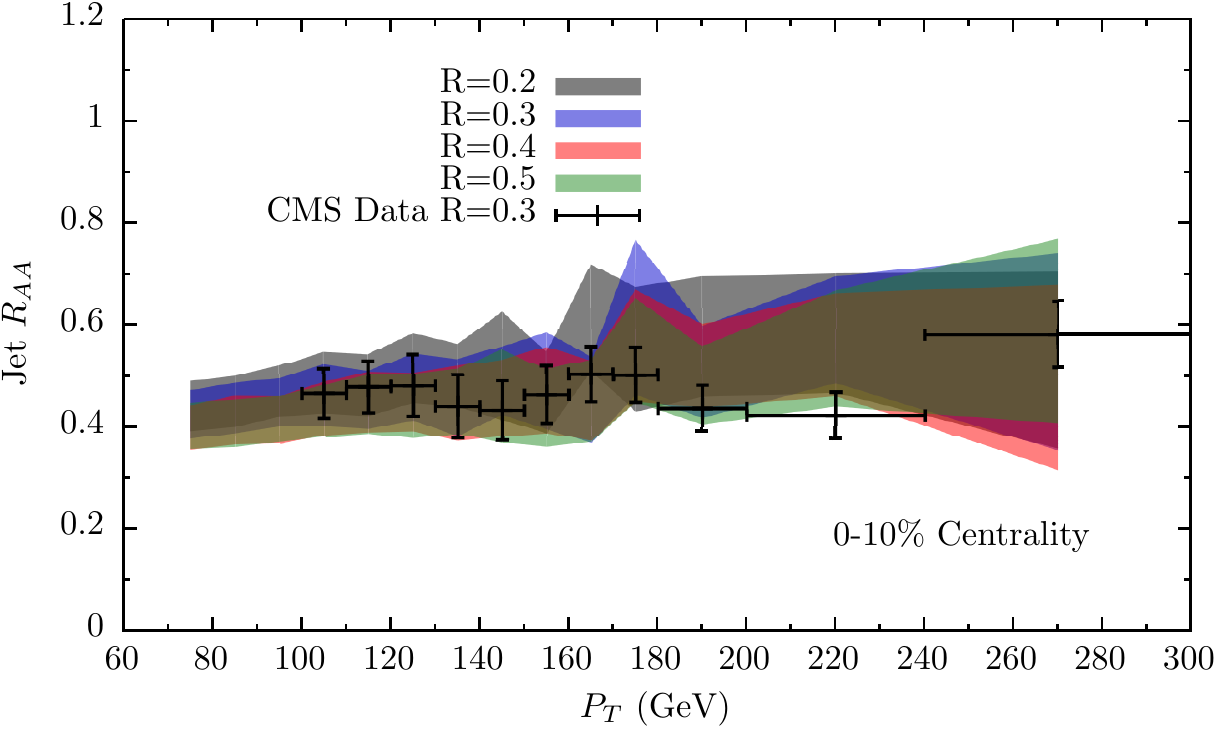}
\end{tabular}
\caption{\label{Fig:backRAA} Dependence of jet $\RAA$ on the anti-$k_t$ parameter $R$ used in reconstructing the jet and on the transverse
momentum of the jet. We have set
$K=0$, turning off transverse momentum broadening which we saw in Section~\ref{sec:Section2} has only very small effects on this observable. We have included the effects of the wake in the hydrodynamic fluid, which is to say the backreaction of the medium to the presence of the jet, and have added a background and implemented two different background subtraction procedures as described in the text and in Appendix~\ref{app:bgsubs}.
The left plot shows this observable for jets with $40<\pt^{\rm jet}<120$ GeV as measured by the ATLAS
collaboration~\cite{Aad:2014bxa}, with their background subtraction procedure, while the right one is for jets with $70<\pt^{\rm jet}<300$ GeV as measured
by the CMS collaboration~\cite{Raajet:HIN}, with their background subtraction procedure. Both panels are for the 0-10\% most central
collisions with $\sqrt{s}=2.76$~ATeV.  We have extended the $\pt$-range in the right panel down to 70 GeV even though present CMS measurements are for jets in the range $100<\pt^{\rm jet}<300$ GeV to make it possible to compare the results from our model in the $\pt$-range where the two panels overlap in order to see the effect on $\RAA$ of choosing between the two different background subtraction procedures.  The difference between the two panels is small, but visible.
}
\end{figure}

As in Section~\ref{sec:Section2}, we first consider the $R$-dependence of jet suppression. 
In \fig{Fig:backRAA} we show  the jet $\RAA$ as a function of $\pt$ for central events for 
a wide range of jet momenta and for different anti-$k_t$ reconstruction parameters $R$.  
We set $K=0$, neglecting transverse momentum broadening since, as we saw in Section~\ref{sec:Section2}, it has little effect on these observables.
In the right panel, we show our results for $70<\pt<300$~GeV in comparison with CMS measurements 
of $\RAA$ for $R=0.3$ jets in the range $100<\pt<300$~GeV.
Note that, as when we included broadening in Section~\ref{sec:Section2}, 
when we incorporate the effect of the backreaction of the medium to the jet
this alters the jet suppression, meaning that we had to retune the energy loss parameter $\aSC$ in our hybrid
model. In this case, we 
only needed to modify the value of $\aSC$ at the percent level, 
which is very much smaller than the theoretical uncertainty corresponding to the
widths of the bands in all our plots in this paper and in our previous publications.  
As we found when we included broadening in Section~\ref{sec:Section2}, for the high energy jets in
the right panel of \fig{Fig:backRAA} the suppression factor
$\RAA$ shows only a very small dependence on $R$, consistent with LHC data~\cite{Raajet:HIN}.


As we saw in Section~\ref{sec:Section2}, the suppression factor $\RAA$ shows a small decrease ({\it i.e.}~increase in suppression) 
with increasing $R$, corresponding to the fact that with increasing $R$ the angular size of the jets that are reconstructed
increases together with the fact that wider jets lose more energy. 
However, this effect is milder here than it was in \fig{Fig:broRAAvsR} because the
effect of the backreaction of the medium that we are incorporating here is that some
particles coming from the wake in the plasma, which is to say some of the energy that the
jet lost, ends up reconstructed as if it were still part of the jet.
%
%
Nevertheless, the wide angular distribution of the spectrum of particles from the wake given by \eq{onebody} implies that, 
even for the relatively large value of $R=0.5$ explored in \fig{Fig:backRAA}, 
the fraction of the energy lost by the jet that is recovered by the reconstruction
procedure is small.
For even larger values of the reconstruction parameter $R$, the full jet energy would in principle be 
recovered and $\RAA$ would approach unity. However, for such large values of $R$ the
fluctuations in the background make the reconstruction algorithm we have implemented unreliable.
(We have tested the recovery of jet energy by embedding our simulations in a homogeneous
background with no fluctuations.)


In the left panel of \fig{Fig:backRAA}, we show the jet $\RAA$ for smaller jet $\pt$ for the different values of $R$. 
The calculated suppression factor shows a mild increase with  $\pt$. It is consistent with all but the
lowest $\pt$ ATLAS data point
within the theoretical and experimental uncertainties, although
the data themselves don't show evidence for such an increase.
In spite of the lower jet energy, the 
effect of including the particles coming from the backreaction of the medium on
$\RAA$ is small for all the values of $R$ we have explored, meaning that the  $R$-dependence of our results for 
jet $\RAA$ is similarly small  to that seen in
the right panel of the figure.
This is in apparent contrast with $\RCP$ measurements reported by ATLAS in Ref.~\cite{Aad:2012vca}, which seem to indicate that 
the suppression of jets decreases with $R$ in this range of momenta, albeit with sizeable systematic uncertainties.
Since ATLAS reports the $R$-dependence of $\RCP$ (which can be thought of as the ratio of $\RAA$ for central
events to that for peripheral events) rather than of $\RAA$, and since our simulations do not 
reproduce $\RAA$ for peripheral events well (since the small amount of energy lost in the hadronic phase,
which we do not include in our hybrid model, is proportionally more relevant in peripheral events where the total energy lost 
is smaller~\cite{Casalderrey-Solana:2014bpa,Casalderrey-Solana:2015vaa}) we will not attempt to make any
quantitative comparisons with these data.
We look forward to anticipated further measurements of the $R$-dependence of $\RAA$ itself, where a direct comparison will
be possible.


\begin{figure}[t]
\centering 
\includegraphics[width=.8\textwidth]{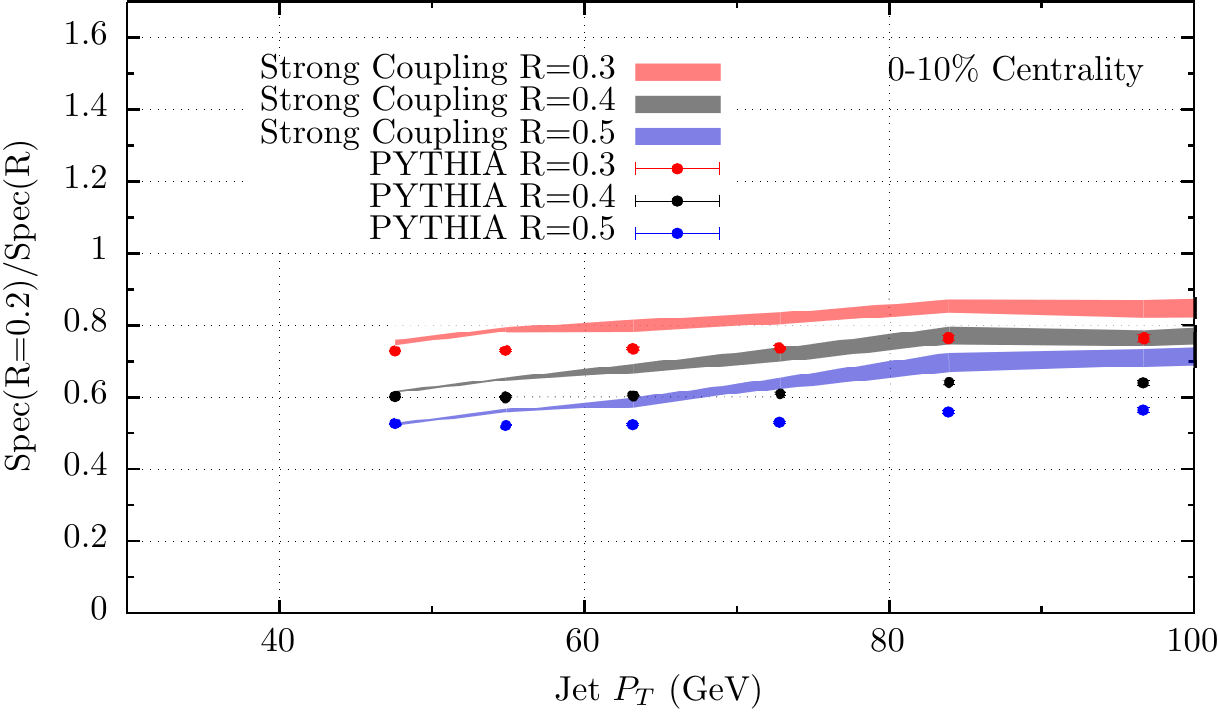}
\caption{\label{Fig:SpecRatio} Ratio of the spectra of jets reconstructed with the anti-$k_t$ parameter $R=0.2$ to the spectrum
of jets reconstructed with various larger values of $R$   as a function of the reconstructed jet $p_t$. The colored bands correspond to the prediction of our 
hybrid model, with no broadening ($K=0$) and including the effects of the backreaction of the medium.  The dots are the predictions
for jets in vacuum, in proton-proton collisions as described by \pythia.
As explained in the text, the fact that the three colored bands are closer together than the three sets of
dots are is due to the fact that wider jets lose more energy than narrower jets do.
}
\end{figure}

To further study the $R$-dependence of jet suppression in our model, and motivated by the ALICE analysis reported in Ref.~\cite{Adam:2015doa}, 
in \fig{Fig:SpecRatio} we show the ratio of spectra of jets in PbPb collisions reconstructed with different values of the anti-$k_t$ parameter $R$  
to that for $R=0.2$ jets. 
For comparison, we show the same ratio in proton-proton collisions as simulated by \pythia. 
An important advantage of this observable is that it is constructed with PbPb data only. 
This leads to a significant reduction in the theoretical uncertainties in our model, 
since the errors in the spectra with different $R$ 
are correlated. This is the reason why the widths of the colored bands depicting our theoretical predictions displayed in \fig{Fig:SpecRatio} 
are significantly narrower than those for our predictions of $\RAA$.  Because of this reduction in the theoretical uncertainty, as we
shall now explain this observable 
is sufficiently discriminating to show that, within the range of $R$ studied, the wider jets that are reconstructed with larger values of $R$
lose more energy than narrower jets.  To see this, let us first understand the behavior of this observable for jets produced in proton-proton
collisions, which evolve in vacuum. Predictions from \pythia for these jets are shown as the colored dots in \fig{Fig:SpecRatio}.
In vacuum, the number of jets with a given $\pt$ increases as $R$ increases, since reconstructing wider jets incorporates 
a larger fraction of the initial partonic energy into the jet. 
Therefore, the ratio of the spectra of jets reconstructed with $R=0.2$ to that of jets reconstructed 
with a given $R$ decreases with increasing $R$.  
In the medium produced in PbPb collisions, 
this general trend is also observed. However, this ratio decreases more slowly  with increasing $R$,
meaning that the number of jets with a given $\pt$ now increases more slowly with increasing $R$ than was the case in vacuum. 
This means that the wider jets reconstructed with larger $R$ have lost more energy than the narrower jets have.


Another interesting feature of this observable is that the deviation between the vacuum and medium ratios increases as the momentum of the jet increases. 
This, too, is a consequence of the fact that wider jets which contain more in-medium partons lose more energy than narrower jets.
Low $\pt$ jets contain, on average, a smaller number of partons propagating simultaneously in medium. 
The extreme case is a jet that consists only of a single parton while it is in the medium, although it may
branch later.  In this case, the energy loss is independent of the reconstruction parameter $R$: if this jet
is reconstructed, its energy loss is the energy lost by that single parton no matter what the value of $R$.
The fact that the three colored bands in \fig{Fig:SpecRatio} come closer together as $\pt$ increases
indicates that the lowest $\pt$ jets are dominated by jets that contain only very few partons while
they are in the medium.  As the $\pt$ increases, the population of jets includes some that are wider, with more
partons, that lose more energy.  And, at any given $\pt$ more of these are reconstructed at larger $R$.
Although a direct comparison with the semi-inclusive jet measurements performed by ALICE~\cite{Adam:2015doa} 
that motivated us to make \fig{Fig:SpecRatio} 
is not possible, our analysis highlights the potential that precise measurements of this type will have 
to discern the mechanisms of jet quenching in future.

\begin{figure}[t]
\centering 
\begin{tabular}{cc}
\includegraphics[width=.5\textwidth]{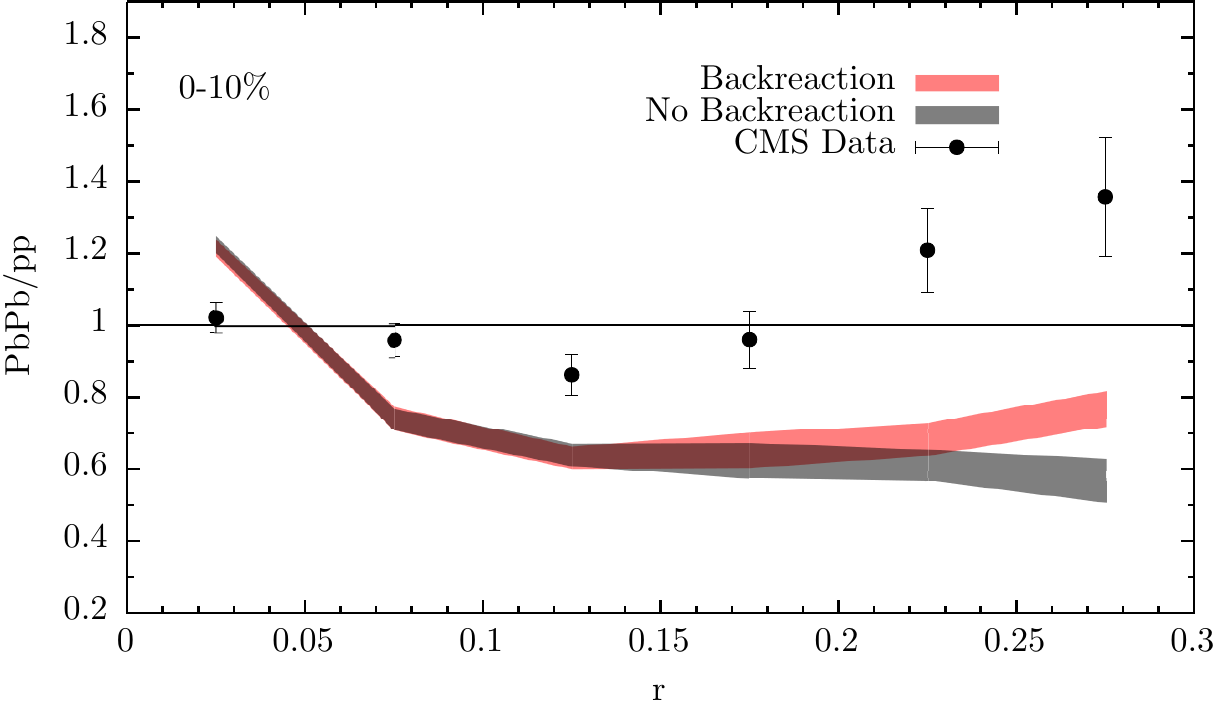}
&
\includegraphics[width=.5\textwidth]{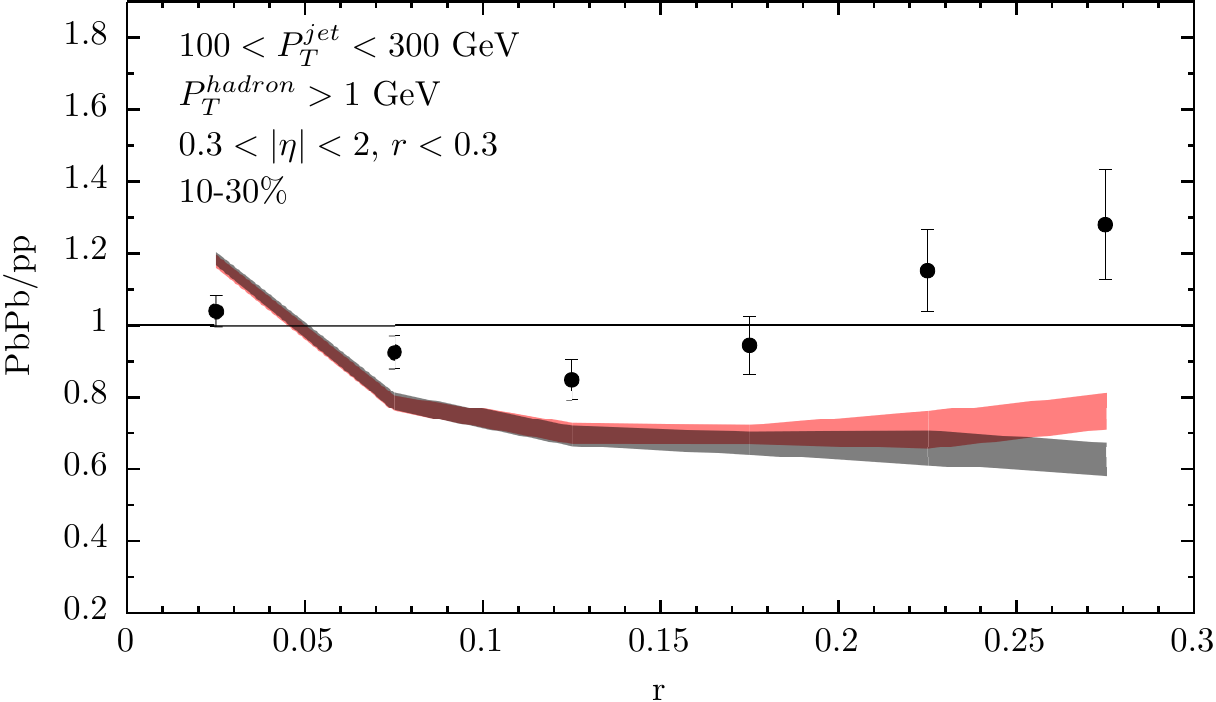}
\end{tabular}
\caption{\label{Fig:backShapes} Ratio of the jet shape in PbPb collisions with $\sqrt{s}=2.76$~ATeV 
with 0-10\% centrality (left) and 10-30\% centrality (right) to the jet shape in proton-proton collisions.
The two colored bands show the results of our hybrid model calculation with no broadening, with both jets and background hadronized,
and with our background subtraction procedure for high-$\pt$ jets applied.  In the calculation shown as  the red band we include the effects of
backreaction, namely the particles coming from a wake in the medium.  We compare our calculation with and without
backreaction to data from CMS~\cite{Chatrchyan:2013kwa}.
}
\end{figure}

Paralleling the discussion in Section~\ref{sec:Section2},
to explore the effects of medium-modification of jets on their angular structure 
we now turn to the jet shapes ratio, shown in \fig{Fig:backShapes} for two different centralities.
We set $K=0$ since, as we saw in Section~\ref{sec:Section2}, transverse momentum broadening has little 
effect on this jet shapes ratio.
We describe the way in which we subtract the thermal background in Appendix~\ref{app:bgsubs}.
Since the fluctuations in our simplified background do not coincide with those of an actual  heavy ion collision, we need to 
correct for the difference in jet energy resolution in order to do a fair comparison with CMS data~\cite{Chatrchyan:2013kwa}.
This amounts to smearing the jet energies with a Gaussian whose width corresponds  
to the difference between the jet energy resolution in the presence of our background and
the jet energy resolution measured by CMS; we describe the procedure in Appendix~\ref{app:bgsubs}.
Last, we subtract background tracks in the jet cone
following a simple procedure from Ref.~\cite{Chatrchyan:2013kwa} 
in which
we subtract the $\eta$-reflection of each event from that event. This procedure does not work for jets near $\eta=0$; this is why 
 $|\eta|<0.3$ is excluded from both our analysis and the measurement reported in~\cite{Chatrchyan:2013kwa}.

To gauge the effects of adding our simplified background, performing the background subtraction procedure, and hadronization on one hand, 
and the effects due to the backreaction of the medium, namely the particles coming from the wake
in the plasma, on the other in both 
panels we show the jet shape ratio computed at the hadronic level with and without backreaction.
As we saw in Section~\ref{sec:Section3}, energy loss serves to narrow the
angular size of jets in a given window of energies in heavy ion collisions relative to that of jets with the same
energies in proton-proton collisions. 
As a consequence, without backreaction the effect of energy loss is to increase the importance of narrow jets in the quenched jet sample, leading to a depletion of the  jet shape at large angles $r$. 
Note that the only differences between the simulations without backreaction in  \fig{Fig:backShapes}  and the $K=0$ simulations displayed in \fig{Fig:broShapes} are: adding the simplified but fluctuating background that we are employing, performing our background subtraction and jet reconstruction, and
adding hadronization. 
The partonic distributions whose ratio is plotted in \fig{Fig:broShapes} 
give rise to narrower distributions that the hadronic ones that go into \fig{Fig:backShapes}, a natural consequence of the non-trivial angular distribution of the Lund strings connecting the hard partons within the jet which means that hadronization
broadens  the jet somewhat. (See for example Ref.~\cite{Ellis:2007ib}.)

Despite the hadronic uncertainties, the jet shape ratio shows a clear increase at larger values of the angular
variable $r$ when we include backreaction, confirming the expectation that some of the particles from
the wake in the plasma do end up reconstructed as part of the jet, and confirming the expectation that 
they are less tightly focused in angle than the jet itself was.  That said, it is also clear 
from \fig{Fig:backShapes} that the data features a much stronger increase in the jet shape ratio
at large angular distance $r$ 
from the jet axis than
we obtain from our calculation.
(Although the effect is large when plotted in this way, it is important to keep in mind that the 
enhancement in the ratio of jet shapes at large $r$ does not mean that in-medium jets are significantly
wider than vacuum jets: in both populations, most of the jet energy is concentrated at small $r$; at large $r$
what is plotted is the ratio of two quantities that are both small.
The data indicate that in PbPb jets the fraction of the jet energy at $r>0.2$ 
is larger than in proton-proton jets, but both are small.) 
Comparing the measured jet shape ratio to our calculations in \fig{Fig:backShapes} 
tells us that our
treatment of backreaction substantially underestimates the amount of  energy that
ends up correlated with the jet axis but separated from it by a relatively large value of $r$.
The hadronization of the wake in the hydrodynamic medium accomplishes this, but
at least in our treatment it does not do enough.
It could be that we are underestimating the amount of energy deposited in the medium
at these angles, or it could be that we are missing energy at these angles that corresponds
to energy lost by the jet that hydrodynamizes only partially or not at all,
or it could
be that our treatment of background fluctuations, background subtraction, and jet reconstruction 
is subtracting away more of the particles originating from the backreaction of the medium
than happens in analyses of experimental data.
And, as we will elaborate in Section~\ref{sec:discussion}, the jet shape analysis is 
sensitive to semi-hard particles in the region of $\pt>2$~GeV, in which the small-perturbation assumption behind \eq{onebody} starts to break down. 
We shall continue this discussion in Section~\ref{sec:discussion}.

 Next, we now turn from the angular jet shape variable to another 
 intrinsically hadronic observable that assesses the longitudinal structure of jets: jet fragmentation functions. 
 These show the distribution of the $z$ of the tracks in a jet, where
 $z \equiv p_{\rm track} \cos \theta /p_{\rm jet}$
is the ratio of the longitudinal momentum of a single charged hadron in the jet (a single track)
to the momentum of the whole jet.  ($\theta$ is the angle between the track and the jet axis.)
Unlike in our previous publications~\cite{Casalderrey-Solana:2014bpa,Casalderrey-Solana:2015vaa},  
where we only  analyzed partonic fragmentation functions, the inclusion of hadronization 
allows us to do a direct comparison with experimental measurements.
However, we must keep in mind that the hadronization process is not under good theoretical control and that we have 
only implemented a simplified prescription which ignores changes in the color flow of jets that shower in a medium. 
We defer the study of effects of 
different prescriptions for hadronization in vacuum and in medium 
within our model to the future.

\begin{figure}[t]
\centering 
\begin{tabular}{cc}
\includegraphics[width=.5\textwidth]{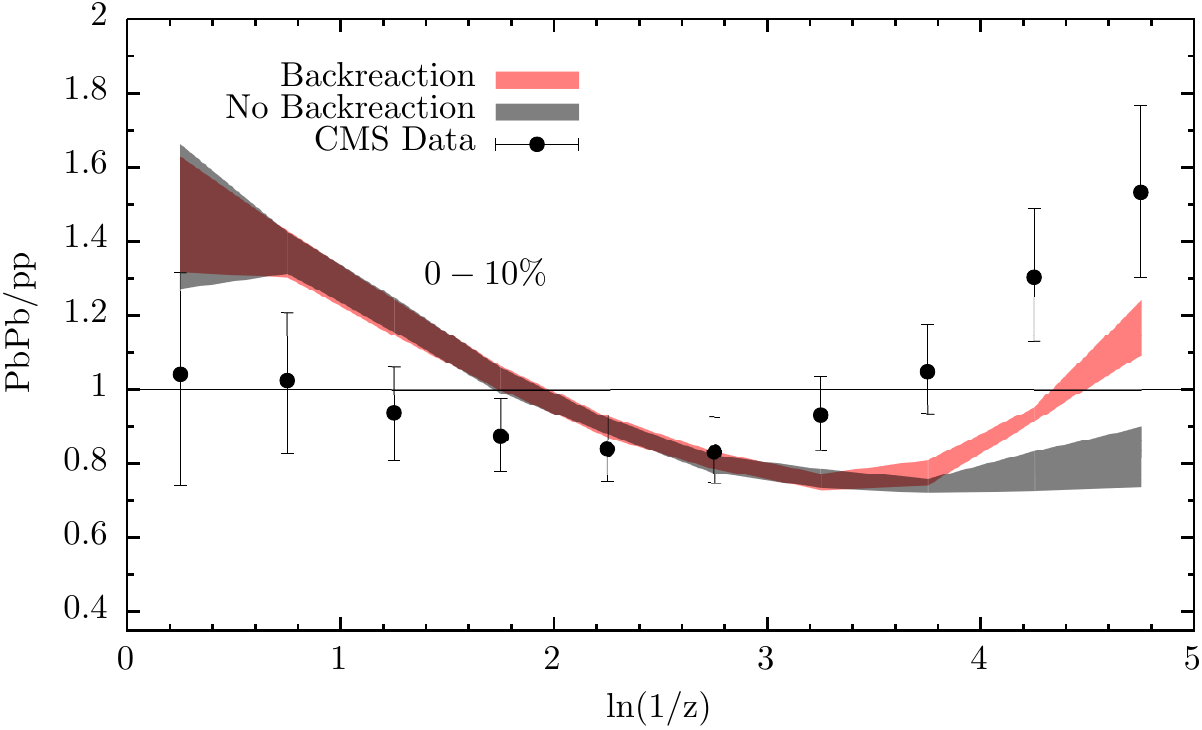}
&
\includegraphics[width=.5\textwidth]{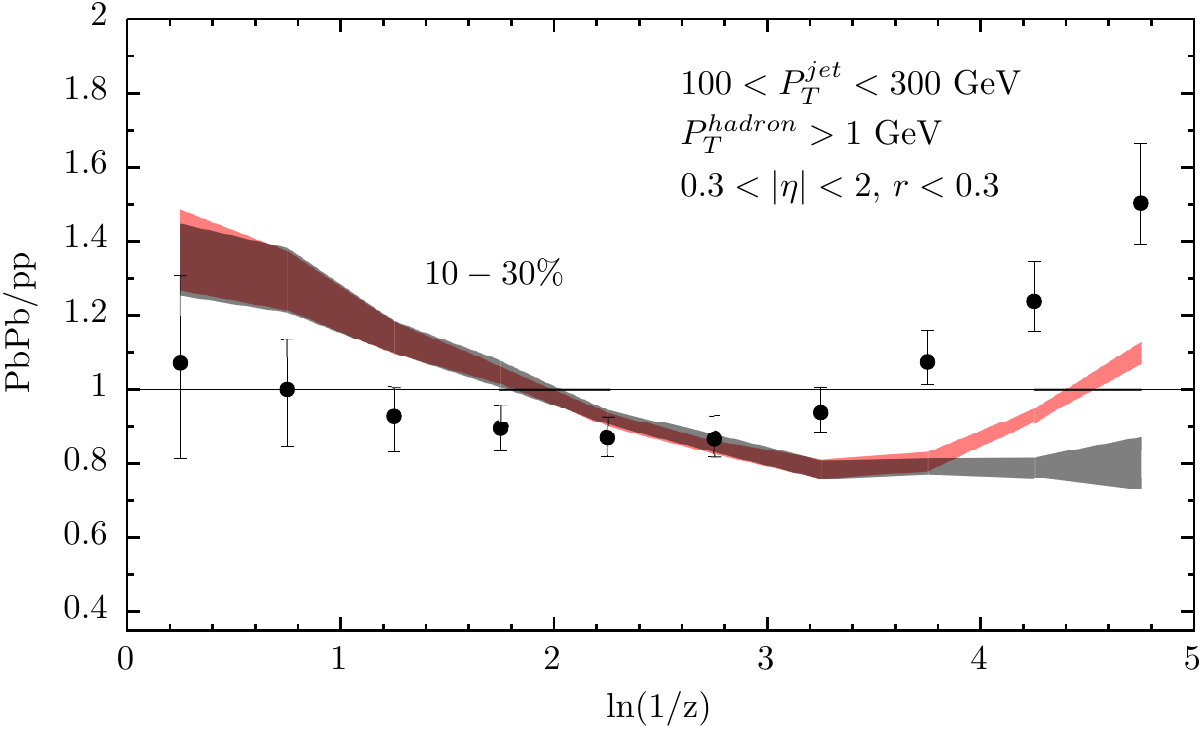}
\end{tabular}
\caption{\label{Fig:backFF} Ratio of the jet fragmentation function in PbPb collisions at $\sqrt{s}=2.76$~ATeV to
that in proton-proton collisions.
As in \fig{Fig:backShapes}, we compare the results of our hybrid model calculation
with and without the inclusion of the  particles coming from the
backreaction of the medium in 0-10\% centrality (left) and 10-30\% centrality (right) collisions
to data measured by CMS~\cite{Chatrchyan:2013kwa}.
}
\end{figure}

Results from our hybrid model calculations of  
jet fragmentation functions are shown in \fig{Fig:backFF} for LHC heavy ion collisions with 0-10\% and 10-30\% centrality. 
Both panels display the ratio between hadronic fragmentation functions in PbPb collisions to 
those in proton-proton collisions compared to experimental measurements of this quantity by CMS~\cite{Chatrchyan:2013kwa}. 
We include two bands, one for the full calculation 
including the effects of the backreaction of the medium, and the other without it.  The overall background subtraction and hadronization
are the same in both calculations, as is the jet energy  resolution correction and the subtraction of
background tracks in the jet cone by $\eta$-reflection.

By comparing the results of our simulations with and without backreaction, the jet fragmentation functions clearly show where the particles resulting from the wake in the plasma  that get reconstructed as part of the jet end up.
The hard part of the jet is practically unaffected by the backreaction of the medium, with an almost identical distribution of hard fragments in the two calculations. (The small differences 
arise from the the small change in reconstructed jet momentum associated with the addition of soft particles from the
backreaction of the medium to the jet.)  Both of the simulations show an enhancement of hard fragments at the largest values of $z$ (smallest $\log 1/z$) in PbPb collisions. 
As we have seen in other ways, wide jets (with more softer fragments) lose more energy, so at any given energy the jets
that remain tend to be narrower, and tend to contain fewer, and therefore more energetic fragments, than in proton-proton collisions.
Such an enhancement therefore seems generic to any mechanism of energy loss which significantly reduces 
the soft, large angle, components of jets. (See Ref.~\cite{Milhano:2015mng} for a similar effect in a perturbative-based jet quenching Monte 
Carlo~\cite{Zapp:2008af,Zapp:2008gi,Zapp:2012ak,Zapp:2013vla}.) 
The small-$z$ region of the fragmentation function is sensitive to backreaction effects. 
The emission of soft particles by the jet-induced wake compensates the suppression of soft fragments due to energy loss and leads to an overall 
enhancement of soft tracks in the PbPb jets relative to proton-proton jets.
The comparison between our calculations with and without the particles coming from the backreaction of
the medium
also shows the range of momenta at which back reaction contributes significantly to the particles reconstructed as part of 
a jet, namely $\pt$ up to $\pt \sim 2.5$~GeV.  At this scale our approximate approach to the wake distribution \eq{onebody} 
underestimates particle production, as we discussed after \eq{eq:deltaNG}.
At softer momenta, by neglecting the effects of viscosity as the sound waves produced by the jet damp out and heat the plasma we are underestimating the particle production also.

Comparing the fragmentation function ratios that we have obtained in our
calculations including the particles from the backreaction of the medium
to the ratios measured by CMS~\cite{Chatrchyan:2013kwa}, we see qualitative
similarities, but not quantitative agreement.  At large $z$ (small $\ln(1/z)$)
our calculated ratio is above 1 while there is no evidence for this in the data.
However, in this regime the uncertainties in both data and theory are significant.
At small $z$ (large $\ln(1/z)$) we see that the particles from the wake in the
plasma that are reconstructed as part of the jet turn the fragmentation function
ratio upwards, but they do not turn it as far upward as in the data, and they
turn it upward at a larger value of $\ln(1/z)$ than in the data.  This suggests
that our treatment of the back reaction of the medium
is missing an increase in the production of few GeV particles.
We have already suggested several possible interpretations of 
this above.  It is quite reasonable to expect that whatever physics it is
that our approximations have missed that would push the large $\ln(1/z)$ 
fragmentation function ratios up would also push the jet shapes up
at larger values of $r$ in \fig{Fig:backShapes}.  We are missing some 
soft to few GeV particles at these values of $r$.



Certainly a more complete description of the backreaction dynamics, from the wake in the plasma to the resulting particles reconstructed
as part of a jet, would be helpful. 
%
It is also worth mentioning, though,
that we have found a  disagreement between the fragmentation functions from \pythia simulations of jets in proton-proton collisions 
and the fragmentation functions measured in small $R$ jets in these collisions, for example with the fragmentation functions from \pythia 
high by as much as 30\% at small $z$, at least with the \pythia tunes we have explored (Monash, 4C and 4C with a modified $Q_0$ parameter). A similar disagreement has also been found between 
state-of-the-art parametrizations of fragmentation functions and proton-proton data~\cite{d'Enterria:2013vba}. These uncertainties are comparable in size to  the size of medium modifications themselves.
This means that understanding fragmentation function ratios more quantitatively will  require a
better understanding of hadronization of jets, both those in vacuum and those that have been
modified by passage through a medium.
Regardless, it is already an inescapable conclusion
that  backreaction dynamics has a significant impact on the jet fragmentation function at small $z$. 

In the next Section we will explore a suite of observables that are even more
sensitive to the soft particles from the wake in the plasma that end up reconstructed
as a part of the jet than the jet shape or the jet fragmentation function.


\subsection{\label{sec:mpt}Recovering the ``Lost'' Jet Energy and ``Missing'' $\pt$ Deposited by a Dijet Pair}

We have seen in the previous subsection that the backreaction of the medium in response
to the passage of a jet, namely the wake in the plasma, contributes to single jet observables
like the jet $\RAA$, jet shape, and jet fragmentation function because some of the particles
that result from the hadronization of the wake in the plasma must of necessity end up counted
as part of the jet after the jet is reconstructed and the background is subtracted.  In these
observables, the effects of the particles coming from hadronizing the wake in the plasma
constitute small corrections overall, although they can play a significant role in certain
kinematic regimes (like in the soft region of the fragmentation function ratio and the large $r$
region of the jet shape ratio).
In this subsection, we focus on a suite of observables which are {\it dominated} 
by effects originating from the backreaction of the medium in response to the
passage of two jets, a dijet, through it.

The principles behind our simplified calculation of the spectrum of particles produced
as the medium (including the wake therein) hadronizes are that the medium acquires
the energy and momentum lost by the jets passing through it, which thermalizes subject
to energy and momentum conservation.  This implies collective motion of the medium, a wake
in the quark-gluon plasma.  After hadronization, this energy and momentum is
recovered in the form of soft particles
with 
a wide angular distribution given by \eq{onebody}.  
This makes it natural for us to calculate the so-called
``missing-$\pt$ observables'' introduced recently by CMS~\cite{Khachatryan:2015lha}, and compare
our calculations to measurements made by CMS. 
These observables are defined to extract the spectrum and angular distribution
of the particles that correspond to the net momentum lost by the two jets in a dijet pair.
Each jet in the pair loses energy and momentum to the plasma, but in general one
will lose more than the other.  By constructing observables designed
to recover the net momentum lost by the pair
of jets we obtain observables that are dominated by the response of the medium to
the jets.  These observables are constructed from {\it all} the tracks in an event, not
just those reconstructed as part of a jet.
We shall set $K=0$ throughout this subsection, neglecting transverse momentum broadening.  We have obtained results with $K\neq 0$,
 but the $K$-dependent effects that we see in these missing-$\pt$ observables are all much smaller in magnitude than the effects of 
 particles coming from the hadronization of the wake in the plasma. As these observable consequences of the backreaction of the medium are our
 focus in this subsection, there is no reason to introduce $K\neq 0$.

The observables that we look at in this subsection are intrinsically hadronic and, therefore, are hard 
to bring under full theoretical control without significant modelling of the differences between hadronization in heavy ion collisions
and proton-proton collisions. 
Nevertheless, we will observe several qualitative features of our results which are similar to what is seen in the experimental data.
This indicates that our treatment, 
although simplified, captures some of the main aspects of the collective reaction of the plasma to the deposited jet energy.

Following Ref.~\cite{Khachatryan:2015lha}, the missing $\pt$ analysis consists in studying the conservation of momentum in heavy ion events which include hard jets reconstructed with a specified anti-$k_t$ reconstruction parameter $R$.  The first step in the analysis is to use the anti-$k_t$ algorithm
to find a sample of events containing at least two jets and to determine the $\pt$ and the jet axis of the leading and subleading jet in each event.
We then define the dijet angle $\phi_{\textrm{dijet}}$ as the bisection between $\phi_{\rm leading}$ and
$-\phi_{\rm subleading}$, where $\phi_{\rm (sub)leading}$ is
the azimuthal angle of the (sub)leading jet.  The only use of the jet reconstruction algorithm is the selection of 
the event sample and the determination of the jet $\pt$'s and axes and hence $\phi_{\rm dijet}$.  In the remainder
of the analysis, all charged tracks in the events in the sample are used.

The next step in the analysis is to project the momenta of each of the charged tracks in 
an event along the dijet axis. This projection is defined by
\begin{equation}
\mpt \equiv - \pt \cos\left(\phi_{\textrm{dijet}}-\phi\right) \, ,
\end{equation} 
where $\pt$ and $\phi$ are the transverse momentum and azimuthal angle of the track.
With this convention, $\mpt$ is positive for tracks in the subleading jet hemisphere and negative for tracks in the leading jet hemisphere. 
By momentum conservation, the sum of all the $\mpt$ for all the tracks   in an event must be zero. 
However, in the experimental analysis only charged tracks with 
$|\eta|<3$ are considered, and therefore, the net $\mpt$ need not cancel identically. 
Nevertheless, by studying the approximate cancellation of this momentum component as a function 
of which tracks we include in the analysis, out to what angular distance from the dijet axis, 
we can extract valuable information about the distribution of the lost jet energy.

\begin{figure}
\centering 
\begin{tabular}{cc}
\includegraphics[width=.5\textwidth]{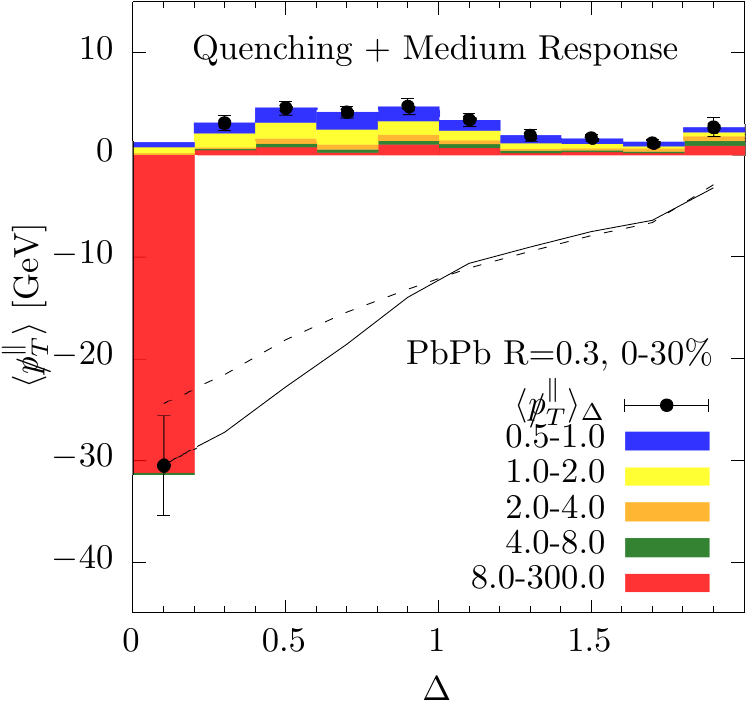}
&
\includegraphics[width=.5\textwidth]{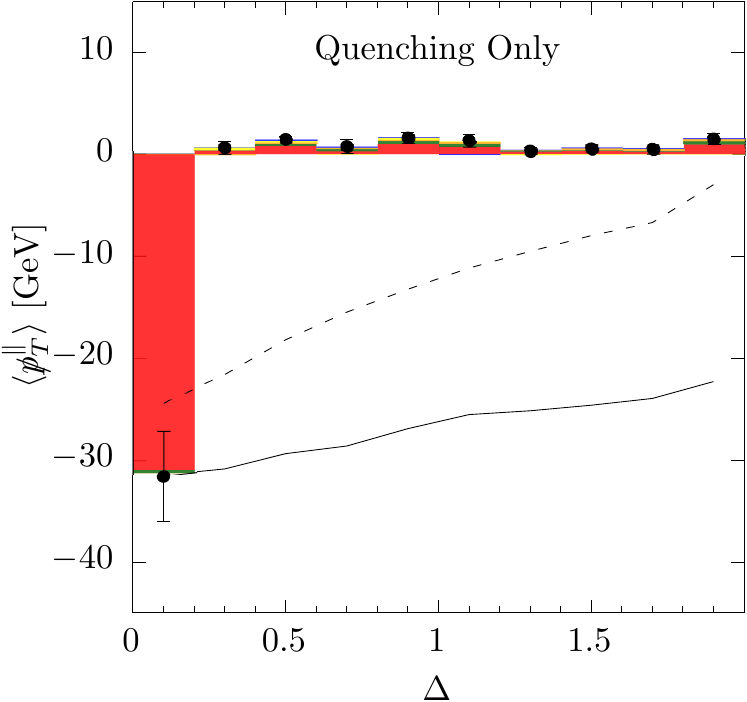}
\end{tabular}
\caption{\label{Fig:mptbacknoback} Results from our calculations of the missing $\mpt$  
for tracks that are $\Delta$ away from the dijet axis in $(\eta,\phi)$-space,
in PbPb collisions containing a pair of $R=0.3$ jets satisfying kinematic criteria
described in the text. 
The first nine bins in $\Delta$ are each 0.2 wide, together covering
$0<\Delta<1.8$.  The tenth bin, plotted at $\Delta=1.9$, contains the missing $\mpt$ for all the tracks
with $\Delta>1.8$.  Note that we limit our analysis to tracks with $|\eta|<3$. 
In the left panel, we show the results of our full calculation, with energy loss calculated
within our hybrid model with no momentum broadening, and with particles coming from the backreaction
of the medium included.  
The right panel shows our results if we leave out the particles coming from
the backreaction of the medium.  
In both plots, the solid black  points show the
net $\mpt$ 
of all charged tracks with $|\eta|<3$
in each $\Delta$ bin.  The contribution to this net momentum of tracks in different momentum bins are codified by the colored histograms. 
In both plots, the solid line shows the cumulative sum of all the $\mpt$ in a $\Delta$ bin and all the $\Delta$ bins to its left.
The black points, the colored histograms, and the solid line are all calculated for 0-30\% central 
PbPb collisions with $\sqrt{s}=2.76$~ATeV.
For comparison, we have repeated the calculation for proton-proton collisions as described by our \pythia tune,
and the dashed line in each plot shows
the same cumulative sum of $\mpt$ for proton-proton collisions that the solid line shows for PbPb collisions.
}
\end{figure}

  We start by computing the $\mpt$ distribution sliced in $\Delta$ bins, where $\Delta$ is the distance in the $(\eta,\phi)$ 
  plane between the 
  track in question and either the leading jet axis or the subleading jet axis, whichever the track is closest to.
  We consider dijet pairs with 
  leading and subleading transverse momenta $\pt^{\rm leading}>120$~GeV and $\pt^{\rm subleading}>50$~GeV respectively,  
  and with both jets within $|\eta|<2$.  We also limit our sample to events in which the two jets are 
  back-to-back, our criterion being $\Delta \phi>5 \pi /6$. 
  Finally, after  identifying the dijet pair, only those dijets in which both jets have $|\eta|<0.6$ are considered.  
  (We start by finding jets within a much larger $|\eta|$ window in order
  to be as confident as we can be that when we have identified a dijet in which both jets
  have $|\eta|<0.6$ these two jets really are the leading and subleading jets in the event.)
   All these specifications of our event sample are the same as those in the experimental
  analysis of Ref.~\cite{Khachatryan:2015lha}.
  The result of this analysis is shown in  \fig{Fig:mptbacknoback} for our full calculations including 
  the medium response (left) and, for comparison, without medium backreaction (right).  
In the right panel, we confirm that momentum is not conserved in our hybrid model
by its construction, when we do not include the particles coming from the wake in the plasma.
The left-most bin in $\Delta$ shows the momentum
imbalance between the high-momentum tracks in the two dijets themselves, with more
momentum in the leading jet direction and less in the subleading jet direction.  In our
hybrid model the two jets have lost different amounts of momentum, and in the right panel
we have neglected the fact that the lost momentum is deposited as two wakes in the medium.
The left panel of \fig{Fig:mptbacknoback} clearly shows how 
the inclusion of the response of the medium, modeled via \eq{onebody}, 
results in a population of soft particles spread broadly in $\Delta$ 
with a net momentum in the subleading
jet hemisphere (positive $\mpt$).  The subleading jet has lost more momentum,
and hence the wake in the plasma moving in the subleading jet direction contains
more momentum than the wake moving in the leading jet direction.


Another way to see the effect of including the backreaction of the medium is to compare
the solid black curves in the two panels of \fig{Fig:mptbacknoback}.
While the accumulated $\mpt$ tends to zero as $\Delta$ increases in the calculations in the left panel, in the right panel the accumulated $\mpt$
is $\sim 20$~GeV after all $\Delta$ bins have been summed. This value, which corresponds to the average imbalance in the energy lost by the 
two jets in the dijet pair, compares well with simple estimates~\cite{CasalderreySolana:2010eh}. 
The medium response transforms the ``lost''  energy from the two jets into two wakes, and hence 
into softer particles in the range of $\sim 0.5-2$~GeV whose net momentum is preferentially
in the hemisphere of the more quenched, subleading, jet.
This is shown by the enhancement of soft $\mpt$ tracks observed in the left panel of the figure. 

The dashed curve, which is the same in both panels, shows the accumulated $\mpt$ for 
proton-proton collision events as described by our \pythia tune.  Recalling that our analysis
employs all the tracks in an event, not just those reconstructed as part of a jet, it is perhaps
not too surprising that the $\mpt$ distribution generated by \pythia that is described
by the dashed curve is not in full agreement with measurements made in proton-proton
collisions reported in Ref.~\cite{Khachatryan:2015lha}.  This small discrepancy
illustrates the difficulty in obtaining full theoretical control over this
inherently hadronic observable. What we shall do in subsequent figures
is to subtract the $\mpt$ distribution in our \pythia calculation of proton-proton
calculations from the $\mpt$ distribution in our PbPb calculation including
black-reaction.  That is, we shall subtract the dashed curve from the solid
curve in the left panel of \fig{Fig:mptbacknoback}.  In this way, we shall focus 
explicitly on effects that are due to medium-induced energy loss, and the
wake in the medium, both of which our model is intended to describe.  

By comparing the solid and dashed curves in the left panel of \fig{Fig:mptbacknoback},
we see that the net momentum lost by the pair of jets is distributed
over a wide angular region, over $\Delta \lesssim 1$. 
We can also see from the similarity between the red histograms in the two panels in the figure that the backreaction
of the medium does not affect the hard tracks in the event, namely the hard components
of the jet.  We also see that almost the entire imbalance in the high-momentum tracks
is an imbalance at small $\Delta$ with the net momentum in the leading jet hemisphere,
whereas almost the entire imbalance in soft tracks is in the hemisphere of the
event in which that corresponds to particles from the (larger) wake of
the subleading jet.   The experimentally measured
distributions~\cite{Khachatryan:2015lha} are quite similar
to what we see in the left panel of  \fig{Fig:mptbacknoback},
but because of the discrepancy between our calculations and
data for proton-proton collisions we will postpone
a quantitative comparison until we have results for
the difference between the missing momentum in 
PbPb and proton-proton collisions, below.

Following Ref.~\cite{Khachatryan:2015lha}, we sharpen the analysis by 
studying how the $\mpt$ distribution 
depends on the dijet asymmetry
$A_J \equiv (\pt^{\rm leading} -\pt^{\rm subleading})/(\pt^{\rm leading}+\pt^{\rm subleading})$. 
In events in which the dijet pair features a larger $A_J$ we should expect a correspondingly
larger momentum asymmetry in the soft particles coming from the hadronization of 
the wakes of the two jets.  As discussed above and as in Ref.~\cite{Khachatryan:2015lha},
we shall also focus on the difference between the $\mpt$ distribution in PbPb collisions
and that in proton-proton collisions containing dijets satisfying the same kinematic criteria.
Jet quenching makes the $A_J$-distribution wider in PbPb collisions, but it is already rather wide in proton-proton
collisions because dijets are often produced in association with a third jet or other radiation, making
it important for us to focus on the difference between PbPb and proton-proton collisions. 

\begin{figure}
\centering 
\vspace{-0.2in}
\includegraphics[width=0.74\textwidth]{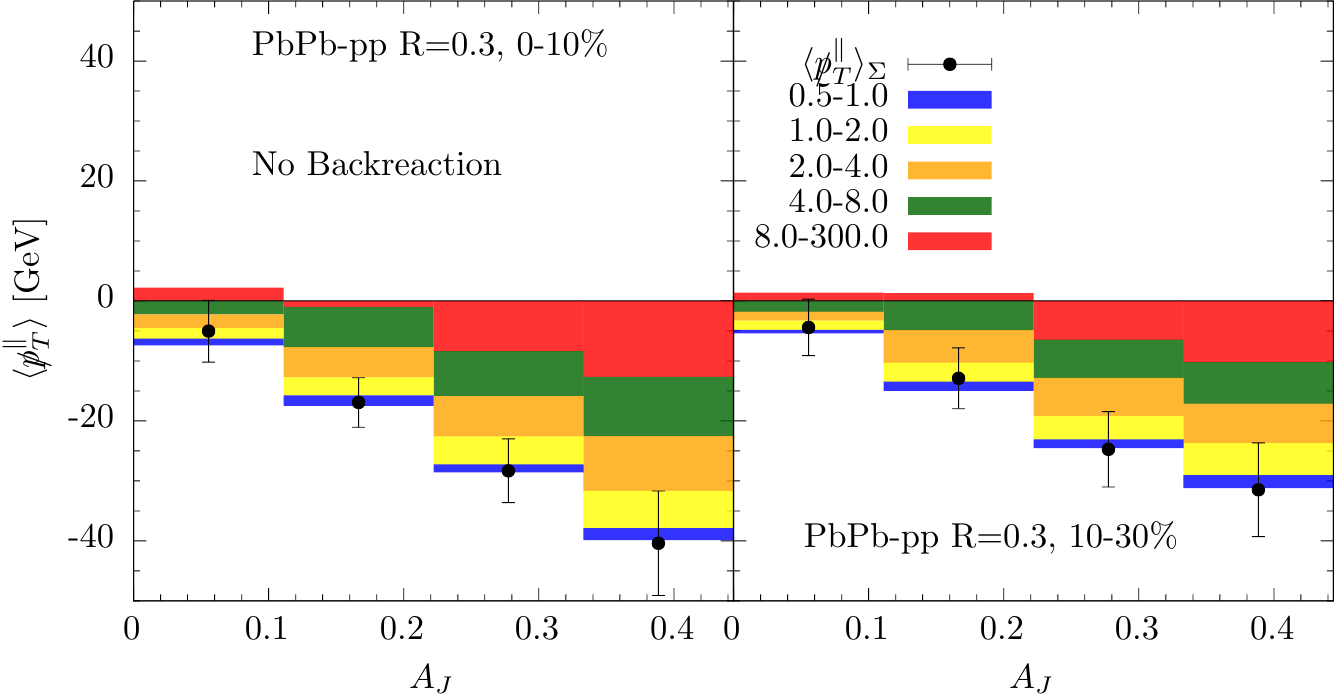}
\vspace{-0.1in}
\includegraphics[width=0.74\textwidth]{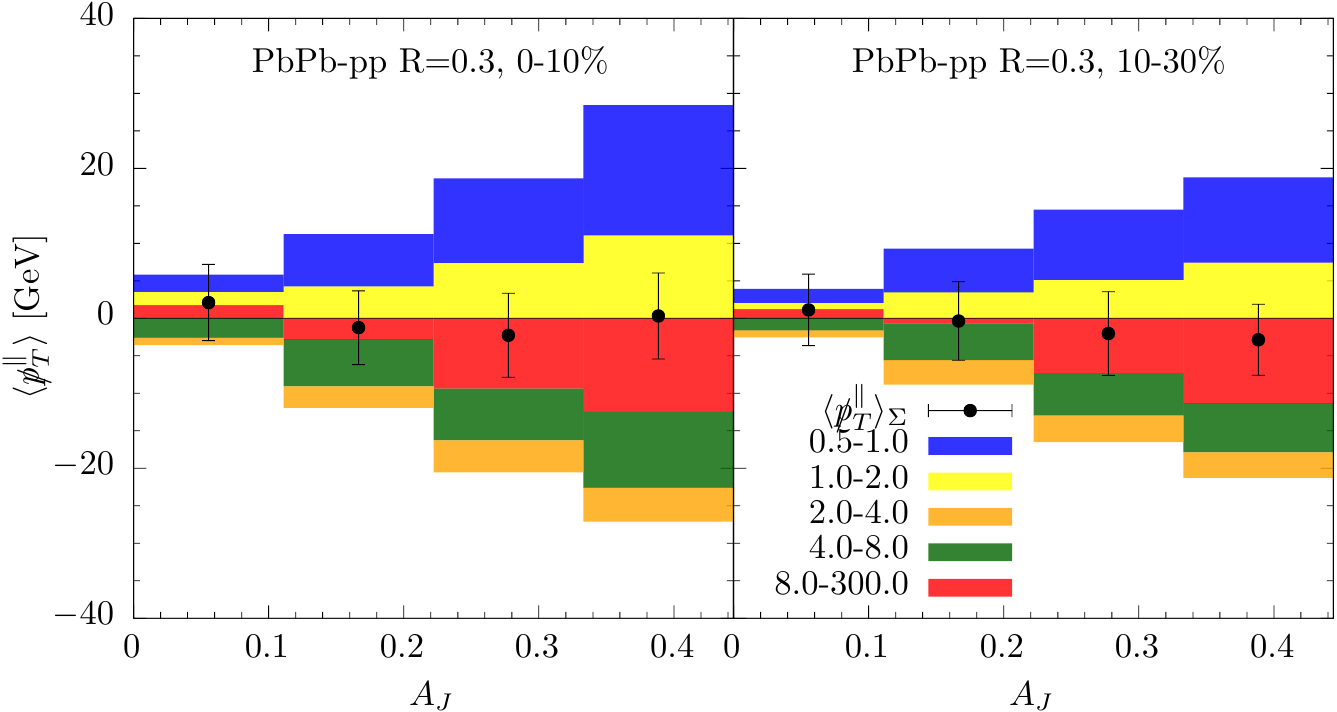}
\includegraphics[width=0.78\textwidth]{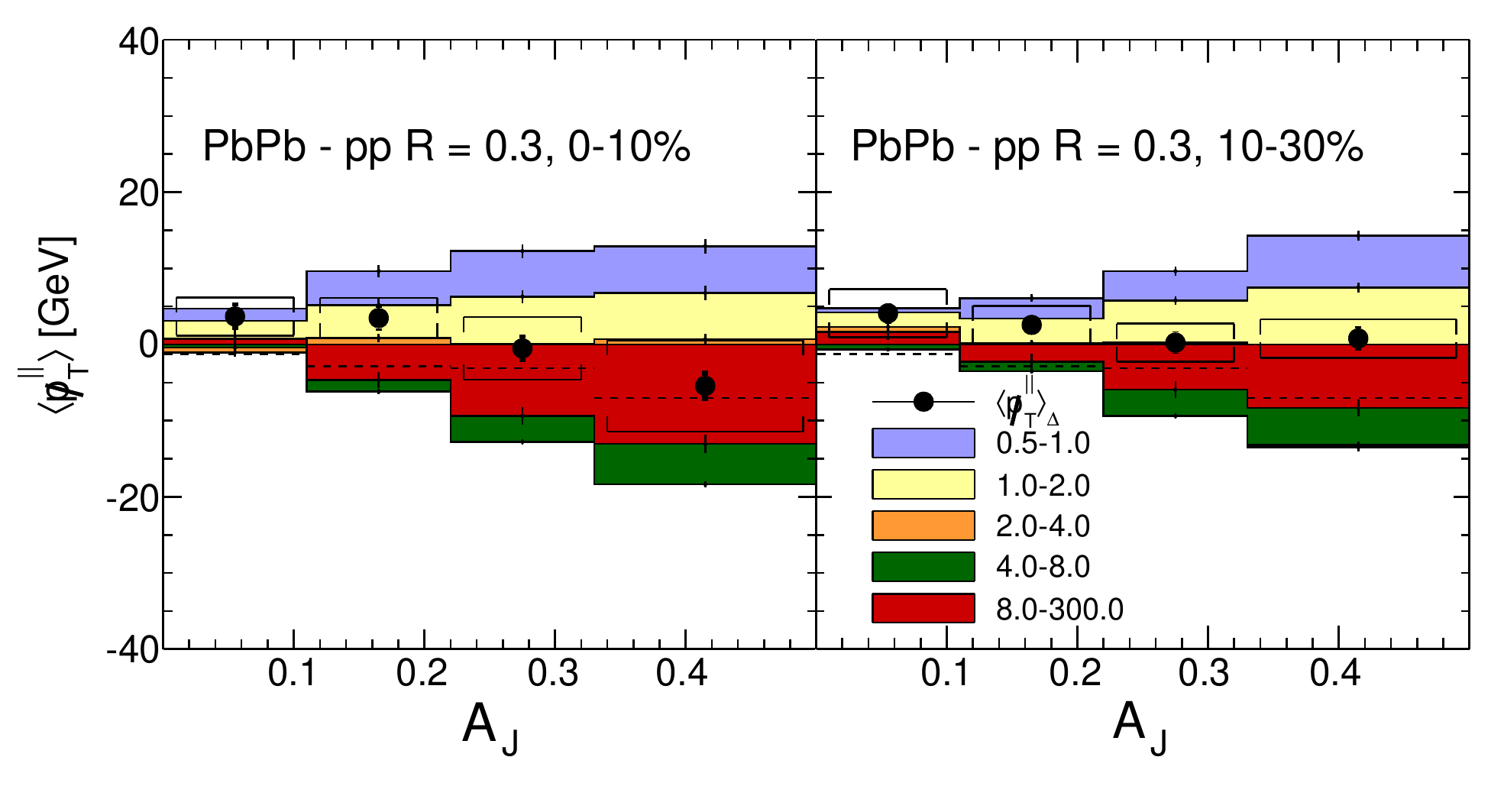}
\vspace{-0.2in}
\caption{\label{Fig:ajcentrality} The middle two panels show the results of our full calculation, while
the upper two panels show the results from our hybrid model for energy loss with the contribution from
the backreaction of the medium left out. Black points with error bars show the difference between the missing $\mpt$  in PbPb and pp collisions, integrated over all $\Delta$, summed over all track momenta, binned in the dijet asymmetry $A_J$,
for events containing $R=0.3$ dijets in collisions with 0-10\% (left) and 10-30\% (right) centrality. The colored histograms show the missing $\mpt$ for tracks whose momenta
is in the indicated range.  The lower two panels, obtained from Ref.~\cite{Khachatryan:2015lha}, show the results from the same analysis performed on experimental
data by the CMS collaboration.
}
\end{figure}

In the upper and middle panels of \fig{Fig:ajcentrality} we show the value of $\mpt$ obtained from our 
calculations upon integrating over all angular separations $\Delta$
in PbPb collisions minus the same obtained in proton-proton collisions, in both cases
for dijet events reconstructed using the anti-$k_t$ algorithm with $R=0.3$ 
in bins of the dijet asymmetry $A_J$, in both cases for collisions with  two different centralities. 
In the upper panels, we include the modifications to the jets because of 
energy loss but leave out the contributions
from the backreaction of the medium.  The middle panels show the results
from our full calculation, including both the modifications to the jets and the
wakes in the plasma.
The black points in the upper panels show the momentum imbalance
in dijet events when the momentum carried by the wakes are neglected.
The black points
in the middle panels confirm that total momentum is conserved regardless of the asymmetry $A_J$ --- once we include
the particles coming from the wakes in the plasma.
We see from the red, green and orange histograms in the upper panels of \fig{Fig:ajcentrality}, though, that  events with a larger and larger dijet asymmetry $A_J$  feature more and more missing $\mpt$, in particular for tracks with
$\pt >2$~GeV, with a sign corresponding to there being more tracks with these (hard) momenta in the leading jet direction.
That is, jet quenching --- here described in our hybrid model --- suppresses the hard particle contribution in the subleading
jet hemisphere, as the subleading jet has lost more energy.  The energy and momenta lost by both jets have been thermalized
as wakes, which in turn are transformed into softer hadrons when the medium including the wakes freezes out.  
We see  the contributions of the momentum carried by hadrons coming from the wakes in the plasma
by comparing the upper and middle panels of \fig{Fig:ajcentrality}.  
In
the blue and yellow histograms in the upper panels of the figure, we see the
missing $\mpt$ for tracks with $\pt <2$~GeV with a sign corresponding to there being more tracks with these (soft) momenta
in the subleading jet direction.  Since the subleading jet lost more momentum, its wake contains more momentum and this shows
up in the missing $\mpt$ of the soft tracks in the event.
All these effects are larger in the central collisions on the
left than in the more peripheral collisions on the right, as expected since due to the smaller size of the fireball 
the dijets are on average less quenched  in more peripheral collisions.

Almost all the features seen in the middle panels of \fig{Fig:ajcentrality} that we have described
above are present in the missing $\mpt$ measurements reported  by the CMS collaboration \cite{Khachatryan:2015lha} that
we show as the lower panels in \fig{Fig:ajcentrality}, at the least at a qualitative level
and in most cases to a degree that constitutes agreement within the uncertainties.

The biggest difference between the results of our calculations and the experimental
results is found for semi-hard particles in the 2~GeV$< \pt <$~4~GeV momentum range.
By comparing the orange histograms in the upper and middle panels of \fig{Fig:ajcentrality},
we see that in our calculation the wake contributes little in this momentum range:
the ``orange contribution'' to the missing $\mpt$ imbalance due to jet quenching that is seen in the upper panels
remains almost unmodified in the middle panels.  Instead, in the data there is almost no
missing $\mpt$ imbalance in this momentum range.  Furthermore, by comparing the upper
and middle panels we see that in our calculation the contribution to the missing $\mpt$
from soft particles with $p_T<2$~GeV is dominated by the effects of the
backreaction of the medium, and then by comparing the middle and lower panels
we see that our calculations yield a greater contribution from these soft particles
than in the data.
So, although
our implementation of medium backreaction does restore momentum conservation --- 
the $\mpt$ excess in hard particles in the direction of the  leading jet due to jet
quenching is balanced by the $\mpt$ excess in soft particles in the direction of the
subleading jet coming from the wake in the plasma --- our calculation of the
particles coming from the wake seems to yield somewhat more particles with $\pt<2$~GeV
than in the data and substantially fewer particles with 2~GeV$~< \pt<4$~GeV.
We will defer further discussion of the origin of this discrepancy to Section~\ref{sec:discussion}.
For the present it suffices to remember from our discussion earlier in this Section
that several of the approximations that
we used in setting up our crude treatment of the particles from the wake
break down for particles with $\pt$ substantially greater than the temperature
of the medium at freezeout, and they break down in a way such that we anticipated underpredicting
the production of semi-hard particles from the wake.

\begin{figure}[!htbp]
 \centering
 \includegraphics[width=1.01\textwidth]{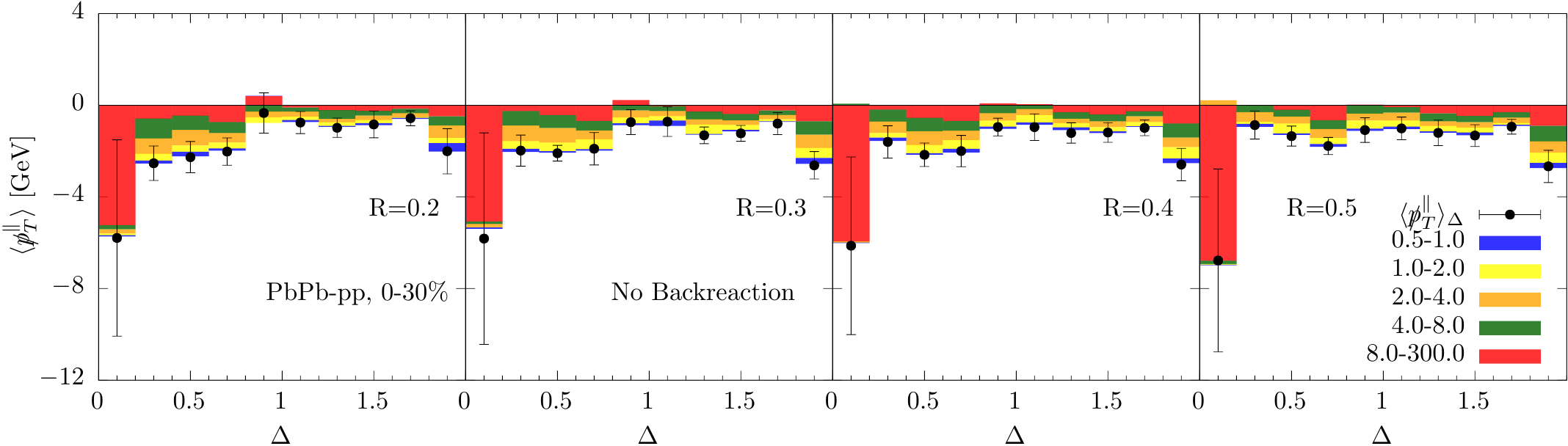}
 \includegraphics[width=1.01\textwidth]{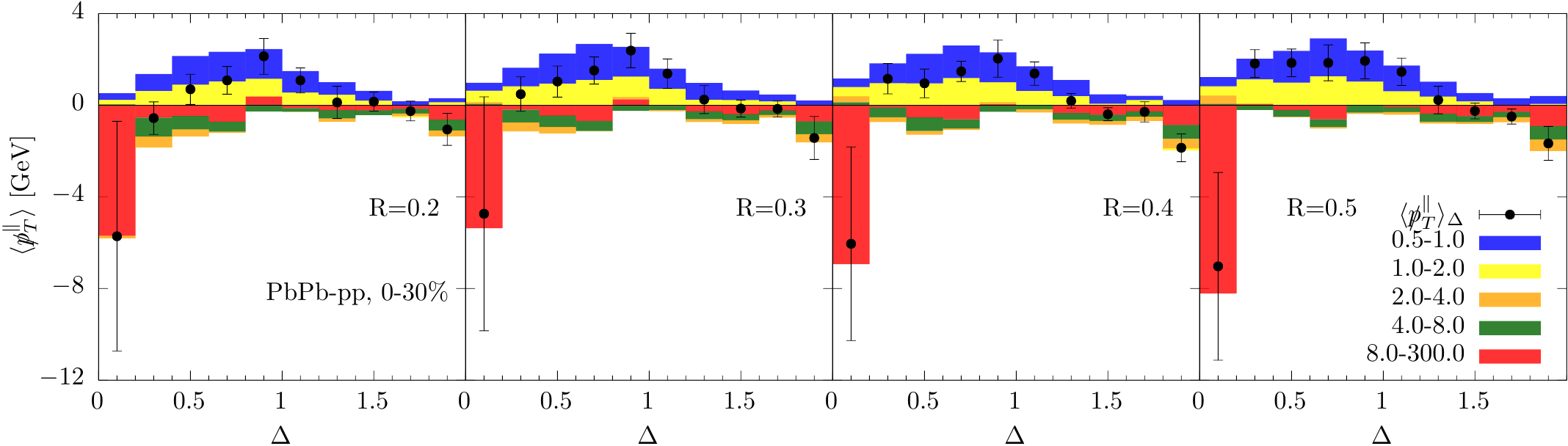}
 \includegraphics[width=1.01\textwidth]{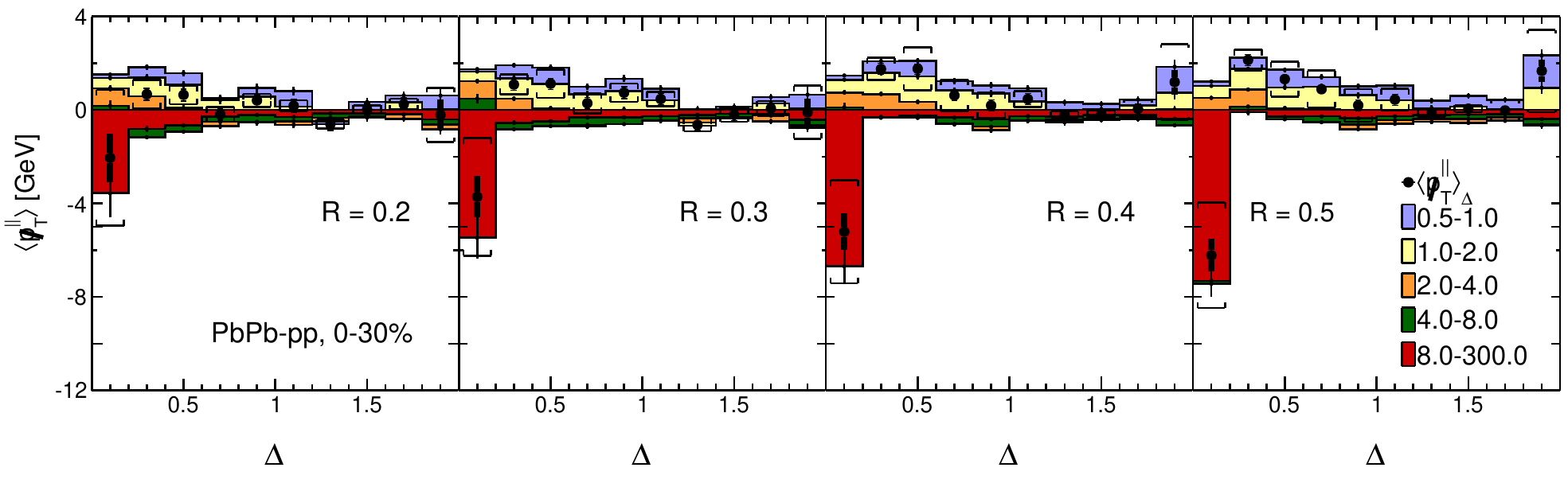}
  
  \caption{\label{Rdependence}Evolution in $R$ of the missing $\mpt$ observable, $A_J$ inclusive and sliced in $\Delta$, for the difference between PbPb and pp collisions. Top four panels are the results of our calculations with no contributions from
  the backreaction of the medium. Middle four panels are the results of our full calculations,
  in which the $\mpt$ imbalance  in hard particles due to jet quenching
  is balanced by that in soft particles coming from the wakes of the two jets.
  The $R=0.3$ panels are the difference between the PbPb results 
  shown in \fig{Fig:mptbacknoback} and the same for proton-proton collisions.
  Bottom four panels come from the analysis of experimental data by
  the CMS collaboration reported in Ref.~\cite{Khachatryan:2015lha}.
  There is some hint of $R$-dependence, despite the big error bars, especially in the first $\Delta$ bin. 
    }
  \label{fig7} 
\end{figure}

Finally, we turn to the dependence of the missing $\mpt$ 
distribution on the value of the parameter $R$ used in finding and reconstructing the jets, recalling that using a large $R$ results
in finding a sample of jets that have a larger average angular size. 
In \fig{Rdependence} we show the $\mpt$ distribution as a function of the angular separation $\Delta$ 
for different values of the jet reconstruction parameter. For the purposes of focusing on the effects of jet quenching and
in order to make a comparison with experimental data, we show the difference between PbPb and proton-proton collisions. 
As in the previous plots, the colored histograms show the contributions of tracks in different $\pt$-ranges,
 and the black points show $\mpt$ for tracks with all values of $p_T$ in a given $\Delta$-bin.  
 (As always when we show results from our calculations, the error bars include the uncertainties coming
 from the error bar on the experimental data point that we used to fix the $\aSC$ parameter
 in our hybrid model and from our estimate of the systematic uncertainty in our model
 that we make by varying the temperature $T_c$ below which we turn off parton energy loss.)
 As in \fig{Fig:mptbacknoback}, in the middle panels of  \fig{Rdependence} 
 momentum conservation is (approximately) recovered after summing over all $\Delta$ bins,
 although not precisely because we only use tracks with $\pt>0.5$~GeV and $|\eta|<3$ in the analysis.

As already mentioned, in this analysis the anti-$k_t$ reconstruction algorithm is only employed to determine the sample of events 
over which the $\mpt$ distribution is computed, as well as the jet and dijet axes in these events. All tracks in
the events are then used in the analysis, whether or not the reconstruction algorithm counted them as being within a jet.
By increasing the anti-$k_t$ parameter  $R$ while keeping the momentum cuts on the dijet selection
fixed, we include jets that are on average wider in angle and that include more, and hence softer, fragments.
%
%
Since wider jets containing a larger number of fragments lose more energy, 
the missing $\mpt$ in the hardest $p_T$ tracks in the smallest $\Delta$ bin grows in magnitude 
with increasing $R$. 
Despite  the large error bars  in our calculations and in data, this enhancement is present in both, as seen in 
\fig{Rdependence}. This is further evidence for the fact that wider jets lose more energy than narrower jets do.

Beyond the first $\Delta$ bin, we see in \fig{Rdependence} that the distribution of $\mpt$ is dominated
by softer particles with net $\mpt$ in the subleading jet direction.
The $\Delta$-dependence of the softest components of the $\mpt$ distribution shows very little sensitivity to $R$.  
This may be understood as a consequence of the wide angular region in which energy is recovered, controlled by \eq{onebody}. 
If we look carefully at the $\mpt$ distribution for the hardest particles in the $0.2<\Delta<0.8$ range of angles,
we do see an $R$-dependence in the results from our calculations and in data: the $\mpt$ imbalance in this kinematic regime
decreases with increasing $R$.  In our calculations and in the data, this imbalance comes from the proton-proton
collisions: there is no significant imbalance in the hard particles at these angles in the PbPb collisions for any $R$.  This
indicates that the effect is a sign of the presence of third jets at these angles.  When $R$ is small, a structure that
would have been reconstructed as a single jet if $R$ were larger can be reconstructed instead as two nearby
jets, which end up being counted as the subleading jet in a dijet and as a third jet.  The third jet results in an
imbalance in the $\mpt$ distribution for hard particles away from the dijet axis in the proton-proton collisions, 
but this weak third jet is greatly quenched in the PbPb collisions.  The result in the difference between PbPb
and proton-proton collisions is as seen in \fig{Rdependence}, and the effect is reduced when jets
are reconstructed with larger $R$.

Remarkably, for all the features in \fig{Rdependence} that we have described to this point
there is good qualitative agreement between the results of our calculations in the middle panels
and the experimental distributions reported in \cite{Khachatryan:2015lha} that we show in the lower panels.
We do wish to note two qualitative disagreements, however.  First, our results for the missing $\mpt$
distribution differ from the experimental results in the last bin in $\Delta$ which, we recall, corresponds
to the $\mpt$ imbalance for particles with all $\Delta>1.8$.  We have checked that this disagreement
is entirely due to a difference between our results and experimental results for proton-proton collisions;
we do not know how to interpret this particular difference between \pythia and experiment.   
Second, as in \fig{Fig:ajcentrality} we see that
our calculation of the $\mpt$ distribution of the particles coming from the backreaction of the plasma to the jets, 
(which we can discern by comparing the middle panels to the top panels) is lacking in ``orange particles''
with 2~GeV~$<\pt <4$~GeV relative to what we see in the data in the lower panels.  Correspondingly,
our calculation restores momentum conservation with a greater imbalance in the softest particles with
$\pt<2$~GeV than in the data.  Our calculation of the particles coming from the wakes in the
plasma has significantly fewer semi-hard particles and somewhat more soft particles.
We will discuss this disagreement further in the next Section.


Before we move on, it is important to look back at the way in which
we defined the new observable that we introduced at the end of Section~\ref{sec:Section2}
that, as we argued in our discussion around \fig{Fig:NewObs} and \fig{Fig:NewObsHad}, is particularly sensitive to the effects of transverse momentum
broadening.  In constructing the hadronic observable in \fig{Fig:NewObsHad}, we only used particles with 
5~GeV~$<\pt <10$~GeV. We can now see clearly why we chose 
{\it not} to use particles with $\pt <5$~GeV in this observable.  We 
want this observable to only be sensitive to (modifications of) the
jets themselves. We want it to be unaffected by the 
wake in the plasma and, more generally, by the energy lost from
the jet wherever that lost energy ends up.  In our calculation, that
lost energy dominates the missing $\mpt$ distribution for $\pt<2$~GeV;
in the data, it is clear that it is important for $\pt<4$~GeV.  In order
to avoid all the physics that is of interest to us here in Section 4,
in \fig{Fig:NewObsHad} in Section~\ref{sec:Section2} we ignored particles with $\pt<5$~GeV.

\section{\label{sec:discussion}Discussion and Outlook}

In this paper we have described the origin and observable consequences of two important effects upon implementing and describing them in our hybrid model of jet quenching: transverse momentum broadening, as the partons in the jet are kicked transverse to their direction of motion during their passage through the medium, and the backreaction of the medium to the momentum and energy deposited in it by the passage of the jet through it.

Our discussion of transverse momentum broadening is based on 
a Gaussian distribution of kicks in transverse momentum experienced by each parton in the jet.
This approximation captures the bulk of the distribution of transverse momentum-transfer but neglects rare scatterings by larger angles that impart more transverse momentum. Quite surprisingly, we have found that adding Gaussian momentum broadening has little impact on typical jet observables such as the jet spectrum, jet shapes and dijet acoplanarity, 
even when we choose unrealistically large values of 
$K\equiv \hat{q}/T^3$, the new second parameter that we add to our model to quantify the strength of the transverse momentum broadening. 
The reason why it is hard to see effects of broadening in these observables 
can be traced back to the parton energy loss and consequent jet quenching, described successfully by our hybrid model.
We find that in our hybrid model parton energy loss results in jets with a given energy observed in PbPb collisions being narrower than jets with the same energy in proton-proton collisions. This happens because jets that are wider in opening angle lose more energy than those that are narrower and because the jet energy spectrum is steeply falling, and has also been seen recently in entirely weakly coupled analyses~\cite{Milhano:2015mng} and in entirely strongly coupled analyses~\cite{Rajagopal:2016uip}.  This narrowing of jets with a given energy because of energy loss turns out to be a larger effect than the broadening of jets due to kicks to the transverse momentum of partons in the jets.  Also, for narrower jets the principal effect of transverse kicks is a change in the overall direction of the jet, rather than changes to its shape.  (As an extreme case, think of a jet containing only one parton.)
For a high energy jet the resulting acoplanarity is small, but for narrow jets the change to their shape is even smaller. (In the extreme case of a single parton, the shape of the resulting jet does not change at all if the direction of that parton changes.)  Although initially surprising, once understood these considerations make it clear that it will be exceedingly difficult to extract the value of $K$, or the jet quenching parameter $\hat q$, directly from the effect of transverse momentum broadening on traditional jet observables.  (In weakly coupled analyses but not in general, $\hat q$ is related to parton energy loss and so can be constrained indirectly via observables that are sensitive to energy loss.)

We have, however, found a new kind of jet shape ratio that is partially differential in the $\pt$ of the tracks entering its definition that does exhibit sensitivity to the value of $K$. 
In \fig{Fig:NewObs} we restrict the $\pt$-range of the partons entering the analysis 
to be $10<\pt^{\textrm{parton}}<20$ GeV, as among the options we have investigated it is for partons in this momentum 
range that we have found the most sensitivity to the value of $K$.  We then need a hadronic observable that is sensitive to the angular distribution of partons in this momentum range,
and in \fig{Fig:NewObsHad} we construct a partially differential jet shape ratio observable using hadrons in the $\pt$-range
$5<\pt^{\textrm{hadron}}<10$ GeV.  
The results we obtain indicate that experimental measurements of this observable, or similar observables, could be used to constrain the value
of the broadening parameter $K$, a key property of the medium.  Restricting attention to $\pt$ above a minimum value (like the 5 GeV we have chosen) is important because it makes the observable insensitive to the other physical effect that we describe in this paper, namely the reaction of the medium to the passage of the jet.  More on that below.  Restricting attention to $\pt$ below a maximum value (like the 10 GeV we have chosen) ensures that we are looking at jets containing many tracks and are looking at tracks that are not too energetic and thus can be affected visibly by transverse momentum broadening.


Although the successes of our model to date~\cite{Casalderrey-Solana:2014bpa,Casalderrey-Solana:2015vaa} motivate
the investments in improving it that we describe in this paper, in the long run the most instructive
use of a simple few-parameter model like our hybrid model is to discover and understand
the ways in which it fails to describe data.  Let us look ahead by imagining one particularly
interesting way in which this could happen in future.
Let us imagine a day when the measurement of 
observables like the partially differential jet shape ratio $\Nob$ plotted in \fig{Fig:NewObsHad} have been used to
discover experimental evidence for medium-induced transverse momentum broadening
and to constrain the value of $K$. What will the next step be?  At such a point in time,
the key next question will be to ask whether the event-by-event and parton-by-parton
distribution of the transverse kicks
contributing to an ensemble-averaged observable like $\Nob$ is indeed Gaussian.
This distribution is Gaussian by hypothesis in our hybrid model, meaning that our model will ultimately serve
as a baseline against which to look for evidence of a non-Gaussian distribution.
If the quark-gluon plasma in QCD were strongly coupled on all length scales, including short length scales,
the distribution of transverse kicks would be Gaussian for all values of the momentum
transfer, including large momentum transfers.  This is not what is expected in 
an asymptotically free theory like QCD.  If the strongly coupled liquid QCD plasma can be 
probed at high enough momentum transfer, the weakly coupled
quarks and gluons present within it at short length scales  should manifest themselves. 
Once the overall strength of the broadening $K$ has been constrained, the way
to look for point-like scatterers within the strongly coupled plasma will be to look
for rare events in which a parton within a jet is scattered by an angle that would be improbably
large if the distribution of transverse momentum kicks were a Gaussian with overall strength specified
by $K$~\cite{D'Eramo:2012jh,Kurkela:2014tla}.  
The hallmark of point-like scatterers is a distribution of momentum transfer that 
has a power-law tail at large momentum transfers.  
If in future our hybrid model, with its Gaussian distribution of transverse kicks with strength $K$, fails to describe data because
it neglects rare (but only power-law-rare) large angle scatterings, their discovery
may potentially be used to determine at which small length scale the fluctuations within strongly coupled quark-gluon plasma behave as weakly 
coupled quark and gluon quasiparticles.

We find that standard intrajet observables, like the conventional jet shape and jet fragmentation function, are sensitive to the presence of the wake in the plasma left behind by the passing jet.  Since the wake carries the momentum lost by the jet, the particles in the final state that come from the hadronization of the medium including the wake must have net momentum in the direction of the jet, meaning that some component of what experimentalists reconstruct as a jet, even after background subtraction, 
must in fact be composed of soft particles coming from the wake in the plasma. Motivated by expectations coming from explicit calculations of
the wake in strongly coupled plasma~\cite{Chesler:2007an,Chesler:2007sv} and by recent weakly coupled calculations in which the energy lost by a jet cascades into multiple soft particles~\cite{Blaizot:2013hx,Kurkela:2014tla,Blaizot:2014ula,Fister:2014zxa,Blaizot:2014rla,Blaizot:2015jea,Iancu:2015uja} we analyze the wake upon making the simplifying assumptions  that all the energy and momentum lost by the jet is incorporated into hydrodynamic motion and that the resulting perturbations to the temperature and velocity of the hydrodynamic fluid are small.
That is, we assume that the lost momentum and energy go into a hydrodynamic wake that is as thermalized as it can be by the time of freezeout, yielding only small corrections to the spectra of the hadrons in the final state, a final state that remembers nothing about the origin of this perturbation other than the momentum and energy dumped into the plasma.
%
%
In our calculation of the corrections to the hadronic spectra coming from the wake in the plasma, we have made further simplifying assumptions (for example assuming boost invariant longitudinal expansion and neglecting transverse expansion)  such that we have been able to determine the corrections to the hadron spectra in the final state coming from the hadronization of the wake in the plasma analytically, and without needing to incorporate any additional parameters into our model.
%
%
By comparing our computations with and without the effects of the backreaction of the medium, we have established that, indeed, these collective effects have important consequences for jet shapes, fragmentation function and missing-$\pt$ observables as discussed in Sections \ref{sec:obs} and Section \ref{sec:mpt}.

Including the effects of the backreaction of the medium results in enhanced jet shapes at large angles and enhanced fragmentation functions at soft momenta. These effects go in the directions favored by experimental data, but in our treatment they are not as large in magnitude as the data seem to require.
It should also be noted that these observables are not dominated by these effects.

The distribution of the missing-$\pt$ observable $\mpt$ constructed from dijet events is particularly interesting because at all but the smallest angles it is dominated by effects coming directly from the backreaction of the medium.  Even with our simplified approach to the backreaction of the medium and to hadronization, we obtain a 
rather good qualitative description of the shape, $\pt$-dependence, angular dependence, and
 $A_J$- and $R$-dependence of the $\mpt$ distributions, see \fig{Fig:ajcentrality} and \fig{fig7}.  This indicates that our simplified analysis of the backreaction of the medium is on the right track, and confirms the absolute necessity of including this physics if one aims to understand the soft and/or large-angle components of jets as reconstructed in heavy ion collisions.
%
%

The main discrepancy that we find between the results of our calculations and the experimentally measured $\mpt$ distributions 
is that our calculation does not yield enough semi-hard particles coming from the wake in the plasma in the $2<\pt<4$~GeV regime at angles less than $\sim 0.5$ radians from the jet axis.  Given that it has momentum conservation built into it, our calculation must overproduce particles in some other regime, and indeed we find that our wake results in somewhat more particles with $\pt<2$~GeV than indicated in the data, over an even broader range of angles.
Although we cannot be sure of this, we expect that the remaining discrepancies between our results and experimental data on jet shapes and fragmentation functions can also be traced to our calculation not producing enough particles with $2<\pt<4$~GeV.  

We are aware of three possible reasons for the discrepancy that we have found in the semi-hard momentum regime, for momenta well above $T$ but not so high that particles coming from the wake can be neglected.  First, we know that our perturbative expression (\ref{eq:deltaNG}) for the correction to the spectrum of particles coming from the plasma due to the wake  is not valid for $\pt$ well above $T$, and furthermore we know that because it is based on a linear approximation to an exponential the expression (\ref{eq:deltaNG}) necessarily underestimates particle production from the wake in this regime.  And, even if the modifications of the flow field induced by the wake are small their effect on the spectrum of particles produced at freezeout can be significant on the tail of the thermal distribution.
Going beyond our simplified expressions (\ref{eq:deltaNG}) and (\ref{onebody}) in order to improve upon our description of the semi-hard momentum regime 
requires taking account of the explicit space-time dependence of the hydrodynamic wake in the plasma~\cite{CasalderreySolana:2006sq}. 
It would be interesting to investigate the effect of the full flow profile as predicted by holographic computations of the hydrodynamic wake~\cite{Chesler:2007an,Chesler:2007sv} on the final particle production spectrum to see how this modifies our results in the semi-hard momentum regime.

The second possible reason for the discrepancy that we have found in the semi-hard momentum regime could be that the energy lost by the jet does not fully hydrodynamize. That is, a part of the discrepancy at semi-hard momenta between our results for the $\mpt$ distribution and the experimental results in \fig{Fig:ajcentrality} and \fig{fig7} could be a manifestation of non-equilbrium, and non-hydrodynamic, physics in quark-gluon plasma that has been disturbed by the passage of the jet.  
This possibility can be thought of in two ways that sound different but are operationally equivalent in the present context.  One way of describing this possibility is to say that the energy and momentum deposited by the jet in the plasma has not all had time to thermalize as fully as we assume in our simplified analysis.  This may be particularly relevant to energy deposited not long before the jet exits the droplet of plasma.  Another way of describing the same possibility is to say that our hybrid model neglects medium-induced gluon radiation:  in our model, the partons in the jet lose energy to a wake in the plasma and are kicked in transverse momentum but the branching structure of the jet shower is not modified in the way that happens in entirely weakly coupled treatments of jet quenching.  To the extent that the medium-induced radiated gluons themselves radiate and cascade down to become soft gluons that form a wake in the plasma, it makes no difference that we neglect medium-modification of the branching structure of the jet shower: our model includes the loss of energy from the partons in the jet and the wake in the plasma.  Where these details will matter is if radiated gluons do not thermalize as fully as we assume, as may be particularly relevant to gluons radiated not long before the jet exits the droplet of plasma.  The first description uses words that arise in strongly coupled analyses of jet quenching.  The second description uses words that arises in weakly coupled analyses.  The two are operationally equivalent here.  

Another physical process which has not been incorporated into our approach is the effect of imperfect resolution. 
In our hybrid model as currently implemented, as soon as one parton in the jet shower splits into two, the two daughter partons are treated as separate losers of energy.
That is, the medium can resolve the separation of one parton into two as soon as it happens, when the two daughters are still very close to each other.  This cannot be correct. In reality, the medium can only resolve two partons as two separate losers of energy after they have separated by some resolution length that can reasonably be expected to be of order the Debye length of the plasma.
%
The main effect that the inclusion of the effect of imperfect resolution would have in our model is that semi-hard partons within jets, in particular within narrow jets, would start losing their energy rapidly on their own later than in our present model, and so would end up less quenched.  Hence, our neglect of these effects of imperfect resolution is a third possible reason for the discrepancy that we have found in the semi-hard momentum regime. 


It would be very interesting to assess how much of the discrepancy that we find in the semi-hard momentum regime is hydrodynamic (and due to our simplifications), or arises because some of the energy lost by the jet does not fully hydrodynamize, or is related to our neglect of resolution effects. Each explanation is quite interesting in its own right and deserves investigation.  We leave this to future work, but note here that this example serves very well to make our point that the most instructive uses of a few-parameter model like the one we have constructed are to discover, and subsequently understand, the ways in which it fails to describe data.  This example also serves to highlight the full power of the suite of experimental data on intrajet structure now available, including jet shapes, fragmentation functions, and the
missing-$\pt$ observables.
%
%
We encourage theorists pursuing other approaches to calculate all these observables and compare to data.  Consider for example the weakly coupled calculations in
which energy is lost to radiated gluons which themselves are quickly degraded down to soft gluons with momenta of order the 
temperature~\cite{Blaizot:2013hx,Kurkela:2014tla,Blaizot:2014ula,Fister:2014zxa,Blaizot:2014rla,Blaizot:2015jea,Iancu:2015uja}, 
which leads to rapid hydrodynamization of the emitted energy~\cite{Iancu:2015uja}.
These approaches yield qualitative agreement with the  $\mpt$-distribution that we have shown in \fig{Fig:mptbacknoback}~\cite{Blaizot:2014ula}, but fail to reproduce the soft enhancement of fragmentation function~\cite{Mehtar-Tani:2014yea}. 
It would be very interesting to see how well they predict the $A_J$-dependence and the $R$-dependence of 
the $\mpt$-distribution as in \fig{Fig:ajcentrality} and \fig{fig7}, in particular in the soft and semi-hard momentum regimes.
Weakly coupled approaches in which it is assumed that the emitted gluons do not further interact at all are able to reproduce the enhancement of jet shapes at large 
angles~\cite{Chien:2015hda}; it would be interesting to analyze the $\mpt$-distributions in these approaches.  

Although the physics of the wake in the plasma is of interest in its own right, if in the longer term our goal is to quantify broadening, 
determine the value of the key medium property $K$, and ultimately to see rare but not-too-rare large angle scattering of partons within jets that may allow us to 
see the length scale at which strongly coupled quark-gluon plasma emerges from the fluctuations at even smaller length scales that behave as weakly coupled quark and gluon quasiparticles, then we will want to focus on observables sensitive to the angular distribution of 10-20 GeV partons within jets as in \fig{Fig:NewObsHad}, where the wake makes no contribution.


\appendix
\section{\label{tkk}Transverse Kicks Kinematics}
As a parton propagates though the plasma, it receives a random transverse kick in the local fluid rest frame that we shall denote by $q$.
In this Appendix, we provide a precise specification of what we mean by this.
Denoting the momenta of the parton in the fluid rest frame before and after  a transverse kick
 by $P^\mu=E_F (1,\w_F)$ and  $P^{' \mu}=E_F (1,\w'_F)$, 
and assuming that the transverse kick serves only to change the direction of $\w_F$, meaning
$\w'^2_F=\w^2_F$, 
the momentum after the kick is given by 
\be
\w'_F=\sqrt{1-\frac{q^2}{E^2_F \w_F^2}} \w_F+ \frac{q}{E_F}  \e_\perp \, ,
\ee
where $\e_\perp$ is a vector perpendicular to the parton velocity  in the rest frame of the fluid.
Note that the virtuality of the parton has not changed. And, in the fluid rest frame, neither has its energy.

Since our computation is performed in the collision center-of-mass frame, we need to 
express the  acquired momentum in that frame.  Let us define the four-vector $W=P/E$, with $E$ the energy of the parton in the collision frame. 
Denoting the fluid velocity in the collision center-of-mass frame by  $u=\gamma_F \left(1,\v\right)$, the energies in the fluid and collision 
frames are related by  
$E_F= E \,\gamma_F \left(1-\w \v\right)$, with $\w$ the velocity of the parton in the collision frame. We 
 construct a four-vector transverse to the fluid velocity  
\be
W_T=\frac{1}{ W^0_F}  \left(W - (W\cdot  u) u \right) \, ,
\ee
which, in the fluid frame, has components $W_T= \left(0, \w_F \right)$. We use this vector to express the change in four-momentum 
associated with the kick as
\be
P^{' \mu}=P^\mu - \beta E_F W_T^{\mu} + q e^\mu_\perp \, \, \,,\qquad\beta\equiv 1- \sqrt{1-\frac{q^2}{E^2_F \w_F^2}}  \, ,
\ee
where $E_F^2 \w_F^2 = E^2_F-E^2(1-\w^2)$  and the four-vector $e^\mu_\perp$ satisfies the conditions
\be
u \cdot \, e_\perp=0 \, , \qquad W\cdot e_\perp=0 \, , \qquad e_\perp^2=-1\,.
\ee
It is then possible to show that  $e_\perp$  can be written as a linear combination of two orthogonal vectors satisfying the conditions
\be
e_1^\mu=(0,\frac{\w \times \v}{\left |\w \times \v\right |})\, , \qquad  e_2^\mu= \frac{1}{\sqrt{N}} \left( l_2^\mu +\alpha \, W_\perp^\mu\right) \,,
\ee
with
\be
l_2^\mu=(0,\frac{\w}{\left | \w\right |}\times\frac{\w \times \v}{\left |\w \times \v\right |}) \,, &\qquad& W_\perp= W - \frac{W^2}{u\cdot W} u \, ,
\\
\alpha= -\frac{(l_2 \cdot u) \, (u\cdot W)}{(u \cdot W)^2-W^2} \, , &\qquad& N= \frac{ (u\cdot W)^2 - W^2 (1 + (l_2\cdot  u)^2)}{(u\cdot  W)^2-W^2}\ .
\ee
These expressions allow us to determine the change in momentum in terms of collision frame quantities alone, once $q$ and the angle in the $(e_1,e_2)$ plane have been chosen. After each time interval $dt$ in the collision frame, we select a random value of $q$ according to a Gaussian probability distribution with width $\Delta Q_\perp^2= \hat q \, d t_F$, where the relation between $d t_F$ and $d t$ is  $dt_F= dt \,\gamma_F \left(1-\w \v\right)$.  The angle is chosen randomly with a uniform distribution. This procedure is repeated for each parton in the shower as long as it is in a location where the temperature of the expanding cooling hydrodynamic fluid still satisfies $T>T_c$.  

The kick that the parton receives is transverse only in the local fluid rest frame.  In the collision center-of-mass frame,
the kick has a transverse component but it also has a longitudinal component which results in energy loss given by
\be
\Delta E= -\beta E \gamma^2_F ({\bf v} {\bf w}-v^2) - q^{(2)}   \frac{ \left| \w\right| \left| \w \times \v \right|}{ \w^2-\v \w} \,, 
\ee
where $q^{(2)}$ is the component of the momentum transfer along the $e_2$ direction. (Note that this formula is badly behaved when the velocities of the particle and the fluid are identical, since in that case it is not possible to define the transverse direction in the fluid frame. We have tested that in our simulation this does not occur over the whole propagation of the parton in the plasma.)

\section{\label{app:bgsubs} Background Subtraction Procedure}

As we have described in Section~\ref{sec:Section3}, in order to analyze the observable
consequences of the wake in the hydrodynamic medium, we need to add a background
of particles that reproduces the measured particle yields and spectra so that we can
then incorporate the effects of the wake as a perturbation on these spectra.
So, our \pythia simulations of jet showers, modified as in our
hybrid model, are now embedded in a (perturbed) background, meaning
that we must run a background subtraction algorithm.
We have used an iterative noise/pedestal subtraction procedure similar to the one implemented by 
the experimental collaborations. Since the algorithms employed by ATLAS and CMS are different, 
we have employed different version of the procedure for the analysis of observables for jets with $p_T>100$~GeV, where
we follow CMS~\cite{Kodolova:2007hd}, 
and for the observables that cover jets with  $p_T<100$~GeV, where we follow ATLAS~\cite{Aad:2012vca}.
See the Figures in Section~\ref{sec:Section3} for the results we have obtained after following
these procedures, procedures which we detail below. 
 
 For high $p_T$ jet observables, our procedure follows the following steps:
\begin{itemize}
\item[$(i)$] Discretize the $(\eta,\phi)$ space in cells of size $0.091 \, \times \, 0.087$. Sum the transverse energy $E_T$ of all particles falling into the same cell.
\item[$(ii)$] Compute the average transverse energy and the standard deviation 
in the transverse energy for all the cells in a strip with a given
rapidity,  {\it i.e.}~$\left\langle E_T(\eta)\right\rangle $ and $\sigma(\eta)\equiv\sqrt{\left\langle E_T^2(\eta)\right\rangle -\left\langle E_T(\eta)\right\rangle ^2}$.
\item[$(iii)$] For each cell, subtract the average for the strip in which the cell is found 
and subtract a contribution proportional to the standard 
deviation for that strip, with the proportionality constant $\beta$ a parameter that we shall return to below. If the result is negative, set it to zero instead. That is, for the $i$'th cell in the strip compute: 
\end{itemize}
\begin{equation}
\hat{E}_T^i\equiv\textrm{max} \, \left[ E_T^i-\left\langle E_T(\eta)\right\rangle -\beta \, \sigma(\eta), \, 0 \right] \,
\end{equation}
\begin{itemize}
\item[$(iv)$] Run the anti-$k_t$ clustering algorithm using all cells 
whose $\hat{E}_T^i$ is
different from zero. 
Each such cell is given to the anti-$k_t$ algorithm as a null four-vector with transverse momentum $\hat{E}_T^i$ 
located at an $\eta$ and $\phi$ corresponding to the geometric center of the cell.
\item[$(v)$] Repeat step $(ii)$ excluding all cells that the anti-$k_t$ algorithm
has already identified as belonging to a jet with transverse energy above $E_T^{\rm{cut}}$.
Jets with energy above this cut are considered signal jets. We now make a second pass,
re-evaluating the background outside these jets.
\item[$(vi)$] Repeat the background subtraction $(iii)$ and the jet finding step$(iv)$ using the values obtained in previous step. 
The jets found by the anti-$k_t$ algorithm on this second pass, which may have
transverse energies above or below $E_T^{\rm{cut}}$, are added to the collection of signal jets
for this event.
\end{itemize}
\noindent
This procedure involves choosing two parameters, namely $\beta$ and $E_T^{\rm{cut}}$. The best choice for these
parameters will in general depend on the anti-$k_t$ reconstruction parameter $R$, the jet $p_T$ range under study,
and the magnitude of the cell-to-cell fluctuations of the background. 
The factor $\beta$ controls the effect of background fluctuations. For a homogeneous background with no fluctuations, 
one should choose $\beta=0$.
For a cell-to-cell fluctuating background, one needs to increase $\beta$ accordingly, 
at the risk of potentially removing some of the signal of interest. 
The value of $E_T^{\rm{cut}}$ determines 
whether a group of cells corresponds to signal and should therefore be excluded from the background estimation.
The choice of these two parameters 
should be guided by the criterion that the $p_T$ of a jet that was artificially embedded into the background 
and then reconstructed upon performing the background subtraction procedure
above is, on average, as close as possible to the original $p_T$ of the artificial jet,
within 2\% for jets with $p_T \sim 100 \,\rm{GeV}$.  This is referred to as ensuring that
one 
has a good jet energy scale (JES).  We have done this test using \pythia jets.
We choose $E_T^{\rm cut}=30$~GeV throughout, and have then
picked a value of $\beta$ for each centrality and each value of the anti-$k_t$ parameter $R$
so as to opimize the JES.  For events with 0-10\% centrality, we choose $\beta=0.48, 0.72, 1.00, 1.17$ for 
$R=0.2, 0.3, 0.4, 0.5$
while for events with 10-30\% centrality the corresponding values that we choose are $\beta=0.42, 0.71, 0.94, 1.13$.

For jets in the lower $p_T$ region, with $p_T<100$~GeV, the procedure that we adopt
begins with the same discretization of $(\eta,\phi)$ space and then continues as follows:
\begin{itemize}
\item[$(i)$] Reconstruct jets using the anti-$k_t$ algorithm for many different values of the reconstruction parameter $R$, in each case
using the uncorrected the $E_T$ in each cell in $(\eta,\phi)$, introducing null four-vectors with transverse momentum $E_T$
with a $\eta$ and $\phi$ corresponding to the geometric center of the cell.  Because in this procedure we start
by reconstructing jets before subtracting background, after subtracting background we must then also remove combinatorial jets.
\item[$(ii)$]Select a set of seed jets with $R=0.2$ which have at least one constituent cell  with $E_T> 3$ ~GeV and whose cell with the maximum transverse energy $E_T^{\textrm{max}}$ satisfies $E_T^{\textrm{max}} > 4\left\langle E_T \right\rangle$, where  $\left\langle E_T \right\rangle$  is the average transverse energy of the cells 	within the seed jet.
\item[$(iii)$] Compute the average transverse energy of each rapidity strip, {\it i.e.}~$\left\langle E_T(\eta)\right\rangle$, but excluding all those cells that belong to a seed jet.  
\item[$(iv)$] Subtract from each cell the average transverse energy for that rapidity strip,  $\hat{E}_T^i=E_T^i-\left\langle E_T(\eta)\right\rangle$. 
This subtraction is applied to all cells in a strip, including those in a seed jet, meaning that it modifies the $p_T$ of the seed jets.
This is the first of two subtractions.
\item[$(v)$] The second subtraction, below, will employ a subset of the seed jets with $R=0.2$ from above whose
transverse momenta after the first subtraction above satisfy $p_T^{\rm{jet}}> 25$~GeV. It will also employ a set of seed ``track jets'' with $p_T^{\rm jet}> 10$ GeV. Track jets are built using only charged tracks with $p_T^{\rm{track}}> 4$~GeV and are reconstructed using the anti-$k_t$ algorithm with $R=0.4$.
\item[$(vi)$] Recompute the average transverse energy $\left\langle E'_T(\eta)\right\rangle$, now excluding all those cells that lie within $\Delta r=0.4$ from the axis of any of the seed jets or seed track jets, where $\Delta R$ is the angular distance in $(\eta,\phi)$ space.
\item[$(vii)$] Subtract the new average energy $\left\langle E'_T(\eta)\right\rangle$
from all cells 
and update the kinematics of all jets reconstructed with all values of $R$. Keep 
only those jets with $E_T> 20$ GeV.
\item[$(viii)$] In order to suppress the contribution of combinatorial jets, we impose that the reconstructed jets have to lie within $\Delta r=0.2$ of a seed track jet (defined above) with $p_T > 7$ GeV, making it probable that the signal jets include one or a few hard particles in them.
\end{itemize}
\vspace{0.2cm}

Depending on which observables we want to compute and compare to data, 
we will need to apply one or more further corrections to the jets in PbPb events
that we have extracted via
the background subtraction and jet reconstruction procedures detailed above.

For all observables, we will follow the experimental analyses and apply a JES correction
which takes into account the remaining average disagreement between 
the energy of a proton-proton jet from \pythia and the energy of that jet after it has
been embedded into one of our PbPb events and reconstructed as
above. 
For the first of the two procedures above, the one that we employ
for jets with $p_T>100$~GeV, the JES correction is less than 2\%, as
we noted above.  For the procedure that we employ for
jets with $p_T<100$~GeV, the JES correction can be as large as 12\% or 20\%
for $R=0.4$ or $R=0.5$ jets with $p_T=40$~GeV.

If we wish to compare to experimental data that has not been unfolded,
as we shall do in the case of several high-$\pt$ observables measured by CMS,
we must include an additional
jet energy resolution (JER) correction. 
We start by doing a Monte Carlo study in which
we insert jets from \pythia into an event,
 fit the distribution of the energy reconstruction efficiency (reconstructed jet energy over generator level jet energy as a function of generator level jet energy) 
 for each jet $p_T$ bin with a Gaussian and extract the corresponding standard deviation $\sigma$, which tells us how much the reconstructed jet energy varies from event to event and jet to jet. (This $\sigma$ is unrelated to the $\sigma$ in the background subtraction procedure above; both are conventionally referred to as $\sigma$.)
 When we do this calculation using the events in our model including its simplified background, 
 we denote what we obtain by $\sigma_{\rm model}$.  CMS has
 done this analysis on data from heavy ion collisions, obtaining $\sigma_{\rm LHC}$ which 
 they have tabulated as a function of jet $\pt$ in Ref.~\cite{Chatrchyan:2012gt}.
We are now ready to correct for the fact that the JER in the real background measured by CMS 
is different from that in our simplified model background. 
We do so by smearing the jet energies in our model calculation
 with a Gaussian whose width corresponds to 
$\sigma_{\rm extra}\equiv \sqrt{\sigma_{\rm LHC}^2-\sigma_{\rm model}^2}$.

If we wish to compare to data that has been unfolded, we will perform the simplest version of the so called 
bin-by-bin unfolding. This affects the jet spectrum measurements. The correction applied consists of
multiplying the spectrum of the medium-modified jets that we have reconstructed via one of the two procedures
above by 
the ratio of two other spectra: the jet spectrum obtained directly from \pythia divided
by the jet spectrum obtained after embedding \pythia jets into heavy ion collision events
and reconstructing them.

\acknowledgments

We thank J.~Brewer, P.~Chesler, Z.~Hulcher, P.~Jacobs, R.~Kunnawalkam Elayavalli, Y.-J.~Lee, Y.~Mehtar-Tani, A.~H.~Mueller, G.~Qin, A.~Sadofyev, K. Tywoniuk,  W.~van~der~Schee and K.~Zapp
for  helpful discussions. 
We also thank J. Balewski, M. Goncharov and C. Paus
for their assistance with computing.
JCS is a Royal Society University Research Fellow.
The work of JCS was also supported by a Ram\'on~y~Cajal fellowship. 
The work of JGM was partly supported by Funda\c c\~ao para a Ci\^encia e a Tecnologia (Portugal) under project CERN/FIS-NUC/0049/2015 and 
contract ``Investigador FCT - Development Grant''.
DP acknowledges the hospitality of the Center for Theoretical Physics, MIT, where part of this work was completed.
JCS and DP acknowledge funding from grants FPA2013-46570 and MDM-2014-0369 of ICCUB (Unidad de Excelencia `Mar\'ia de Maeztu') 
 from the Spanish MINECO,  from grant 2014-SGR-104 from the 
Generalitat de Catalunya
and from the Consolider CPAN project. KR acknowledges the hospitality of the CERN Theory division during the completion of this work.
The work of KR was supported by the U.S. Department of Energy under grant Contract
Number DE-SC0011090.

\end{document}